\documentclass[a4paper,11pt]{article}
\pdfoutput=1 

\usepackage{jheppub} 
\usepackage{hyperref}
\usepackage{caption}
\usepackage{subcaption}
\usepackage[T1]{fontenc} 
\usepackage{listings}
\usepackage{tabularx}
\usepackage{cleveref}
\usepackage{tikz}
\usepackage{float}
\usepackage{booktabs}
\usepackage{shuffle}

\def\eps{\epsilon}

\def\tW{\widetilde{W}}
\def\ie{\begin{equation}\begin{aligned}}
\def\fe{\end{aligned}\end{equation}}

\definecolor{mybrown}{RGB}{171,6,44}
\def\omegaTwoLadd{{\begin{tikzpicture}[scale=1.50,baseline=(current bounding box.south)]

    \draw[mybrown, very thick] (0,0) -- (1,0);

   \filldraw (0,0) circle (2pt);
   \filldraw (1,0) circle (2pt);

\end{tikzpicture}}}
\def\omegaTwoLaddfrz{{\begin{tikzpicture}[scale=1.50,baseline=(current bounding box.south)]

    \draw[mybrown, very thick] (0,0) -- (1,0);

   \draw[fill=white] (0,0) circle (2pt);
    \draw (0,0) -- (45:2pt);
    \draw (0,0) -- (135:2pt);
    \draw (0,0) -- (-45:2pt);
    \draw (0,0) -- (-135:2pt);
   \filldraw (1,0) circle (2pt);

\end{tikzpicture}}}
\def\omegaThreeLadd{{\begin{tikzpicture}[scale=1.50,baseline=(current bounding box.south)]

    \draw[mybrown, very thick] (0,0) -- (0.6,0) -- (1.2,0);

   \filldraw (0,0) circle (2pt);
   \filldraw (0.6,0) circle (2pt);
    \filldraw (1.2,0) circle (2pt);
\end{tikzpicture}}}
\def\omegaThreeLaddfrz{{\begin{tikzpicture}[scale=1.50,baseline=(current bounding box.south)]

    \draw[mybrown, very thick] (0,0) -- (0.6,0) -- (1.2,0);

   \draw[fill=white] (0,0) circle (2pt);
    \draw (0,0) -- (45:2pt);
    \draw (0,0) -- (135:2pt);
    \draw (0,0) -- (-45:2pt);
    \draw (0,0) -- (-135:2pt);
   \filldraw (0.6,0) circle (2pt);
    \filldraw (1.2,0) circle (2pt);
\end{tikzpicture}}}
\def\omegaThreeLaddMidfrz{{\begin{tikzpicture}[scale=0.8,rotate=90,baseline=(current bounding box.center)]
    \coordinate (A) at (0.8,0);
    \coordinate (B) at (-0.8,0);
    \coordinate (C) at (0,1); 
    
    \draw[mybrown, very thick] (B) -- (C);
    \draw[mybrown, very thick] (C) -- (A);

 \filldraw (A) circle (3.5pt);

    \filldraw (B) circle (3.5pt);

    \draw[fill=white] (C) circle (3.5pt);
    \draw (C) -- ++(45:3.5pt);
    \draw (C) -- ++(135:3.5pt);
    \draw (C) -- ++(-45:3.5pt);
    \draw (C) -- ++(-135:3.5pt);
\end{tikzpicture}}}
\def\omegaThreeLoop{{\begin{tikzpicture}[scale=0.8,baseline=(current bounding box.center)]
    \coordinate (A) at (1,0);
    \coordinate (B) at (-1,0);
    \coordinate (C) at (0,1); 
    
    \draw[mybrown, very thick] (A) -- (B);
    \draw[mybrown, very thick] (B) -- (C);
    \draw[mybrown, very thick] (C) -- (A);

 \filldraw (A) circle (3.5pt);

    \filldraw (B) circle (3.5pt);

    \filldraw (C) circle (3.5pt);
\end{tikzpicture}}}
\def\omegaThreeLoopfrz{{\begin{tikzpicture}[scale=0.8,rotate=90,baseline=(current bounding box.center)]
    \coordinate (A) at (0.8,0);
    \coordinate (B) at (-0.8,0);
    \coordinate (C) at (0,1); 
    
    \draw[mybrown, very thick] (A) -- (B);
    \draw[mybrown, very thick] (B) -- (C);
    \draw[mybrown, very thick] (C) -- (A);

 \filldraw (A) circle (3.5pt);

    \filldraw (B) circle (3.5pt);

    \draw[fill=white] (C) circle (3.5pt);
    \draw (C) -- ++(45:3.5pt);
    \draw (C) -- ++(135:3.5pt);
    \draw (C) -- ++(-45:3.5pt);
    \draw (C) -- ++(-135:3.5pt);
\end{tikzpicture}}}
\def\dlog{{\rm dlog}}
\def\i{\textup{i}}
\def\laddonepic{{\begin{tikzpicture}[baseline=(current bounding box.south)]
 
    \draw[mybrown, thick] (0,0) -- (0.3,0);
 
    \draw[fill=white] (0,0) circle (2pt);
    \draw (0,0) -- (45:2pt);
    \draw (0,0) -- (135:2pt);
    \draw (0,0) -- (-45:2pt);
    \draw (0,0) -- (-135:2pt);
 
    \filldraw (0.3,0) circle (2pt);
\end{tikzpicture}}} 

\def\laddtwopic{{\begin{tikzpicture}[baseline=(current bounding box.south)]

    \draw[mybrown, thick] (0,0) -- (0.3,0) -- (0.6,0);

    \draw[fill=white] (0,0) circle (2pt);
    \draw (0,0) -- (45:2pt);
    \draw (0,0) -- (135:2pt);
    \draw (0,0) -- (-45:2pt);
    \draw (0,0) -- (-135:2pt);

   \filldraw (0.3,0) circle (2pt);
    \filldraw (0.6,0) circle (2pt);
\end{tikzpicture}}}

\def\laddladdpic{{\begin{tikzpicture}[baseline=(current bounding box.south)]
    \draw[mybrown, thick] (0,0) -- (0.3,0) -- (0.6,0);
    \filldraw (0,0) circle (2pt);
    \draw[fill=white] (0.3,0) circle (2pt);
    \draw (0.3,0) -- ++(45:2pt);
    \draw (0.3,0) -- ++(135:2pt);
    \draw (0.3,0) -- ++(-45:2pt);
    \draw (0.3,0) -- ++(-135:2pt);
    \filldraw (0.6,0) circle (2pt);
\end{tikzpicture}}}

\def\laddthreepic{{\begin{tikzpicture}[baseline=(current bounding box.south)]

    \draw[mybrown, thick] (0,0) -- (0.3,0) -- (0.6,0) -- (0.9,0);

    \draw[fill=white] (0,0) circle (2pt);
    \draw (0,0) -- (45:2pt);
    \draw (0,0) -- (135:2pt);
    \draw (0,0) -- (-45:2pt);
    \draw (0,0) -- (-135:2pt);

   \filldraw (0.3,0) circle (2pt);
    \filldraw (0.6,0) circle (2pt);
    \filldraw (0.9,0) circle (2pt);
\end{tikzpicture}}}

\def\looppic{{\begin{tikzpicture}[baseline=(current bounding box.center)]
    \coordinate (A) at (0,0);
    \coordinate (B) at (0,0.3);
    \coordinate (C) at (-0.2,0.15); 
    
    \draw[mybrown, thick] (A) -- (B);
    \draw[mybrown, thick] (B) -- (C);
    \draw[mybrown, thick] (C) -- (A);

    \draw[fill=white] (C) circle (2pt);
    \draw (C) -- ++(45:2pt);
    \draw (C) -- ++(135:2pt);
    \draw (C) -- ++(-45:2pt);
    \draw (C) -- ++(-135:2pt);

    \filldraw (B) circle (2pt);

    \filldraw (A) circle (2pt);
\end{tikzpicture}}}

\newcommand{\p}[1]{(\ref{#1})}

\newcommand \vev [1] {\langle{#1}\rangle}
\newcommand \ket [1] {|{#1}\rangle}
\newcommand \bra [1] {\langle {#1}|}

\newcommand{\pa}{\partial}
\newcommand{\ep}{\epsilon}

\renewcommand{\a}{\alpha}

\newcommand{\da}{{\dot\alpha}}

\newcommand{\tr}{\mbox{tr}}

\newcommand{\la}{\langle}
\newcommand{\ra}{\rangle}

\usepackage{ragged2e}

\title{Positivity properties of five-point two-loop Wilson loops with Lagrangian insertion}

\author[a]{Dmitry Chicherin}
\author[b]{Johannes Henn}
\author[c]{Jaroslav Trnka}
\author[b]{Shun-Qing Zhang}

\affiliation[a]{LAPTh, Université Savoie Mont Blanc, CNRS, B.P. 110, F-74941 Annecy-le-Vieux, France}

\affiliation[b]{Max-Planck-Institut für Physik, Werner-Heisenberg-Institut, Boltzmannstr. 8, 85748 Garching, Germany}

\affiliation[c]{Center for Quantum Mathematics and Physics (QMAP), University of California, Davis, 95616 CA, USA}

\emailAdd{chicherin@lapth.cnrs.fr}
\emailAdd{henn@mpp.mpg.de}
\emailAdd{trnka@ucdavis.edu}
\emailAdd{sqzhang@mpp.mpg.de}

\preprint{
\begin{tabular}{l}MPP-2024-194\\
LAPTH-052/24
\end{tabular}}

\abstract{In this paper we discuss the geometric integrand expansion of the five-point Wilson loop with one Lagrangian insertion in maximally supersymmetric Yang-Mills theory. We construct the integrand corresponding to an all-loop class of ladder-type geometries. We then investigate the known two-loop observable from this geometric viewpoint. 
To do so, we evaluate analytically the new two-loop integrals corresponding to the negative geometry contribution, using the canonical differential equations method. 
Inspecting the analytic result, we present numerical evidence that in this decomposition, each piece has uniform sign properties, when evaluated in the Amplituhedron region. Finally, we present an alternative bootstrap approach for the ladder-type geometries. We find that certain minimal bootstrap assumptions can be satisfied at two loops, but lead to a contradiction at three loops. This suggests to us that novel alphabet letters are required at this loop order. Indeed studying planar three-loop Feynman integrals, we do identify novel pentagon alphabet letters.}

\begin{document} 
\maketitle

\newpage

\section{Introduction}

Scattering amplitudes are central ingredients in the description of particle interactions, for example at collider experiments. Starting from the Lagrangian of a quantum field theory Feynman rules provide a definition of scattering amplitudes in perturbation theory. However, in recent years, alternative ways of thinking about scattering amplitudes have been found. In fact, a host of new formulations and surprising dualities are known (for reviews, see \cite{Arkani-Hamed:2022rwr,Travaglini:2022uwo}). This is interesting for both conceptual and practical reasons. 

One of the new formulations takes a geometric starting point. Based on Hodges' initial observation that certain tree-level six-particle amplitudes can be viewed as the canonical form of a polytope defined in kinematic space \cite{Hodges:2009hk}, Arkani-Hamed and Trnka proposed the Amplituhedron \cite{Arkani-Hamed:2013jha}, which applies to all planar scattering amplitudes in maximally supersymmetric Yang-Mills theory (sYM). Their finding defines tree-level amplitudes, and loop-level integrands in that theory, as canonical forms associated to the Amplituhedron geometry.

Once one thinks of (the integrand of) a scattering amplitude as (the canonical form of) some geometric object, it becomes natural to consider triangulations, as well as other ways of decomposing that object in terms of smaller building blocks. One example are the Britto-Cachazo-Feng-Witten recursion relations \cite{Britto:2005fq}, which may be thought of as such a triangulation. Similarly, there are different possible decompositions and representations of loop-level integrands that may be derived from geometry. The important example of that are \emph{local triangulations} where the individual building blocks are local integrands without any spurious poles \cite{Arkani-Hamed:2010wgm,Arkani-Hamed:2014dca,Herrmann:2020oud,Herrmann:2020qlt}. In that case the triangulation is external, so the individual terms do overlap but regions outside the Amplituhedron cancel.

An important conceptual and practical question, once a loop integrand is known, is that of carrying out the loop integrations. This gives rise, in general, to transcendental functions (of the kinematic variables). While there are well-developed beautiful techniques for Feynman integral computations, the latter are often one of the bottlenecks of state-of-the-art perturbative computations. We think that leveraging the underlying positive geometry properties when evaluating the integrals could lead to significant progress. 

A technical obstacle is that scattering amplitudes typically have infrared divergences. Although those are known in principle, and only the (suitably-defined) finite part of a given scattering amplitude is truly new, dealing with the infrared-divergent parts of the amplitudes is an important practical concern. Since our main focus in this paper is on exploring properties of transcendental functions associated to positive geometries, we choose as objects of study directly a suitably defined finite version of a scattering amplitude. 
(We work in maximally supersymmetric Yang-Mills theory, so the scattering amplitudes do not require UV renormalization.)

The objects we study in this paper can be conceptualized in different ways. One definition is as that of the correlation function of a Wilson loop -- defined on polygonal contours -- with a Lagrangian, normalized by the vacuum expectation value of the Wilson loop. Thanks to taking the ratio, this object is free from divergences. 
This object is related (at least, conjecturally) to correlation functions of local operators, and to scattering amplitudes in that theory. A good way of thinking about this triality is in terms of the {\it integrands} of the three objects. To define an integrand of an $L$-loop observable, one uses the Lagrangian insertion technique. This naturally gives a formulation where the integrand is given by a rational function. Assuming the triality, if one starts with the $(L+1)$-loop {\it integrand} of the logarithm of the maximally-helicity-violating (MHV) amplitude, and performs $L$ of the integrations, then this is equivalent to the above ratio of Wilson loops.

In summary, Wilson loops with a Lagrangian insertion are finite observables in maximally supersymmetric Yang-Mills theory. In the large $N_c$ limit, their integrand is given directly by the MHV loop Amplituhedron. Therefore the question of what transcendental functions arise from those positive geometries can be formulated in a transparent and direct way, without the need of infrared regularization of subtractions.

Wilson loops with a Lagrangian insertion have been first studied in references \cite{Alday:2011ga,Engelund:2011fg,Engelund:2012re}.
Perturbative results for four and five points are known to three and two loops from references \cite{Alday:2012hy,Alday:2013ip,Henn:2019swt}, and \cite{Chicherin:2022bov,Chicherin:2022zxo}, respectively.
The integrated functions obtained from canonical forms associated to geometries have a number of interesting properties. 
Due to the dual conformal symmetry of sYM, they depend on $(3n-11)$ cross-ratios \cite{Alday:2011ga}. This is the same number of variables as $n$-particle on-shell scattering amplitudes in generic theories depend on. One may use dual conformal transformations to send the Lagrangian insertion point to infinity, which suggests that a connection to Wilson loops in non-dual conformal theories, without Lagrangian insertion. The latter are kinematically equivalent to non dual-conformal scattering amplitudes. Indeed, the function space encountered in perturbative calculations so far matches that of generic scattering amplitudes.
Being finite, the results have a similarity with infrared finite parts of scattering amplitudes. 

The results have a structure reminiscent of next-to-MHV (NMHV) scattering amplitudes: they can be expanded in terms of transcendental functions and leading singularities. The leading singularities were studied in \cite{Chicherin:2022bov}. It was conjectured that they are given by a Grassmannian formula that enjoys both a dual conformal and a conformal symmetry (in a special dual conformal frame). The latter does not automatically follow from the Yangian symmetry of sYM \cite{Drummond:2009fd}, since the Lagrangian integration point is not integrated over.

The underlying Amplituhedron geometry motivates studying positivity properties of the integrated answers, as in the case of scattering amplitudes \cite{Dixon:2016apl}. While in the latter case one needs to choose an infrared subtraction scheme, the Wilson loops with Lagrangian insertion are infrared finite. 
Very interestingly, the results computed so far
 have been observed to be positive, when evaluated inside certain kinematic regions that are suggested by the geometry \cite{Arkani-Hamed:2021iya,Chicherin:2022zxo}. This positivity comes about as a non-trivial cancellation of different contributions, and involves an interplay of the leading singularities and the transcendental functions.

One of the motivations of the present paper is to explore further the connection between geometry of integrands and posivity properties of integrated functions.
We therefore wish to provide more detailed perturbative data. Indeed, a ``negative geometry'' expansion of the Wilson loop has been proposed in reference \cite{Arkani-Hamed:2021iya}, which further decomposes the answer in terms of building blocks that each have a geometric interpretation. 
This decomposition in general is different from a Feynman diagram expansion. 
At two loops, there are three contributions: a factorized, one-loop squared contribution, which is trivially known; a ``ladder-type'' contribution (in terms of the geometry), and a ``loop-type'' contribution. Since the full five-point answer is already known \cite{Chicherin:2022zxo}, it is sufficient to compute the ``ladder-type'' contribution, in order to be able to provide the full decomposition. This is the goal of the present paper.

The geometric decomposition goes beyond standard Feynman diagram expansions. Therefore we define more general Feynman integrals, and compute them via the differential equations method, generalizing the work done in references \cite{Drummond:2010cz}.
The ladder-type geometries are also known to satisfy a particular d`Alembertian differential equation \cite{Arkani-Hamed:2021iya,Brown:2023mqi}. In a complementary analysis, we study how this equation may be used to perform a symbol bootstrap of the answer. This may have higher-loop applications. However, as we also discuss, further information about the relevant function space is required to implement this program.

The outline of this paper is as follows: In section \ref{sect:negativegeometry}, we recall the definition of the Wilson loop with Lagrangian insertion, and how it may be decomposed in terms of ``negative'' geometries. In section \ref{sect:mtintegrand}, we construct an infinite class of ladder-type geometries at five points. We then proceed, in section \ref{sect:mtintegrandkinemtics}, to discuss the structure of integrated loop corrections at five points. Sections \ref{sect:2LNPFI} and \ref{sect:integratednegativegeometries} are devoted to evaluating the ladder-type geometries via differential equations, and investigate positivity properties of Wilson loop observable and negative geometries in the Amplituhedron region. Sections \ref{sect:boxing} and \ref{sect:bootstrap} provide an alternative, bootstrap approach to evaluating ladder-type negative geometries. We derive, in section \ref{sect:boxing}, a powerful d`Alembert-type differential equation, and in section \ref{sect:bootstrap}, we combine this equation with a suitable ansatz for the pentagon function space, which allows one to uniquely fix the answer. We demonstrate this to two loops for a minimal ansatz, and find that at three loops novel alphabet letters are required, for which we make a proposal.
We present a summary and conclusion in section \ref{sect:summary}. Appendix \ref{sect:AppLetters} contains the relevant information on the function spaces used in this paper. 
Appendix \ref{sect:pf_review} reviews pentagon functions and their derivatives.
Appendix \ref{sect:limits} explores various kinematic limits of the integrated negative geometries, including the soft/collinear limits, where the observables reduce to the four-cusp case recalled in \cref{sect:App4cusp}.

\section{Two-loop negative geometry decomposition of the Lagrangian insertion in the Wilson loop}
\label{sect:negativegeometry}


We consider an $n$-cusp polygon $[x_1,\ldots,x_n]$ with light-like edges, 
\begin{align}
(x_i - x_{i+1})^2 = 0 \,,\qquad i= 1,\ldots,n \,,
\end{align}
embedded in Minkowski space, and $x_{n+1} \equiv x_1$. Along this polygonal contour, we define the Wilson loop in the fundamental representation of the gauge group $SU(N_c)$,
\begin{align}
W_{\rm F}[x_1,\ldots,x_n] = \tr_{\rm F} \,{\rm Pexp} \left( \i\, g_{\rm YM} \,  t^a \oint A^a_\mu d x^\mu\right).
\end{align}
The main object of our interest is the following ratio of the correlation functions,
\begin{align}
\frac{1}{\pi^2} F_n(x_0;x_1,\ldots,x_n)  = \frac{\vev{W_{\rm F}[x_1,\ldots,x_n] {\cal L}(x_0)}}{\vev{W_{\rm F}[x_1,\ldots,x_n]}} \,, \label{eq:F}
\end{align}
which we refer to as the Lagrangian insertion in the Wilson loop. The composite operator ${\cal L}$ is the chiral Lagrangian of ${\cal N} = 4$ sYM, which is a conformal operator of dimension $4$. Due to the ultra-violet finiteness of the theory and the conformal nature of the Lagrangian, the correlators in \p{eq:F} are free from ultra-violet divergences. However, they do contain cusp divergences \cite{Korchemskaya:1992je,Drummond:2007aua},  which come from gluon exchanges in the vicinity of the Wilson loop cusps. The cusp divergences cancel out in the ratio \p{eq:F}, so $F_n$ is finite in four space-time dimensions. 

We consider the weak coupling $g^2 \equiv \frac{g_{\rm YM}^2 N_c}{16}$ perturbative expansion of $F_n$ in the large color $N_c \to \infty$ limit,
\begin{align}
F_n = \sum_{L \geq 0} (g^2)^{1+L} F_n^{(L)} \,, \label{eq:Fl}
\end{align}
which starts at order $g^2$. The simplest nondegenerate light-like polygonal contour has $n=4$ cusps. The Born-level contribution $F^{(0)}_{n}$ and one-loop correction $F^{(1)}_n$ have been calculated for any number of cusps $n$ in \cite{Chicherin:2022bov}. In the four-cusp case, the perturbative corrections $F^{(L)}_4$ are known up to order $L=3$ \cite{Alday:2012hy,Alday:2013ip,Henn:2019swt}. In the five-cusp case, the two-loop correction $F^{(2)}_5$ is available \cite{Chicherin:2022zxo}. 

The cancellation of divergences in the ratio \p{eq:F}, conformal invariance of the quantum theory, and conformal nature of the involved operators lead to conformal covariance of  $F_n$. Due to the nontrivial conformal weight of the Lagrangian, $F_n$ carries conformal weight $+4$ with respect to $x_0$ and zero conformal weight with respect to the cusp coordinates. The conformal symmetry severely restricts the kinematic dependence of $F_n$. Because of relations to scattering amplitudes, we adopt the amplitude terminology and refer to the space-time conformal symmetry as the {\em dual-conformal} symmetry. In the four-cusp case, the dual-conformal symmetry implies that $F_4$ is a nontrivial function of one variable. In what follows we are mainly interested in the five-cusp case, $n=5$, and we tacitly omit index the $n$. Up to a rational prefactor that absorbs the  dual-conformal weight, $F_5$ is a function of four independent dual-conformal cross-ratios, which we can choose as follows,
\begin{align}
{\bf u} =\left\{ \frac{x_{10}^2 x_{40}^2 x_{25}^2}{x_{20}^2 x_{50}^2 x_{14}^2} ,\,
\frac{x_{40}^2 x_{13}^2}{x_{30}^2 x_{14}^2} ,\,
\frac{x_{10}^2 x_{24}^2}{x_{20}^2 x_{14}^2} ,\,
\frac{x_{10}^2 x_{40}^2 x_{35}^2}{x_{30}^2 x_{50}^2 x_{14}^2} \right\} \,, \label{eq:u}
\end{align}
where $x_{ab}^2 := (x_a-x_b)^2$.

The correlator ratio $F_n$ is intimately related to MHV scattering amplitudes and their four-dimensional integrands. The kinematics of amplitudes is specified by $n$ light-like momenta ($p_i^2 = 0$) satisfying the momentum conservation,
\begin{align}
p_1 = x_n - x_1 \;,\quad  p_i = x_i - x_{i-1} \;,\quad i=2,\ldots,n\,. 
\end{align}
The vacuum expectation values of the null Wilson loop, $\vev{W_{\rm F}[x_1,\ldots,x_n]}$, are equal to $n$-particle MHV amplitudes in the large color limit \cite{Alday:2007hr,Drummond:2007aua,Brandhuber:2007yx} provided dimensional regularizations $D=4-2\ep$ of the two objects are properly identified. Then, applying the Lagrangian insertion formula \cite{Eden:2000mv,Alday:2011ga,Engelund:2011fg,Engelund:2012re}, we obtain the following relation between $F_n$ and the logarithm of the MHV amplitude,
\begin{align}
g^2 \pa_{g^2} \log\vev{W_{\rm F}} = \int \frac{d^D x_0}{\i \,\pi^{\frac{D}{2}}} F_n(x_0)\,. \label{eq:intF}
\end{align}
Namely, according to \cref{eq:intF}, $L$-loop correction $F_n^{(L)}$ is obtained from the $(L+1)$-loop integrand of the logarithm of the MHV amplitude where $L$ loop integrations are carried out. The remaining loop integration is divergent and requires a regularization in \cref{eq:intF}.  It corresponds to cusp divergences of $\log\vev{W_{\rm F}}$ which manifest themselves as poles $1/\ep^2$.


In order to describe the loop integrand of the Lagrangian insertion in the Wilson loop and negative geometries, and relate it to the Amplituhedron construction, we employ the momentum twistors \cite{Hodges:2009hk}. The momentum twistor variables are very convenient for massless scattering since they automatically resolve the momentum conservation and take into account the light-like nature of the momenta. A space-time coordinate (dual momenta) is equivalent to a line in momentum twistor space, which can be specified by a pair of momentum twistors belonging to it. For example, the Lagrangian coordinate is represented as follows, $x_0 \sim Z_A Z_B$ where $Z_A = (\lambda^{\a}_A, x^{\da \a}_0 \lambda_{A\,\a})$ and $Z_B = (\lambda^{\a}_B, x^{\da\a}_0 \lambda_{B\,\a})$, and $\lambda_A,\lambda_B$ is a pair of helicity spinors. Similarly, each loop variable is represented by a pair momentum twistors. In what follows, we label the loop variables of the integrands as $AB_i$, $i = 1,\ldots,L$, and $x_0 \sim  AB_0$. The cusps of the Wilson loop contour are light-like separated that is equivalent to  intersection of the corresponding momentum twistor lines. These $n$ intersecting momentum twistors lines are specified by $n$ momentum twistors $\{ Z_i \}_{i=1}^{n}$ located at their intersections and $x_i \sim Z_i Z_{i+1}$, with $i=1,\ldots,n$ and $n+1 \equiv 1$. They have the following explicit expressions, $Z_i = (\lambda^{\a}_i, x_i^{\da\a} \lambda_{i\,\a})$ where $\lambda_i,\tilde\lambda_i$ is a pair of helicity spinors, $p_i^{\da\a} = \tilde{\lambda}_i^{\da}{\lambda}^{\a}_i$.


Let us briefly recall the Amplituhedron construction of the MHV loop integrand in the planar ${\cal N}=4$ sYM theory, and the connection to the Wilson loop with the Lagrangian insertion. In the Amplituhedron picture \cite{Arkani-Hamed:2013jha,Arkani-Hamed:2017vfh} we consider a space of $L$ lines $AB_i$, $i=1,2,\dots,L$ which are subject to a set of inequalities:
\begin{align}
\mbox{For each loop:} \quad & \la AB_i\,j\,j{+}1\ra > 0 \quad \mbox{for $j=1,2,\dots,n{-}1,$\,\,and} \quad \la AB_i 1n\ra>0 \nonumber \\
& \mbox{sequence}\,\,\{\la AB_i1j\ra\} \,\,\mbox{for $j=2,\dots,n$ has 2 sign flips} \label{eq:cond1}\\
\mbox{For each pair of loops:} \quad & \, \la AB_iAB_j\ra>0 \label{eq:cond2}
\end{align}
The $n$-point MHV $L$-loop integrand then corresponds to the canonical differential form ${\Omega}_L$ with logarithmic singularities on the boundaries of this space. We call such integrands ``dlog'' forms. 

A variation of this picture leads to the definition of \emph{negative geometries}. To specify a particular negative geometry we use a graphic representation where the vertices correspond to loop lines $AB_i$, each of them satisfying the one-loop Amplituhedron conditions \eqref{eq:cond1}, while links represent \emph{mutual negativity conditions} $\la AB_iAB_j\ra<0$. Each negative geometry is then equipped with a canonical form $\Omega$ with logarithmic singularities on the boundaries of the space.

As was proven in \cite{Arkani-Hamed:2021iya}, we can construct the integrand for the \emph{logarithm of the amplitude} $\widetilde{\Omega}_L$ as a sum of dlog forms on all connected negative geometries:
\begin{equation}\label{eq:Omega_L}
   \widetilde{\Omega}_L =\sum_{\boldsymbol{g} \,{\rm with }\,L\,{\rm nodes}} (-1)^{E(\boldsymbol{g})} \,\widetilde{\Omega}_{\boldsymbol{g}}\,,
\end{equation}
where the subscript $\boldsymbol{g}$ sums all connected graphs at $L$ loops, i.e. graphs with $L$ nodes, and
$E(\boldsymbol{g})$ denotes number of edges in a graph.
The full integrand for logarithm of the amplitude as an expansion in $g$ is given by
\begin{align} \label{eq:Omega_expansion}
   \widetilde{\Omega} =\sum_{L=1}^{\infty} \,\,g^{2L}\;\widetilde{\Omega}_{L}\,.
\end{align}
Note that the individual terms $\widetilde{\Omega}_{\boldsymbol{g}}$ have unit leading singularities but the overall sign is not fixed (because both $\widetilde{\Omega}_{\boldsymbol{g}}$ and $-\widetilde{\Omega}_{\boldsymbol{g}}$ have correct singularity properties). The ambiguity of these signs in (\ref{eq:Omega_L}) is completely fixed once we require that the integrand of the logarithm of the amplitude can be also expressed in terms of products of ordinary loop integrands -- these are represented as certain special positive geometries: graphs with $L$ nodes and positive links, where nodes are divided into subgroups that each separately form a complete graph (this is equivalent to having products of amplitudes). As result, the only sign ambiguity is the overall sign of $ \widetilde{\Omega}_L$. 

For example, for $L=2$ there is only one negative geometry, i.e. one graph, while for $L=3$ we have two different negative geometries:
\begin{align}
    &\widetilde{\Omega}_2=-\;\;\omegaTwoLadd \;\;,\\[0.4cm]    &\widetilde{\Omega}_3=\quad\omegaThreeLadd\;\;-\;\omegaThreeLoop \;\;.
\end{align}
In all these pictures the complete symmetry in all $AB_i$ is implied. 

The integrand for the amplitude logarithm has very special infrared (IR) properties (which are equivalent to the ultraviolet, cusp properties, of the Wilson loop discussed above): when integrated over all loop momenta $AB_i$ at any loop order $L$, its leading divergence is a $1/\epsilon^2$ pole, as opposed to a naive $1/\epsilon^{2 L}$ one. Furthermore, if one of the loops is kept frozen (i.e, not integrated over), the result is finite. The resulting function is equal to $F_n^{(L)}$, see \cref{eq:F,eq:Fl}, the $L$-loop contribution to the Wilson loop with Lagrangian insertion. The role of the insertion point is played by the frozen loop, which we denote by $AB_0$. 
We introduce a graphical notation where the marked point 
$AB_0$ is indicated by a crossed circle, while all other $AB_i$, $i=1,2,{\dots},L$ are indicated by black vertices, as before. 
The following graph serves as an example,
\begin{equation}\label{example_integrated_negative}
\begin{aligned}
\includegraphics[scale=0.4]{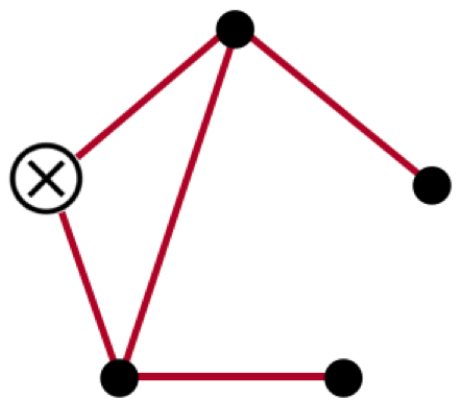} \;\quad.
\end{aligned}    
\end{equation}
In this picture all loops, except for the one corresponding to the marked point $AB_0$, are integrated over. This results in certain transcendental functions (with rational prefactors) in $AB_0$ and in the external kinematic variables). We will refer to objects such as eq. (\ref{example_integrated_negative}) as integrated negative geometries.

Note that because of the total symmetry of the integrand in all $AB$s, the integrated negative geometries pick up symmetry factors. For example, at the first two loop orders,
\begin{align}
   \omegaTwoLadd \;\; & \;\;\longrightarrow\;\; \omegaTwoLaddfrz\quad,\\[0.4cm]
    \omegaThreeLadd \;\; & \;\;\longrightarrow\;\; \omegaThreeLaddfrz\;\;
    +\;\;\frac{1}{2}\;\;\omegaThreeLaddMidfrz\quad,\\
    \omegaThreeLoop \;\; & \;\;\longrightarrow\;\; \omegaThreeLoopfrz \quad.
\end{align}
We see in the middle equation that a single negative geometry leads to two different contributions to the Wilson loop, according to where the frozen loop is located. The contributions to the $L$-loop function in eq. (\ref{eq:Fl}) is obtained by performing $L$ integrations on $\widetilde{\Omega}_{L+1}$ with $AB_0$ frozen
\footnote{To avoid clutter of notation, we refrain from introducing $\mathcal{F}_L$, used in \cite{Arkani-Hamed:2021iya}, to denote the integrated loop form. Instead, we directly define the $F^{(L)}$ in \eqref{eq:Fl} via performing $L$ integrations on $\widetilde{\Omega}_{L+1}$. Strictly speaking, the RHS of eqs. (\ref{eq:definitionF}) is a differential form in $AB_0$. For simplicity of notation, in this and in the following, we will tacitly drop a measure factor when identifying integrated negative geometries with $F^{(L)}$.} 
\begin{equation} \label{eq:Fint}
    F^{(L)}=-g^2\,\int_{AB_1,\ldots,AB_L} \widetilde{\Omega}_{L+1}(AB_{0},AB_{1},..,AB_{L})\,.
\end{equation}
In this notation, the one-and two-loop functions $F^{(1)}$ and $F^{(2)}$ are expressed as follows,
\begin{align} \label{eq:Fdecomp1}
F^{(1)} =& \, F^{(\laddonepic)}\,, \\
F^{(2)} =& - F^{(\laddtwopic)} - \frac{1}{2} F^{(\laddladdpic)} +\frac{1}{2} F^{(\looppic)} \,, \label{eq:Fdecomp2}
\end{align}
where $F^{(\boldsymbol{g})}$ denotes the contribution from a specific graph,
\begin{equation}\label{eq:definitionF}
    F^{(\boldsymbol{g})}=-g^2\,\int_{AB_1,\ldots,AB_L} \widetilde{\Omega}_{\boldsymbol{g}}(AB_{0},AB_{1},..,AB_{L})\,.
\end{equation}
Note that there is a shift in the loop order: an $(L+1)$-loop integrand is associated to an $L$-loop integrated result  $F^{(L)}$, as we are freezing one of the loops.

The detailed discussion of the four-point case is presented in \cite{Arkani-Hamed:2021iya}. In that case the integrands for all \emph{tree negative geometries}, i.e. graphs with no cycles, were found in a very compact form. A special form of these integrands allowed to derive a differential equation for the integrated negative geometries and found the result at all loops. This also allowed for the strong coupling expansion and surprisingly good qualitative agreement for the cusp anomalous dimension. Negative geometries with internal cycles are more complicated: canonical forms for all geometries with one cycles were found in \cite{Brown:2023mqi}, but the same differential method does not work. The evaluation and resummation of a subclass of these diagrams is work in progress \cite{DOPT}.

Most of this remarkable progress is limited to $n=4$.  The decomposition of the integrand of the amplitude logarithm as the sum of dlog forms on negative geometries is valid for any $n$, as well as the equality between the Wilson loop with a Lagrangian insertion and the integrated negative geometries with a marked point. However, the integrand for the negative geometries is more complicated, and also the integrated results are more complex: they depend on multiple cross-ratios and we also have non-trivial prefactors (leading singularities) unlike in the four-point case. 
In this paper we focus on the $n=5$ case. The next section is dedicated to deriving canonical forms for its geometric integrand decomposition.


\section{Momentum-twistor integrands for the negative geometries}
\label{sect:mtintegrand}

Having reviewed the geometric expansion of the Wilson loop with Lagrangian insertion, we now discuss what this implies in terms of loop integrands.
To find the dlog form for complicated negative geometries is a challenging open problem.
A conceptually clear procedure to find the integrands for negative geometries is to triangulate the associated spaces. In principle, this is a straightforward procedure of solving inequalities, but in practice it becomes very complicated even for a low number of loops. 
In this paper we will use a hybrid triangulation / unitarity-based method to calculate the integrand. 
As we will discuss presently, this is particularly efficient for the \emph{ladder geometries}, which are as follows,
\begin{equation}
\begin{aligned}
\includegraphics[scale=0.22]{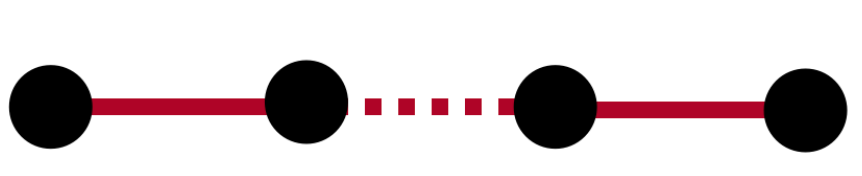}
    \end{aligned} \quad.
\end{equation}
Note that this is very different from the ladder Feynman diagrams. 
This was also the first approximation considered for $n=4$ \cite{Arkani-Hamed:2021iya}. 

We consider a general \emph{ladder negative geometry} 
\begin{equation}
\begin{aligned}
\includegraphics[scale=0.42]{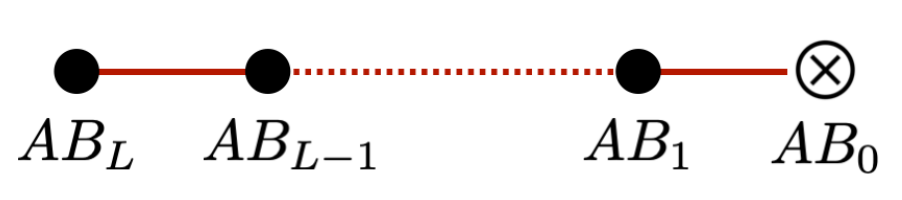}
\end{aligned} \quad,
\end{equation}
where here we labeled all loops: $AB_k$, $k=1,2,{\dots},L$ and a marked point $AB_0$. While we will construct here the integrand for this geometry, we will later integrate over all loops except for $AB_0$. 

The marked point $AB_0$ can be also located in the middle of the chain. We refer to this as \emph{product ladder negative geometry},
\begin{equation}
\begin{aligned}
\includegraphics[scale=0.42]{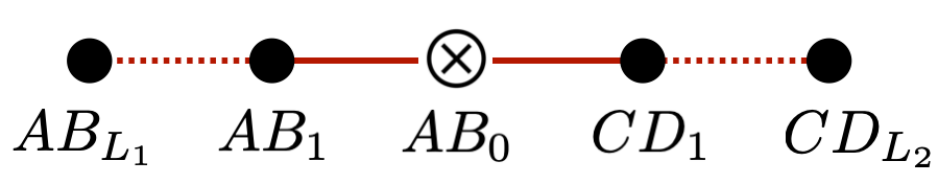}
\end{aligned}\quad,
\end{equation}
where here we denoted the loops on one side $AB_k$ for $k=1,{\dots},L_1$ and from the other side $CD_l$ for $l=1,{\dots},L_2$ such that $L_1+L_2=L$. Note that the integrands for both of these pictures are the same (up to relabeling of the loops) but we consider them separately because the integration procedure obviously distinguishes between them.

\subsection{Generalized-unitarity-inspired derivation of the ladder integrand}

Our goal is to present the integrand for the ladder negative geometry in the following way:
\begin{equation}
\Omega_L = \sum_k \Omega_k \times \Omega_0^{(k)}(AB_0)\,, \label{genOmega}
\end{equation}
where $\Omega_k$ can be written as a sum of dlog forms in $AB_1\dots AB_k$, i.e. the integrand has unit leading singularities when we localize all $AB_k$ on cuts. The ``coefficients'' $\Omega_0^{(k)}(AB_0)$ are dlog forms in $AB_0$ and do not depend on any of the $AB_k$ loops.
As we do not think about $AB_0$ as the loop to integrate over but a fixed point, we later tacitly drop a measure in $AB_{0}$. 

This expansion (\ref{genOmega}) is reminiscent of the generalized unitarity approach when we write the integrand for the amplitude as a sum of basis integrands multiplied by coefficients (leading singularities) which only depend on external kinematics. Here $\Omega_k$ are not planar diagrams in the usual sense, but they can be written in the momentum twistor space. 
This organization of the result is motivated by the structure after integration,
\begin{equation}
F_L^{\rm ladder} = \sum_k f_k \times \Omega_0^{(k)}(AB_0)\,,
\end{equation}
where the $f_k$ are transcendental functions, and $\Omega_0^{(k)}(AB_0)$ are rational prefactors\footnote{We have tacitly dropped a measure factor in $AB_0$.} (both depend on cross ratios of external kinematics and on $AB_0$). 
The latter can be interpreted as leading singularities of the integrand.
The classification of the complete set of all such factors that can appear in any negative geometry or in the full observable is an important question \cite{Chicherin:2022bov}, which will be further addressed in \cite{THMT}.

In order to write down the form $\Omega$ without the need of triangulation of the space we use the chiral box expansion \cite{Arkani-Hamed:2010pyv,Bourjaily:2013mma,Bourjaily:2015jna}. This is a refinement of the generalized unitarity where we pre-diagonalize the basis according to the list of cuts we wish to match. It is a one-loop version of the prescriptive unitarity approach to construct higher-loop integrands \cite{Bourjaily:2017wjl,Bourjaily:2019iqr,Bourjaily:2020qca}. The chiral box expansion uses a convenient infrared-friendly basis which nicely separates IR finite and IR divergent integrals, which is especially advantageous when dealing with infrared-finite objects. In fact, we will write $\Omega_L^{\rm ladder}$ in a form where each term is infrared finite. 

According to the generalized unitarity strategy, we ensure that the terms in this expansion match exactly one physical cut of interest and vanishes on all other cuts. Once all the physical cuts are matched, we check that all spurious cuts cancel. This then concludes the proof that the obtained formula is correct. 
Let us first show a few low loop examples, before formulating the general procedure.

\subsubsection*{Two-loop integrand}

Let us first demonstrate it the simplest case of $L=1$, i.e. the two-loop integrand $\Omega_1$ which is given as the canonical form on the negative geometry,
\begin{equation} \label{Omega1}
\begin{aligned}
\includegraphics[scale=0.4]{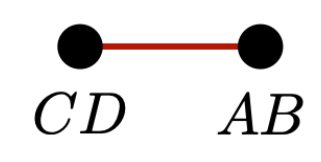}
\end{aligned}\quad.
\end{equation}
For simplicity we denote $CD\equiv AB_1$ and $AB\equiv AB_0$.
The space of all $CD$ lines is just a one-loop Amplituhedron with the extra condition $\la CDAB\ra<0$. When we perform the quadruple cut on $CD$ (not cutting $\la ABCD\ra$), the list of all allowed leading singularities is 
\begin{equation}
    CD = 13, 24, 35, 14, 25  \,.\label{LS5}
\end{equation}
Note that all leading singularities of the type $CD=i\,i{+}1$ are not allowed because they would imply $\la AB\,i\,i{+}1\ra<0$ which would violate the one-loop Amplituhedron inequalities for $AB$. The absence of these leading singularities is also related to infrared finiteness. 
Each of the five allowed leading singularities in eq. (\ref{LS5}) can be matched by a chiral pentagon integral.
Namely, for the leading singularity at $CD=13$, we write down
\begin{equation}
    {\cal C}_{13}(CD,AB) =  \frac{\la CD(512){\cap}(234)\ra\la AB13\ra}{\la CD12\ra\la CD23\ra\la CD34\ra\la CD15\ra\la ABCD\ra} \,,\label{chiral}
\end{equation}
and similar for the other ${\cal C}_{i\,i{+}2}$. 
In total we have five cyclically related terms, and each of them gives a unit leading singularity on one of the quadruple cuts (\ref{LS5}), and vanishes on all other four cuts.

Note that eq. (\ref{chiral}) has support on singularities of other quadruple cuts involving the $\la ABCD\ra$ propagator, but these are not in our list of cuts to match (they are redundant in this logic and must be matched automatically). In principle, we could also consider an integrand of the form
\begin{equation}
    \frac{\#}{\la CD15\ra\la CD12\ra\la CD34\ra\la ABCD\ra} \,,\label{spur}
\end{equation}
which has no support on any of the leading singularities (\ref{LS5}) and has non-vanishing residues on quadruple cuts involving $\la ABCD\ra=0$. However, this integral is necessarily IR divergent -- in the cut structure this is manifest by the presence of a spurious leading singularities $CD = (512)\cap(134)$. This singularity is obviously absent in all ${\cal C}_{i\,i{+}2}$ integrands, and is also absent in the form $\Omega_1$ for the negative geometry (\ref{Omega1}). Hence, (\ref{spur}) has no place in the expansion for $\Omega_1$ and we can write
\begin{equation}
\Omega_1 = \sum_{\rm cycl} {\cal C}_{13}(CD,AB)\times \Omega^{13-}_0 (AB)\,.
\end{equation}
The coefficient $\Omega_0^{13-}(AB)$ is the dlog form on the remaining geometry when we localize $CD$ on the leading singularity, $CD=13$. We get a one-loop Amplituhedron with an extra condition $\la AB13\ra<0$ which originates from $\la ABCD\ra<0$ for $CD=13$. For us this coefficient is the leading singularity of the form $\Omega_1$ as the loop $AB$ is frozen for us (and treated as external data).

In order to calculate the form $\Omega^{13-}_0$ we have to go back to the triangulation of the five-point one-loop Amplituhedron. According to eq. (\ref{eq:cond1}), we have to impose $\la AB\,i\,i{+}1\ra>0$ and the series$\{\la AB1i\ra\}$ for $i=2,3,4,5$ has two sign flips. Because now we impose explicitly $\la AB13\ra<0$ we get a more stringent sign flip condition,
\begin{equation}
\left(\begin{array}{cccc} \la AB12\ra & \la AB13\ra & \la AB14\ra & \la AB15\ra \\
+ & - & \ast & +\end{array}\right) \,. \label{ABcell}
\end{equation}
As a result, we get two terms depending on the sign of $\la AB14\ra$. Each of them is just a simple kermit form \cite{Arkani-Hamed:2010zjl} (eq. (32)) with known dlog form,
\begin{align}
\ast = +: &\quad [123,134] = \frac{\la1234\ra^2}{\la AB12\ra\la AB23\ra\la AB34\ra\la AB14\ra} \equiv  - B_5 \label{eq:B5}\\
\ast = -: &\quad [123,145] = \frac{\la AB(123)\cap(145)\ra^2}{\la AB12\ra\la AB13\ra\la AB23\ra\la AB14\ra\la AB15\ra\la AB45\ra}\,.
\end{align}
While the first term is a box integral (we denoted it as $B_5$), the second term is a general kermit. For our purpose it is useful to choose the following basis of the $AB$-forms \cite{Chicherin:2022bov,Chicherin:2022zxo},
\begin{equation}
{\cal B} = \{B_1,B_2,B_3,B_4,B_5,A^{\rm tree}_5\} \label{eq:Bbasis}
\end{equation}
where $B_j$ are box integrands: $B_1(2345)$, $B_2(3451)$, $B_3(4512)$, $B_4(5123)$, $B_5(1234)$ where the propagator structure is obvious from $B_5$ above. We also denoted 
\begin{equation}
A^{\rm tree}_5\equiv {\cal I}_5^{\rm 1-loop}(AB) = [123,134] + [123,145] + [134,145]
\end{equation}
which is the one-loop five-point MHV integrand for the loop $AB$. We rewrite the kermit term as 
\begin{equation}
\begin{aligned}
[123,145] =& A^{\rm tree}_5 - [123,134] - [134,145] \\
=& A^{\rm tree}_5 + B_5 + B_2\,,
\end{aligned}
\end{equation}
and we get for $\Omega_0^{13-}(AB)$,
\begin{equation}\label{eq:Om13}
\begin{aligned}
\Omega_0^{13-}(AB) =& [123,134] + [123,145]\\
=& A_5^{\rm tree} + B_2\,. 
\end{aligned}
\end{equation}
The other forms are related by cyclic shifts. 
\begin{equation}\label{eq:Om}
\begin{aligned}
&\Omega^{14-}_0(AB) =& A_5^{\rm tree}+B_5, \qquad
\Omega^{24-}_0(AB) = A_5^{\rm tree}+B_3,\nonumber\\
&\Omega^{25-}_0(AB) =& A_5^{\rm tree}+B_1, \qquad
\Omega^{35-}_0(AB) = A_5^{\rm tree}+B_4 \,.
\end{aligned}
\end{equation}
Note that, interestingly, only five linear combinations of the six basis elements ${\cal B}$ appear in the expansion for $\Omega_1$. To conclude, we can write the form $\Omega_1$ for the two-loop ladder
in the following way,
\begin{equation}
    \Omega_1 = \sum_{ab} {\cal C}_{ab}(CD,AB)\times\Omega^{ab-}_0(AB) \,,\label{eq:Om1Ladder}
\end{equation}
where the sum is over $ab=13,24,35,14,25$.

\subsubsection*{Three-loop integrand}

Let us consider the next case, which is an integrand for the three-loop ladder
corresponding to the two-loop problem, 
\begin{equation}
\begin{aligned}
\includegraphics[scale=0.42]{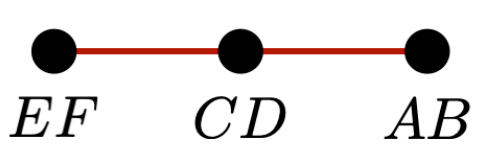}
\end{aligned}\quad.
\end{equation}
We are supposed to integrate over $CD$, $EF$ keeping $AB$ fixed. We start with the chiral box expansion on the $EF$ loop. We get
\begin{equation}
\Omega_{2} = \sum_{\rm cycl} {\cal C}_{13}(EF,CD)\times \Omega_1^{13^-}(CD,AB) \,,\label{Omega2}
\end{equation}
where the leading singularities in $EF$ are matched by chiral boxes. Using the same argument as before, no other $EF$-dependent term can appear in $\Omega_2$ (otherwise we would introduce spurious singularities that do not cancel). Next, we want to do the chiral box expansion on the object $\Omega_1^{13^-}(CD,AB)$ which is just a two-loop negative ladder
with an extra condition $\la CD13\ra<0$, i.e. 
\begin{equation}
\begin{aligned}
\includegraphics[scale=0.26]{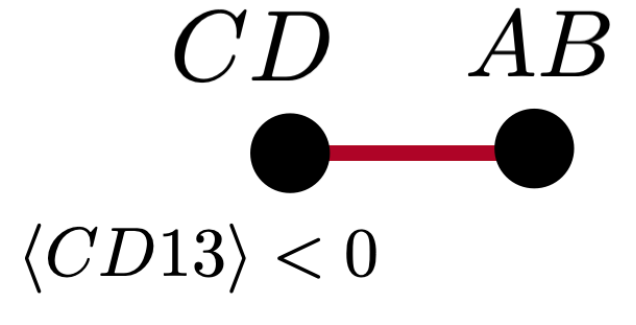}
\end{aligned}\quad.
\end{equation}
This space now has an extra boundary $\la CD13\ra=0$ which shows up as the pole in the denominator. This makes the chiral box expansion tricky, since a pole $\la CD13\ra$ is always spurious in the amplitude, but we can use a small trick to avoid the problem. We perform the chiral box expansion on the $AB$ loop (this is a regular one-loop Amplituhedron). This gives us the correct form for $ \Omega_1^{13^-}(CD,AB)$. Once this is obtained, we use it as a reference formula, and rewrite the result as a chiral box expansion on $CD$ with coefficients in AB. 

The chiral box expansion on $AB$ is easily obtained: it is the usual one with five leading singularities,
\begin{align}
 \Omega_1^{13^-}(CD,AB) &= \sum_{ab} {\cal C}_{ab}(AB,CD) \times \Omega_0^{13-,ab-}(CD)\,. \label{Omega113}
 \end{align}
%
Now we need to calculate the coefficients, i.e. the $\Omega_0(CD)$ terms. Note that the superscript indicates which conditions $\la CDij\ra<0$ are imposed on $CD$ on the top of just being in the one-loop Amplituhedron. Some of these terms are just regular kermits and sums of kermits, namely
\begin{align}
\Omega_0^{13-}(CD) &= \frac{\la CD(123)\cap(145)\ra^2}{\la CD12\ra\la CD23\ra\la CD13\ra\la CD14\ra\la CD15\ra\la CD45\ra} \nonumber\\ &\hspace{0.5cm} + \frac{\la 1234\ra^2}{\la CD12\ra\la CD23\ra\la CD34\ra\la CD14\ra} \,,\\
\Omega_0^{13-,35-}(CD) &= \frac{\la CD(123)\cap(345)\ra^2}{\la CD34\ra\la CD35\ra\la CD45\ra\la CD13\ra\la CD23\ra\la CD12\ra} \,,\\
\Omega_0^{13-,14-}(CD) &= \frac{\la CD(123)\cap(145)\ra^2}{\la CD12\ra\la CD13\ra\la CD23\ra\la CD14\ra\la CD15\ra\la CD45\ra} \,.
\end{align}
Note that the term $\Omega_0^{13-,35-}$ can be obtained directly as one kermit from the sign flip series starting with 3:
\begin{equation}
\left(\begin{array}{cccc} \la CD34\ra & \la CD35\ra & \la CD13\ra & \la CD23\ra \\
+ & - & - & +\end{array}\right) \quad.
\end{equation}
the term $\Omega_0^{13-}$ is just a sum of two kermits: $\Omega_0^{13-,35-}$ and 
\begin{equation}
\left(\begin{array}{cccc} \la CD34\ra & \la CD35\ra & \la CD13\ra & \la CD23\ra \\
+ & + & - & +\end{array}\right) \quad.
\end{equation}
and $\Omega_0^{13-,14-}(CD)$ is again just a single kermit,
\begin{equation}
\left(\begin{array}{cccc} \la CD12\ra & \la CD13\ra & \la CD14\ra & \la CD15\ra \\
+ & - & - & +\end{array}\right) \quad.
\end{equation}
The last two terms $\Omega_0^{13-,ab-}(CD)$ for $ab=24,25$ in (\ref{Omega113}) are more interesting because they do not correspond to a kermit. Let us start with $ \Omega_0^{13-,24-}(CD)$. This form can be obtained by starting with $\la CD13\ra<0$ space -- that is a sum of two kermits $[123,134]$ and $[123,145]$, and impose additional $\la CD24\ra<0$ condition. The first kermit incorporates this condition automatically, and the form is given by $B_5$, see \cref{eq:B5}. The other kermit geometry gets effectively ``chopped'' by the $\la CD24\ra<0$ condition. To obtain the canonical forms for these spaces we start in the kermit space and solve extra inequalities. As a result the space has six boundaries, and the associated canonical form has six poles and two distinct numerator factors,
\begin{align}
\Omega_0^{13-,24-}(CD) &= \frac{\la CD(451)\cap(123)\ra\la CD(234)\cap(451)\ra}{\la CD14\ra\la CD23\ra\la CD13\ra\la CD15\ra\la CD45\ra\la CD24\ra}\nonumber\\
& \hspace{0.5cm} + \frac{\la 1234\ra^2}{\la CD12\ra\la CD23\ra\la CD34\ra\la CD14\ra}\,,
\end{align}
and similarly,
\begin{align}
\Omega_0^{13-,25-}(CD) &= \frac{\la CD(123)\cap(345)\ra\la CD(345)\cap(512)\ra}{\la CD35\ra\la CD12\ra\la CD13\ra\la CD34\ra\la CD45\ra\la CD25\ra}\nonumber\\
& \hspace{0.5cm} + \frac{\la 1235\ra^2}{\la CD12\ra\la CD23\ra\la CD15\ra\la CD35\ra} \,.
\end{align}
This finishes the construction of the form $\Omega_2$. However, we want to reorganize the expansion for $\Omega_1^{ij-}(CD,AB)$ and write it as an expansion in $CD$ with coefficients in $AB$, rather than the other way around. Using eq. (\ref{Omega113}) as a reference result, together with a simple hybrid method of ansatz, and imposing cuts we can rewrite it in the following form,
\begin{align}
\Omega_1^{13-}(CD,AB) = \sum_{ab} {\cal C}_{13,ab} (CD,AB) \times \Omega^{ab-}_0(AB)\,, \label{Omega1a}
\end{align}
where we denoted
\begin{align}
    {\cal C}_{13,13} &= \frac{\la 1235\ra\la AB13\ra}{\la CD51\ra\la CD13\ra\la CD23\ra\la ABCD\ra}-\frac{\la 1234\ra\la AB13\ra}{\la CD34\ra\la CD13\ra\la CD12\ra\la ABCD\ra}\,,\nonumber\\
    {\cal C}_{13,24} &= \frac{\la CD(123)\cap(345)\ra\la AB24\ra}{\la CD12\ra\la CD23\ra\la CD34\ra\la CD45\ra\la ABCD\ra}\,, \nonumber\\
    {\cal C}_{13,35} &= -\frac{\la CD(123)\cap(451)\ra\la AB35\ra}{\la CD23\ra\la CD13\ra\la CD45\ra\la CD51\ra\la ABCD\ra}\,,\nonumber\\
    {\cal C}_{13,14} &= - \frac{\la CD(123)\cap(345)\ra\la AB14\ra}{\la CD12\ra\la CD13\ra\la CD34\ra\la CD45\ra\la ABCD\ra}\,, \nonumber\\
    {\cal C}_{13,25} &= \frac{\la CD(123)\cap(451)\ra\la AB25\ra}{\la CD12\ra\la CD23\ra\la CD45\ra\la CD51\ra\la ABCD\ra}\,. \label{eq:Cabcd}
\end{align}
The $AB$ leading singularities $\Omega_0^{ij-}(AB)$ have been calculated before, see \cref{eq:Om13,eq:Om}. Note that (\ref{Omega113}) and (\ref{Omega1a}) represent the same expression, just organized differently. It was easy for us to write (\ref{Omega113}) using chiral box expansion as there were not extra conditions imposed on the line $AB$, so we can treat it as a one-loop Amplituhedron and expanded the integrand in terms of building blocks. On the other hand, in (\ref{Omega1a}) we expand in $CD$ but there is an additional condition $\la CD13\ra<0$ imposed so the standard chiral box expansion naively does not work as we can not treat the $CD$ space as the one-loop Amplituhedron. Our result (\ref{Omega1a}) does provide an extension of the chiral box expansion to the space where additional condition $\la CD13\ra<0$ is imposed. 

As a result, this allows us to write the final result for the canonical form $\Omega_2$ of the three-loop ladder
as
\begin{equation}
    \Omega_2 = \sum_{ab,cd} {\cal C}_{ab}(EF,CD)\times {\cal C}_{ab,cd}(CD,AB)\times \Omega_0^{cd-}(AB)\,, \label{Omega2}
\end{equation}
where the sum runs over $ab,cd=13,24,35,14,25$ and 
${\cal C}_{ab,cd}$ are related with ${\cal C}_{13,ij}$ by cyclic shifts. Note that ${\cal C}_{13,cd}(CD,AB)$ has a pole $\la CD13\ra$ in the denominator, but it is canceled by the numerator of ${\cal C}_{13}(EF,CD)$.

Just for convenience, we collect together all terms in eq. (\ref{Omega2}) that multiply the $AB$-basis elements in $\Omega_2$, and reorganize the sum as 
\begin{align}
\Omega_2 = \Omega_2^{\rm tree} \times A_5^{\rm tree} + \sum_{k=1}^5 \Omega_2^{(k)} \times B_k \,.\label{Omega2res}
\end{align}
Writing now
\begin{equation}
\Omega_2^{\rm tree} = \Omega_A + \Omega_B +\Omega_C +\Omega_D + \Omega_E \,,
\end{equation}
where we have
\begin{align}
\Omega_A &= \frac{\la CD13\ra\la CD(123)\cap(345)\ra\la EF(512)\cap(234)\ra\la AB24\ra}{\la EF51\ra\la EF12\ra\la EF23\ra\la EF34\ra\la CDEF\ra\la CD12\ra\la CD23\ra\la CD34\ra\la CD45\ra\la ABCD\ra}\nonumber\\
\Omega_B &= \frac{\la CD13\ra\la CD(123)\cap(451)\ra\la EF(512)\cap(234)\ra\la AB25\ra}{\la EF51\ra\la EF12\ra\la EF23\ra\la EF34\ra\la CDEF\ra\la CD12\ra\la CD23\ra\la CD45\ra\la CD51\ra\la ABCD\ra}\nonumber\\
\Omega_C &= -\frac{\la CD(123)\cap(451)\ra\la EF(512)\cap(234)\ra\la AB35\ra}{\la EF51\ra\la EF12\ra\la EF23\ra\la EF34\ra\la CDEF\ra\la CD23\ra\la CD45\ra\la CD51\ra\la ABCD\ra}\nonumber\\
\Omega_D &= -\frac{\la CD(123)\cap(345)\ra\la EF(512)\cap(234)\ra\la AB14\ra}{\la EF51\ra\la EF12\ra\la EF23\ra\la EF34\ra\la CDEF\ra\la CD12\ra\la CD34\ra\la CD45\ra\la ABCD\ra}\nonumber\\
\Omega_E &= +\frac{\la EF(512)\cap(234)\ra\la AB13\ra\la1235\ra}{\la EF51\ra\la EF12\ra\la EF23\ra\la EF34\ra\la CDEF\ra\la CD51\ra\la CD23\ra\la ABCD\ra}\nonumber\\
&\hspace{0.4cm}-\frac{\la EF(512)\cap(234)\ra\la AB13\ra\la1234\ra}{\la EF51\ra\la EF12\ra\la EF23\ra\la EF34\ra\la CDEF\ra\la CD12\ra\la CD34\ra\la ABCD\ra}\,,
\end{align}
and the $B_1$ coefficient is 
\begin{equation}
\Omega_2^{(1)} = \Omega_A (+3) + \Omega_B (0) +\Omega_C (+2) +\Omega_D (+1) + \Omega_E (+4)\,,
\end{equation}
where we denoted $\Omega(+x)$ a cyclic shift of external momentum twistors of $\Omega$ by $+x$. Interestingly, the five integrals $\Omega_A$, $\Omega_B$, $\Omega_C$, $\Omega_D$, $\Omega_E$ we need to calculate have pretty compact nice forms. 

\subsubsection*{General problem}
The procedure outlined above generalizes to ladders of  arbitrary length. 
We outline the general setup here.
Consider the $L$-loop ladder,
\begin{equation}
\begin{aligned}
\includegraphics[scale=0.42]{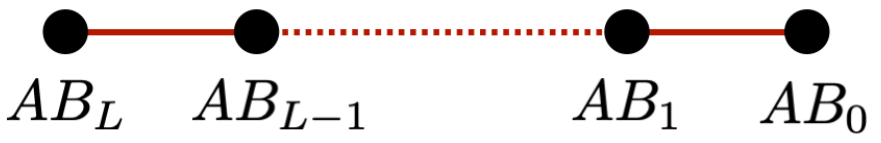}
\end{aligned} \quad,
\end{equation}
where each point is in the $n$-pt MHV one-loop Amplituhedron, and for neighboring points we have $\la AB_i AB_j\ra<0$. The strategy to write down the canonical form for this space is the following:
\begin{enumerate}
\item Start from the left and write down the chiral box expansion for the $AB_L$ loop (no other integrals are allowed because of the absence of spurious cuts). This matches all leading singularities in $AB_L$. The prefactors are functions which depend on the other $L-1$ loops.
\item Each term in this expansion has support on one leading singularity $AB_L=ij$, so the form of the remaining $L-1$ loops corresponds to a negative geometry with an extra condition $\la AB_{L{-}1} ij\ra < 0$. 
\begin{equation}
\Omega_L = \sum_{ij} {\cal C}_{ij}(AB_L,AB_{L{-}1}) \times \Omega_{L{-}1}^{ij-}(AB_{L{-}1},\dots AB_0)\,,
\end{equation}
where  ${\cal C}_{ij}$ are chiral boxes, see \cref{chiral}. 
\item Continue this procedure recursively until one reaches $AB_1$. Then the coefficients $\Omega_0^{ab-}$ depend on $AB_0$ only, and can be expressed in the basis ${\cal B}$ of leading singularities, see \cref{eq:Bbasis,eq:Om13,eq:Om}.
\end{enumerate}
Collect all pieces for the final result, we find
\begin{align}
    \Omega_L &= \sum {\cal C}_{i_1j_1}(AB_L,AB_{L{-}1}) \times 
    {\cal C}_{i_1j_1,i_2j_2}(AB_{L{-}1},AB_{L{-}2}) 
    \times 
    {\cal C}_{i_2j_2,i_3j_3}(AB_{L{-}2},AB_{L{-}3}) \dots \nonumber \\
    & \hspace{2cm} \dots
    \times {\cal C}_{i_{L{-}1}j_{L{-}1},i_Lj_L}(AB_{1},AB_0) \times \Omega_0^{i_Lj_L-}(AB_0) \,,\label{genladder}
\end{align}
where all the building blocks ${\cal C}_{ab}$, ${\cal C}_{cd,ef}$ and $\Omega_0^{i_Lj_L-}(AB_0)$ have been calculated above in (\ref{chiral}), \eqref{eq:Cabcd}, and \eqref{eq:Om13}. 
This simple expansion gives us the integrand for a general ladder of an arbitrary length. 

Note that each term in (\ref{genladder}) is manifestly absent of spurious poles which appear in individual ${\cal C}_{ab}$, ${\cal C}_{cd,ef}$, because of pairing of indices. The term ${\cal C}_{cd,ef}(AB_k,AB_{k{-}1})$, which has a spurious pole $\la AB_kcd\ra$ is always multiplied by ${\cal C}_{ab,cd}(AB_{k{+}1},AB_k)$ which has the same factor in the numerator. Hence each term in (\ref{genladder}) has only physical poles $\la AB_k j\,j{+}1\ra$, $\la AB_k AB_l\ra$.

\subsection{Product of ladders}

The second negative geometry topology is when the marked point $AB_0$ is in the middle of the ladder. As mentioned before, the integrand is the same (up to relabeling of the loops) as the integrand for the ladder with $AB_0$ as the endpoint. 

However, for our purposes, we wish to present the result in the form (\ref{genOmega}), where the leading singularities in $AB_{0}$ are explicit. This requires a reorganization of the integrand, which we discuss presently.

We start with the $L=2$ example, i.e. the three-loop ladder,
\begin{equation}
\begin{aligned}
\includegraphics[scale=0.42]{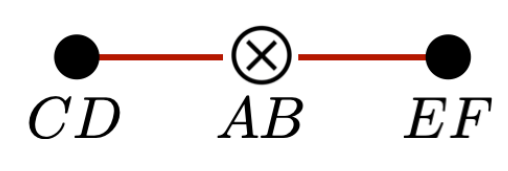}
\end{aligned}\quad.
\end{equation}
We perform a double chiral box expansion on $CD$ and $EF$. As these two loops do not ``communicate'' with each other (via the mutual inequality on $\la CDEF\ra$), we can do these expansions independently,
\begin{equation}
    \Omega_2 = \sum_{ab,cd} {\cal C}_{ab}(CD,AB)\times {\cal C}_{cd}(EF,AB)\times \Omega_0^{ab-,cd-}(AB)\,. \label{eq:Om2prod}
\end{equation}
Here the $AB$-geometry is the one-loop Amplituhedron with additional conditions $\la ABab\ra<0$ and $\la ABcd\ra<0$. We have already encountered these spaces above, in the context of the chiral box expansion of $\Omega_1^{13-}(CD,AB)$ in $AB$ (rather than $CD$). Hence we can immediately write down the result,
\begin{align}
\Omega_0^{13-,13-}(AB) = \Omega_0^{13-}(AB) &= \frac{\la AB(123)\cap(145)\ra^2}{\la AB12\ra\la AB23\ra\la AB13\ra\la AB14\ra\la AB15\ra\la AB45\ra} \nonumber\\ &\hspace{0.5cm} + \frac{\la 1234\ra^2}{\la AB12\ra\la AB23\ra\la AB34\ra\la CD14\ra}\,,\\
\Omega_0^{13-,35-}(AB) &= -\frac{\la AB(123)\cap(345)\ra^2}{\la AB34\ra\la AB35\ra\la AB45\ra\la AB13\ra\la AB23\ra\la AB12\ra}\,,\\
\Omega_0^{13-,14-}(AB) &= -\frac{\la AB(123)\cap(145)\ra^2}{\la AB12\ra\la AB13\ra\la AB23\ra\la AB14\ra\la AB15\ra\la AB45\ra}\,,\\
\Omega_0^{13-,24-}(AB) &= \frac{\la AB(451)\cap(123)\ra\la AB(234)\cap(451)\ra}{\la AB14\ra\la AB23\ra\la AB13\ra\la AB15\ra\la AB45\ra\la AB24\ra}\nonumber\\
& \hspace{0.5cm} + \frac{\la 1234\ra^2}{\la AB12\ra\la AB23\ra\la AB34\ra\la AB14\ra}\,,\\
\Omega_0^{13-,25-}(AB) &= \frac{\la AB(123)\cap(345)\ra\la AB(345)\cap(512)\ra}{\la AB35\ra\la AB12\ra\la AB13\ra\la AB34\ra\la AB45\ra\la AB25\ra}\nonumber\\
& \hspace{0.5cm} + \frac{\la 1235\ra^2}{\la AB12\ra\la AB23\ra\la AB15\ra\la AB35\ra}\,.
\end{align}
The first three terms can be written in terms of the leading singularities ${\cal B}$, cf. eq.  \p{eq:Bbasis}, as follows,
\begin{equation}
\begin{aligned}
     \Omega_0^{13-,13-}(AB) &= A_5^{\rm tree} + B_2, \\  \Omega_0^{13-,35-}(AB) &= A_5^{\rm tree} + B_2 + B_4\,,\\  
     \Omega_0^{13-,14-}(AB) &= A_5^{\rm tree} + B_2 + B_5\,.
\end{aligned}
\end{equation}
However, the other two terms introduce new leading singularities.  We can write 
\begin{equation}
\begin{aligned}
    \Omega_0^{13-,24-}(AB) =& B_5 + C_5\,, \\ \Omega_0^{13-,25-}(AB) =& B_4 + C_4\,,
    \end{aligned}
\end{equation}
where we introduced 
\begin{equation}
    C_4 \equiv \frac{\la AB(123)\cap(345)\ra\la AB(345)\cap(512)\ra}{\la AB35\ra\la AB12\ra\la AB13\ra\la AB34\ra\la AB45\ra\la AB25\ra}\,, \label{eq:C4}
\end{equation}
and similarly for four other terms $C_k$, which are obtained by cyclic rotations. Hence our leading singularity basis has now a total of 11 terms, compared to the 6 terms in ${\cal B}$ \p{eq:Bbasis}. They appear in 10 combinations (6 terms from ${\cal B}$ and 5 combinations $B_i+C_i$). 

The construction for a general ladder is analogous. 
\begin{equation}
\begin{aligned}
\includegraphics[scale=0.42]{figs/ladder2.pdf}
\end{aligned}\quad,
\end{equation}
where $L_1+L_2=L$. The form $\Omega_L$ can be then written as 
\begin{align}
    \Omega_L = &\sum {\cal C}_{i_1j_1}(AB_{L_1},AB_{L_1{-}1}) \times {\cal C}_{i_1j_1,i_2j_2}(AB_{L_1{-}1},AB_{L_1{-}2}) \dots {\cal C}_{i_{L_1-1}j_{L_1-1},i_{L_1}j_{L_1}}(AB_1,AB_0)\nonumber\\
    & \times {\cal C}_{k_1l_1}(CD_{L_2},CD_{L_2{-}1}) \times {\cal C}_{k_1l_1,k_2l_2}(CD_{L_2{-}1},CD_{L_2{-}2}) \dots {\cal C}_{k_{L_2-1}l_{L_2-1},k_{L_2}l_{L_2}}(CD_1,AB_0)\nonumber\\
    & \times \Omega_0^{i _{L_1}j_{L_1}-,k_{L_2}l_{L_2}-}(AB_0) \,.\label{eq:ladderProdGen}
\end{align}
The logic of this formula is straightforward: we just apply twice the procedure from the previous subsection, once from the left on $AB_{L_1},\dots,AB_1$ and once from the right on $CD_{L_2}\dots CD_1$. The leading singularity is now the dlog form on the $AB_0$ one-loop Amplituhedron with two extra conditions $\la AB_0 i _{L_1}j_{L_1}\ra <0$ and $\la AB_0 k_{L_2}l_{L_2}\ra<0$. 

To summarize, in this section we have obtained the integrand for all five-point ladder geometries. 
Furthermore, as the main result of this section, eqs. (\ref{genladder}) and (\ref{eq:ladderProdGen}) are written in a way that makes the leading singularities manifest. This is useful when discussing the structure of the integrated results in terms of transcendental functions, and prefactors (given by the leading singularities). 
This will be explored in the following sections.

In addition to the six leading singularities known from references \cite{Chicherin:2022bov,Chicherin:2022zxo}, we saw that further leading singularities may arise from product-type geometries. However, it is known from the above references that at two loops, these additional leading singularities drop out in the full observable. This topic will be explored in more detail in reference \cite{THMT}.


\section{Integrated negative geometries in five-particle kinematics}
\label{sect:mtintegrandkinemtics}

In the previous section we have obtained the loop integrands for ladder-type geometries, and identified their leading singularities. In this section, we discuss the structure one expects after integration.  

\subsection{Five-particle kinematics}
\label{sect:5part}

The Lagrangian insertion in the Wilson loop \p{eq:F} and the individual negative geometries in its decomposition are dual-conformal covariant. In particular, the momentum twistor expressions for the integrands of the ladder geometries \p{genladder} and \p{eq:ladderProdGen} make the dual conformal symmetry manifest. In the following, we find it convenient to fix the frame $x_0 \to \infty$ that corresponds to identifying $AB_0$ with the infinity bi-twistor. In this frame, $F$ \p{eq:F} and the negative geometries have residual Lorentz covariance.

Then we switch from the dual-momenta, which are space-time coordinates of the polygonal contour cusps, to momenta variables. We assign five light-like momenta $\{ p_i^\mu \}_{i=1}^{5}$ to the edges of the Wilson loop contour,  
\begin{align}
\label{eq:dual_momentum}
p_i = x_i - x_{i-1} \,,\qquad
(p_i)^2 = 0 \,,\qquad i=1,\ldots,5\,,
\end{align}
where we assume that the labels take cyclic values from $\{ 1,\ldots,5 \}$.
Thus, the kinematics in the five-cusp case is the same as for the five-particle massless scattering, e.g. the kinematics of five-gluon scattering amplitudes in QCD. We choose bi-particle adjacent Mandelstam variables to specify the kinematic configuration,
\begin{align}
X:= \left(s_{12} = x_{25}^2 \,,\, s_{23} = x_{13}^2 \,,\, s_{34}= x_{24}^2 \,,\, s_{45}= x_{35}^2 \,,\, s_{15} = x_{14}^2 \right) \label{eq:s}
\end{align}
where $s_{ij} := (p_i+p_j)^2$, and the non-adjacent bi-particle Mandelstam variables are linear combinations of \p{eq:s},
\begin{align}
\label{eq:non-adj-s}
s_{i \, i+2} = s_{i+3\,i+4} - s_{i\, i+1} - s_{i+1\, i+2} \,,\quad i =1,\ldots, 5 \,,  
\end{align}
where we assume the cyclicity of the labels. Also, the parity-odd Lorentz invariant is required to distinguish kinematic configurations of opposite parity,
\begin{align}
\ep_5 = 4 \i \ep_{\mu \nu \rho \sigma} p_1^\mu p_2^\nu p_3^\rho p_4^\sigma  = \tr \left( \gamma_5 \widehat{p}_1 \widehat{p}_2 \widehat{p}_3 \widehat{p}_4\right) \label{eq:ep5}
\end{align}
which value is fixed up to sign by the parity-even Mandelstam variables, 
\begin{align}
(\ep_5)^2 = \Delta_5 \equiv \det\left( s_{ij}|_{i,j=1,\ldots,4} \right) . \label{eq:del5}
\end{align}
Although the Wilson loop is parity-even, the Lagrangian operator is chiral. Thus, the parity-odd $\ep_5$ appears in the expression for the correlator ratio $F$ \p{eq:F}.

\subsection{Leading singularities in momentum space notation}

In the four-cusp case, all negative geometries are proportional to the unique leading singularity, see \cref{sect:App4cusp}. The five-cusp case is much more nontrivial, and eleven rational prefactors (leading singularities) are required to describe the two-loop negative geometries. Indeed, constructing the momentum-twistor integrands of the ladder negative geometries in \cref{sect:mtintegrand}, we introduced the ${\cal B}$ basis \p{eq:Bbasis} consisting of six elements, and we
extended it with five $\{ C_i \}_{i=1}^{5}$ \p{eq:C4} considering product ladders. 

In order to establish a connection with definitions in \cite{Chicherin:2022zxo}, we introduce the following basis of 11 rational prefactors in the frame $x_0 \to \infty$,
\begin{align}
r_0,r_1,\ldots,r_5, \overline{r}_1,\ldots, \overline{r}_{5}\,, \label{eq:r0r11}
\end{align}
which have the following explicit expressions in momentum variables, 
\begin{align}
r_0 & = \tr_{-} \left( (\widehat{p}_1 + \widehat{p}_2) (\widehat{p}_2 + \widehat{p}_3) (\widehat{p}_3 + \widehat{p}_4) (\widehat{p}_4 + \widehat{p}_5) \right) , \label{eq:r0}\\[0.2cm] 
r_i & = \frac{s_{i+2\, i+3}}{s_{i+1\, i+4}} \,\tr_-\left(\widehat{p}_{i+1}\, \widehat{p}_{i+2}\, \widehat{p}_{i+3}\, \widehat{p}_{i+4}\right) , \label{eq:r1} \\[0.2cm]
\overline{r}_{i} & = \frac{s_{i\,i+1} s_{i\,i+4}}{s_{i+1\,i+3} s_{i+2\, i+4}} \, \tr_-\left(\widehat{p}_{i+1}\, \widehat{p}_{i+2} \,\widehat{p}_{i+4} \,\widehat{p}_{i+3}\right), \qquad i=1,\ldots,5\,,\label{eq:r6}
\end{align}
where we assume cyclicity of the momenta and Mandelstam labels, e.g. $p_6 \equiv p_1$, and we use the shorthand notation for the chiral trace
\begin{align}
\tr_{-} \left(\widehat{p}_i \widehat{p}_j \widehat{p}_k \widehat{p}_l \right) := \frac{1}{2} \tr  \left((1-\gamma_5)\widehat{p}_i \widehat{p}_j \widehat{p}_k \widehat{p}_l \right) . \label{eq:trm}
\end{align}
The chiral traces in \crefrange{eq:r0}{eq:r6} evaluate as follows in terms of the Mandelstam  variables and the parity-odd $\ep_5$ \p{eq:ep5},
\begin{align}
& 2\tr_{-} \left( (\widehat{p}_1 + \widehat{p}_2) (\widehat{p}_2 + \widehat{p}_3) (\widehat{p}_3 + \widehat{p}_4) (\widehat{p}_5 + \widehat{p}_1) \right)  = - s_{12} s_{23} - s_{23} s_{34} - s_{34} s_{45} - s_{15} s_{12} - \ep_5 \,,\notag \\[0.2cm]
& 2\tr_-\left(\widehat{p}_{i+1}\, \widehat{p}_{i+2}\, \widehat{p}_{i+3}\, \widehat{p}_{i}\right) = s_{i+1\, i+2} s_{i+3\, i+4} -s_{i+1\,i+3} s_{i+2\,i+4} + s_{i+1\, i+4} s_{i+2\, i+3} -\eps_5 \,, \notag \\[0.2cm]
& 2\tr_-\left(\widehat{p}_{i+1}\, \widehat{p}_{i+2} \,\widehat{p}_{i+4} \,\widehat{p}_{i+3}\right) = s_{i+1\, i+2} s_{i+3\, i+4} - s_{i+1\, i+4} s_{i+2\, i+3} + s_{i+1\, i+3} s_{i+2\, i+4} + \eps_5 \,. \label{eq:trmeval}
\end{align}

Relations between the momentum-twistor basis of the leading singularities  ${\cal B}$ \p{eq:Bbasis} and $\{ C_i \}_{i=1}^{5}$ \p{eq:C4} in the frame $x_0 \to \infty$ and the momentum-space basis \p{eq:r0r11} are as follows 
\begin{align}
A_5^{\rm tree} \to -r_0 \,,\qquad \, B_i \to r_i  \,,\qquad C_i \to r_0 -r_i -r_{i+2} - r_{i+3} - \overline{r}_i  \;,\qquad i=1,\ldots,5\,. \label{eq:LStox0inf}
\end{align}

The Lagrangian insertion in the Wilson loop \p{eq:F} and the negative geometries are invariant under the discrete group of dihedral transformations, which is generated by the cyclic shift transformation $\tau$ and the inversion $\rho$. For the five-cusp contour, they act on the momenta variables, which are the edges of the contour, and on the dual momenta, which are cusps of the contour, as follows,   
\begin{align}
& \tau(p_i) = p_{i+1} \,,\qquad 
\rho(p_{i}) = p_{6-i} \,,\notag\\ 
& \tau(x_i) = x_{i+1} \,,\qquad 
\rho(x_{i}) = x_{6-i} \,, \qquad i=1,\ldots,5 \,, \label{eq:tau}
\end{align}
where we recall cyclicity of the labels, $p_6 \equiv p_1$ and $x_6 \equiv x_1$. The rational prefactor $r_0$ is dihedral invariant 
\begin{align}
& \tau(r_0) = \rho(r_0) = r_0\,,
\end{align}
whereas $\{r_i\}_{i=1}^{5}$ and $\{\overline{r}_i\}_{i=1}^{5}$ form the cyclic orbits,
\begin{align}
& \tau(r_5)=r_1 \,, \quad
\tau(r_i) = r_{i+1}\,, \quad \tau(\overline{r}_{5})=\overline{r}_1 \,, \quad \tau(\overline{r}_{i}) = \overline{r}_{i+1} \,, \quad i=1,\ldots,4\,, \\[0.2cm]
& \rho(r_i) = r_{6-i} \,,\quad \rho(\overline{r}_{i}) = \overline{r}_{6-i} \,, \quad i=1,\ldots,5\,.
\end{align}

As observed in \cite{Badger:2019djh,Henn:2019mvc,Chicherin:2022bov}, the six rational prefactors $\{ r_i \}_{i=0}^{5}$ are conformal invariant in momentum space when normalized by the Parke-Taylor prefactor,
\begin{align}
{\rm PT} = \frac{1}{\prod_{j=1}^{5} \vev{j j+1}}
\end{align}
where $\vev{ij} := \lambda_{i\, \alpha} \lambda^{\alpha}_j$ are spinor brackets of the helicity spinors, and $p_i^\mu \sigma^{\dot\alpha\alpha}_{\mu} = \lambda^{\alpha}_i \tilde\lambda^{\dot \alpha}_i$.
The nontrivial part of the conformal symmetry statement is that the rational prefactors are annihilated by the conformal boost generator, 
\begin{align}
\mathbb{K}_{\alpha \dot\alpha} = \sum_{j=1}^{5}\frac{\pa^2}{\pa \lambda^{\alpha}_j \pa \tilde\lambda^{\dot\alpha}_j}
\,,\qquad \mathbb{K}_{\alpha \dot\alpha} \left( {\rm PT} \, r_i \right) = 0\,,\qquad i=0,\ldots,5\,.
\end{align}
Let us note that the remaining five $\{ \overline{r}_i \}_{i=1}^{5}$ are not conformal.

\subsection{The structure of the five-point integrated negative geometries}

The Lagrangian operator is conformal and carries weight $(+4)$. Then, using the amplitude terminology, the Lagrangian insertion in the five-cusp Wilson loop $F$ \p{eq:F} carries the dual-conformal weight $(+4)$ with respect to the Lagrangian coordinate $x_0$. This weight has to be canceled out when fixing the frame $x_0 \to \infty$, 
\begin{align}
F^{(L)}(X) \equiv \lim_{x_0\to \infty} (x_0^2)^4 \,F^{(L)}(x_0;x_1,\ldots,x_5) \,. \label{eq:Fx0inf}
\end{align}
In the following, we tacitly assume that the frame $x_0 \to \infty$ is chosen and the loop corrections $F^{(L)}$ are functions of five Mandelstam variables $X$ \p{eq:s}, as well as of the parity-odd $\ep_5$.
In the five-cusp case, $F$ is known up to the two-loop order \cite{Chicherin:2022zxo}. The perturbative corrections $F^{(L)}$ are expanded in the basis of the rational prefactors $\{r_i\}_{i=0}^{5}$, introduced in eqs.~\p{eq:r0},~\p{eq:r1}, as follows,
\begin{align}
& F^{(0)} = r_0 \,, \label{eq:F0}\\
& F^{(1)} = \sum_{i=1}^{5}  (r_i-r_0) g^{(1)}_i \,, \label{eq:F1L}\\
& F^{(2)} = r_0 \,g^{(2)}_0 + \sum_{i=1}^{5}  r_i \,g^{(2)}_i\,, \label{eq:F2L}
\end{align}
where $g^{(L)}_i$ are pure polylogarithmic functions of the transcendental weight $2L$. More precisely, they are expressed in terms of the planar pentagon functions \cite{Gehrmann:2018yef,Chicherin:2020oor}, whose definitions are recalled in \cref{sect:pf}. The one-loop functions $g^{(1)}_i$ have simple expressions as the classical dilogarithms, whereas the two-loop functions $g^{(2)}_i$ are known as iterated integrals with $\dlog$ kernel.

In the present work, we are interested in the analogous expressions for their negative geometry decomposition.
According to \cref{sect:mtintegrand}, the ladder geometries involve five leading singularities $\Omega^{ab-}_0(AB_0)$ only, see \cref{genladder}. Carrying out the loop integration over $AB_1,\ldots,AB_L$, the $L$-loop ladder takes the form
\begin{align}
\int_{AB_1,\ldots,AB_L} \Omega_{L} (AB_{0},AB_{1},..,AB_{L}) = \sum_{ab} \Omega^{ab-}_0(AB_0) \, h_{ab}^{(L)}(AB_0)\,, \label{eq:inthL}
\end{align}
where the summation is over $ab=13,24,35,14,25$, and where $h^{(L)}_{ab}$ are pure functions. According to \cref{eq:Om1Ladder,Omega2}, at $L=1,2$ they are given by the following integrals,
\begin{align}
& h_{ab}^{(1)}(AB_0) = \int\limits_{AB_1} {\cal C}_{ab}(AB_1,AB_0) \,,\label{eq:h1}\\
& h_{ab}^{(2)}(AB_0) =  \sum_{cd} \int\limits_{AB_1,AB_2}{\cal C}_{cd}(AB_2,AB_1)\; {\cal C}_{cd,ab}(AB_1,AB_0) \,.\label{eq:h2}
\end{align}
The building blocks ${\cal C}_{ab}$ and ${\cal C}_{ab,cd}$ of the integrands are defined in \cref{chiral,eq:Cabcd}.

Similar to $F^{(L)}$ in \cref{eq:Fx0inf}, we consider the ladders in the frame $x_0 \to \infty$. Their leading singularities are linear combinations of the rational prefactors $\{r_i\}_{i=0}^{5}$,
\begin{align}
& \Omega^{25-}_0 = r_1-r_0\,, \quad
\Omega^{13-}_0 = r_2-r_0\,, \quad
\Omega^{24-}_0 = r_3-r_0\,, \notag \\
& \Omega^{35-}_0 = r_4-r_0\,, \quad
\Omega^{14-}_0 = r_5-r_0\,,
\end{align}
where we tacitly imply that the dual-conformal weight of $\Omega^{ab-}_0$ at point $x_0$ is canceled out and $x_0 \to \infty$ as in \cref{eq:Fx0inf}. These relations follow from \cref{eq:Om13,eq:Om,eq:LStox0inf}. We can label the five leading singularities of the ladders either by pair of indices $ab$ as in \cref{eq:inthL}, or by a single index $i=1,\ldots,5$. Introducing an analogous labeling for the pure functions of the ladders, we define
\begin{align}
h^{(L)}_{25} \equiv h^{(L)}_1\,, \quad
h^{(L)}_{13} \equiv h^{(L)}_2\,, \quad
h^{(L)}_{24} \equiv h^{(L)}_3\,, \quad h^{(L)}_{35} \equiv h^{(L)}_4\,, \quad
h^{(L)}_{14} \equiv h^{(L)}_5\,.  \label{eq:hab-hi}  
\end{align}
Then the integrated one-loop and two-loop ladders have the following form,
\begin{align}
& F^{(\laddonepic)} = \sum_{i=1}^{5} (r_i-r_0)\, h^{(1)}_i \,, \label{eq:F1Ll} \\
& F^{(\laddtwopic)} = \sum_{i=1}^{5} (r_i-r_0)\, h^{(2)}_i \,. \label{eq:F2Ll}
\end{align}
Since the one-loop negative geometry decomposition involves only the ladder-type negative geometry, i.e. $F^{(1)} = F^{(\laddonepic)}$ according to \cref{eq:Fdecomp1}, then the pure functions in \cref{eq:F1L,eq:F1Ll} coincide,
\begin{align}
g^{(1)}_{i} = h^{(1)}_i   \,,\qquad i = 1,\ldots, 5\,. \label{eq:h1-g1} 
\end{align}
Performing the one-loop integrations of the integrand \p{eq:h1},
we find \cite{Arkani-Hamed:2010pyv,Chicherin:2022bov},
\begin{align}
g_i^{(1)} = \frac{\pi^2}{6} - \log\left(\frac{s_{i\,i+1}}{s_{i+2\,i+3}}\right) \log\left(\frac{s_{i\,i+4}}{s_{i+2\,i+3}}\right) - {\rm Li}_2 \left( 1 - \frac{s_{i\,i+1}}{s_{i+2\,i+3}}\right) - {\rm Li}_2 \left( 1 - \frac{s_{i\,i+4}}{s_{i+2\,i+3}}\right). \label{eq:g1}
\end{align}
The integration of the two-loop ladder \p{eq:h2}, to be discussed in the following seciton,
constitutes one of the main goals of the present work. 

The factorizable two-loop negative geometry $\laddladdpic$ is easy to evaluate. Indeed, its integrand \p{eq:Om2prod} requires only one-loop integrations, which are the same as for the one-loop ladder in \cref{eq:h1},
\begin{align} F^{(\laddladdpic)}(AB_0) = \sum_{ab,cd} \Omega_0^{ab,cd-}(AB_0) h^{(1)}_{ab} (AB_0) h^{(1)}_{cd}(AB_0) \,.
\end{align}
In the frame $x_0 \to \infty$, it results in the sum of products of the one-loop pure functions $\{ g^{(1)}_i \}_{i=1}^{5}$ \p{eq:g1}. As compared to the ladders and non-decomposed $F^{(L)}$, this negative geometry involves all eleven rational prefactors \p{eq:r0r11},
\begin{align}
F^{(\laddladdpic)} = -r_0 \left(\sum_{j=1}^5 g_j^{(1)} \right)^2 + \sum_{i=1}^{5} r_i \left(- g_i^{(1)}  + 2\sum_{j=1}^5 g_j^{(1)} \right) g_i^{(1)}  + 2\sum_{i=1}^{5} \overline{r}_{i}\, g_{i+2}^{(1)} g_{i+3}^{(1)}\,, \label{eq:F2Lfact}
\end{align}
where we tacitly assume cyclicity of the summation indices, i.e. $6 \equiv 1$, and use relations \p{eq:LStox0inf}.

Once the two-loop ladder $\laddtwopic$ \p{eq:F2Ll}, the factorized negative geometry $\laddladdpic$ \p{eq:F2Lfact}, and the nondecomposed $F^{(2)}$ \p{eq:F2L} are known, the decomposition equation \p{eq:Fdecomp2} 
immediately provides an expression for the ``loop'' negative geometry $\looppic$. The latter involves all eleven rational prefactors \p{eq:r0r11},  
\begin{equation}\label{eq:F2Lloop}
\begin{aligned} F^{\left(\looppic\right)} = & r_0\left[ 2 g_0^{(2)} - \left(\sum_{j=1}^{5}g_j^{(1)}\right)^2 - 2 \sum_{j=1}^{5} h_j^{(2)} \right] \notag\\ & + \sum_{i=1}^{5} r_i \left[ 2g_i^{(2)}+2h_i^{(2)}- \left(g_i^{(1)}\right)^2  + 2 g_i^{(1)} \sum_{j=1}^5 g_j^{(1)} \right] + 2\sum_{i=1}^{5} \overline{r}_{i}\, g_{i+2}^{(1)} g_{i+3}^{(1)} \,. 
\end{aligned}
\end{equation}
This completes the discussion of the general structure of the two-loop corrections. The only unknown terms are the functions $h_j^{(2)}$. The next section is devoted to their computation.

\section{Two-loop nonplanar Feynman integrals for the negative geometries}
\label{sect:2LNPFI}

We are going to perform loop integrations in the integrand of the two-loop ladder $\laddtwopic$\,, see \cref{eq:h2}. In order to achieve this goal, we rely on a conventional Feynman integral calculation. Namely, we rewrite the momentum-twistor integrand as a linear combination of two-loop scalar Feynman integrals in the dimensional regularization, and then we calculate the contributing Feynman integrals. This is a universal approach of perturbative calculations in a QFT. However, a considerable drawback of this universal approach is that all nice properties of the negative geometry (e.g. its finiteness) are not manifest at the intermediate steps of the calculation. 

In \cite{Chicherin:2022zxo}, the Lagrangian insertion in the Wilson loop at two-loop order, $F^{(2)}$ \p{eq:Fl}, also has been calculated from the Feynman integrals in dimensional regularisation. In that case, the relevant family of Feynman integrals is the planar penta-box \cite{Gehrmann:2015bfy,Gehrmann:2018yef}. The planar penta-box family is depicted in \cref{sect:integr2LL}. It turns out that the two-loop ladder integrand from \cref{sect:mtintegrand} involves a larger family of two-loop Feynman integrals. One can also easily understand it from the momentum twistor expression for the integrand \p{Omega2}, which is essentially nonplanar -- its propagators cannot be drawn as planar graphs in the dual-momentum space. In this section, we identify a family of two-loop Feynman integrals which is required in our calculation of the two-loop ladder $\laddtwopic$\,, and we analytically calculate these Feynman integrals using a differential equation approach \cite{Henn:2013pwa,Henn:2014qga}. 

\subsection{Two-loop five-point eleven-propagator family of Feynman integrals}
\label{sect:11propfamily}

We consider the following family of two-loop Feynman integrals in $D=4-2\ep$ dimensions which have kinematics of the five-particle massless scattering, see \cref{sect:5part},
\begin{align}
\label{eq:integral_def}
G_{\vec a}(X,\ep)  := \int \frac{d^{D}k_{1} d^{D}k_{2}}{(i \pi^{D/2})^2} \frac{1}{\prod_{i=1}^{11} D^{a_i}_i} =  \int\frac{d^{D}x_{6} d^{D} x_{7}}{(i \pi^{D/2})^2} \frac{1}{\prod_{i=1}^{11} {\cal D}^{a_i}_i}\,,
\end{align}
where we recall that $X$ denotes the set of five-particle Mandelstam variables, ${\vec a} := \{ a_1, \ldots a_{11} \}$ is a vector of integer numbers, and the 11 propagators are defined as follows
in terms of momenta and space-time coordinates (dual momenta), see \cref{eq:dual_momentum},  
\begin{center}
\begin{tabular}{ccc}
\toprule
     &  $D_i$ in $p$-space & ${\cal D}_i$ in $x$-space \\     
\midrule
   $1$ & $k_{1}^2$ & $x_{57}^2$\\
   $2$ & $(k_{1}+p_{1})^2$ & $x_{17}^2$\\
   $3$ & $(k_{1}+p_{1}+p_{2})^2$ & $x_{27}^2$\\
   $4$ & $(k_{1}-p_{4}-p_{5})^2$ & $x_{37}^2$ \\
   $5$ & $k_{2}^2$ & $x_{56}^2$ \\
 & &   \\
   \bottomrule
\end{tabular} \qquad\qquad 
\begin{tabular}{ccc}
\toprule
     &  $D_i$ in $p$-space & ${\cal D}_i$ in $x$-space \\     
\midrule
   $6$ & $(k_{2}-p_{4}-p_{5})^2$ & $x_{36}^2$\\
   $7$ & $(k_{2}-p_{5})^2$ & $x_{46}^2$\\
   $8$ & $(k_1-k_2)^2$ & $x_{67}^2$\\
   $9$ & $ (k_{1}-p_{5})^2$ & $ x_{47}^2$ \\
   $10$ & $(k_{2}+p_{1})^2$ & $x_{16}^2$ \\
   $11$ & $(k_{2}+p_{1}+p_{2})^2$ & $x_{26}^2$ \\
   \bottomrule
\end{tabular}
\end{center}
The family \p{eq:integral_def} is closed under dihedral permutations \p{eq:tau}, which act on the dual momenta $x_1,\ldots,x_5$. Namely, a dihedral permutation $\sigma$ of $G_{\vec{a}}$ generates a permutation $\sigma(\vec{a})$ of the list $\vec{a}$,
\begin{align}
\sigma\left(  G_{\vec a}  \right) = G_{\sigma(\vec a)} \,.  \label{eq:sigmaG}
\end{align}

\begin{figure}[t]
\centering
\subfloat[]{
\includegraphics[scale=0.325]{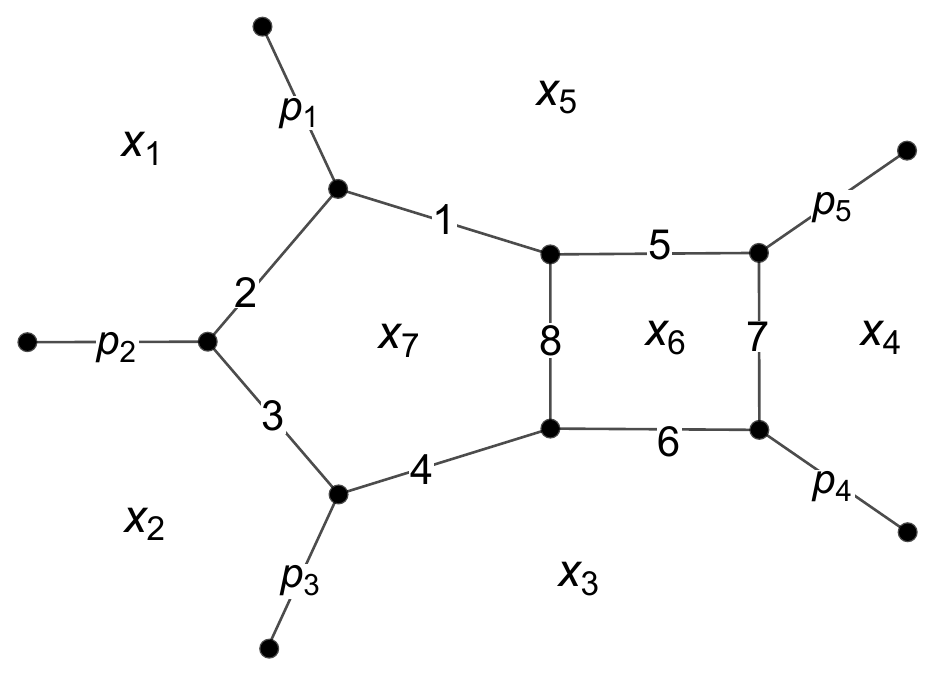}
\label{fig:I_np_prop11_a}
}
\;
\subfloat[]{ 
\includegraphics[scale=0.33]{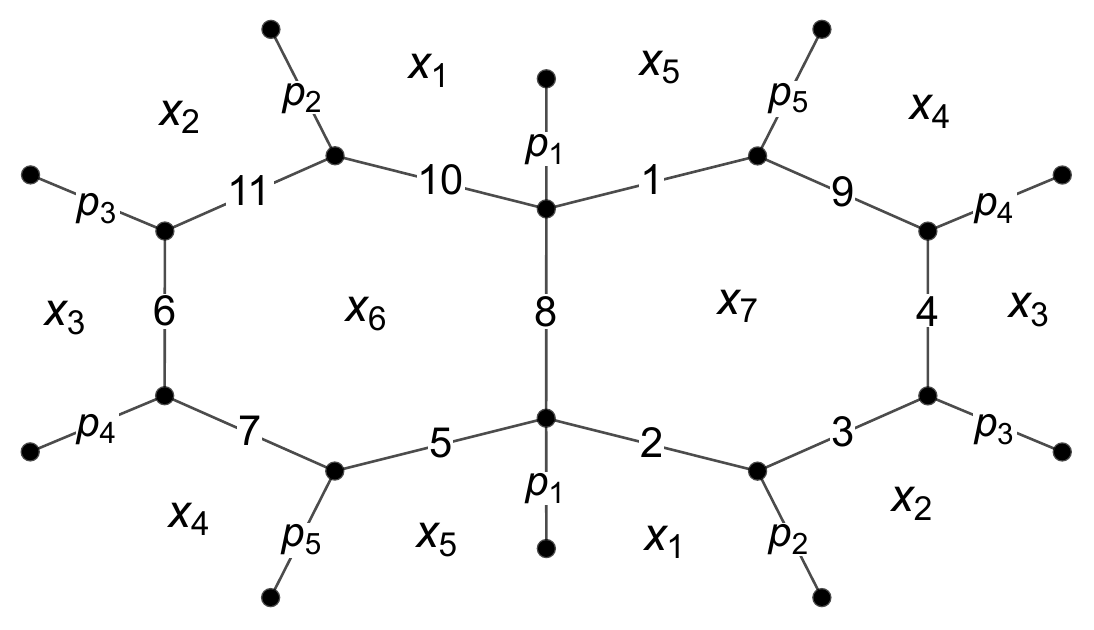}
	\label{fig:I_np_prop11_b}
}
\caption{Diagram (a) represents a  
penta-box diagram contained in integral family \eqref{eq:integral_def} in the dual-momentum variables. 
Diagram (b) represents the complete 11-propagator integral family \eqref{eq:integral_def} in the dual-momentum variables.}
\end{figure}

The planar penta-box family \cite{Gehrmann:2018yef}, which involves 8 propagators, is contained in \cref{eq:integral_def}. For example, if $a_{9},\,a_{10},\,a_{11} \in \mathbb{Z}_{\leq 0}$ then $D_9, \, D_{10},\, D_{11}$ could appear only in the numerator, and the corresponding graph is depicted in \cref{fig:I_np_prop11_a}. The graph is planar both in momentum and dual-momentum space. The kinematics is that of five-particle massless scattering, i.e. the momentum space graph has five legs which carry light-like momenta $p_1,\ldots,p_5$. Let us note that \cref{fig:I_np_prop11_a} is not the only way to identify the planar penta-box among \cref{eq:integral_def}. Acting with dihedral permutations on \cref{fig:I_np_prop11_a} we find further penta-box subtopologies in the 11-propagator family \p{eq:integral_def}.

\begin{figure}[t]
\centering
\subfloat[]{
\includegraphics[scale=0.33]{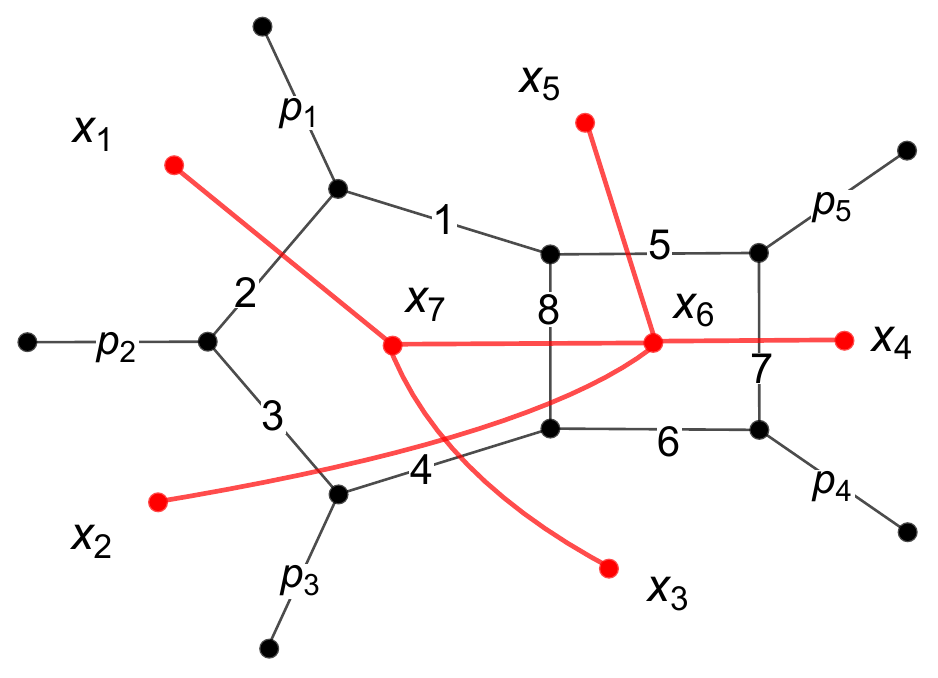}
\label{fig:neg_geom_int_a}
}
\;
%
\subfloat[]{ 
\includegraphics[scale=0.32]{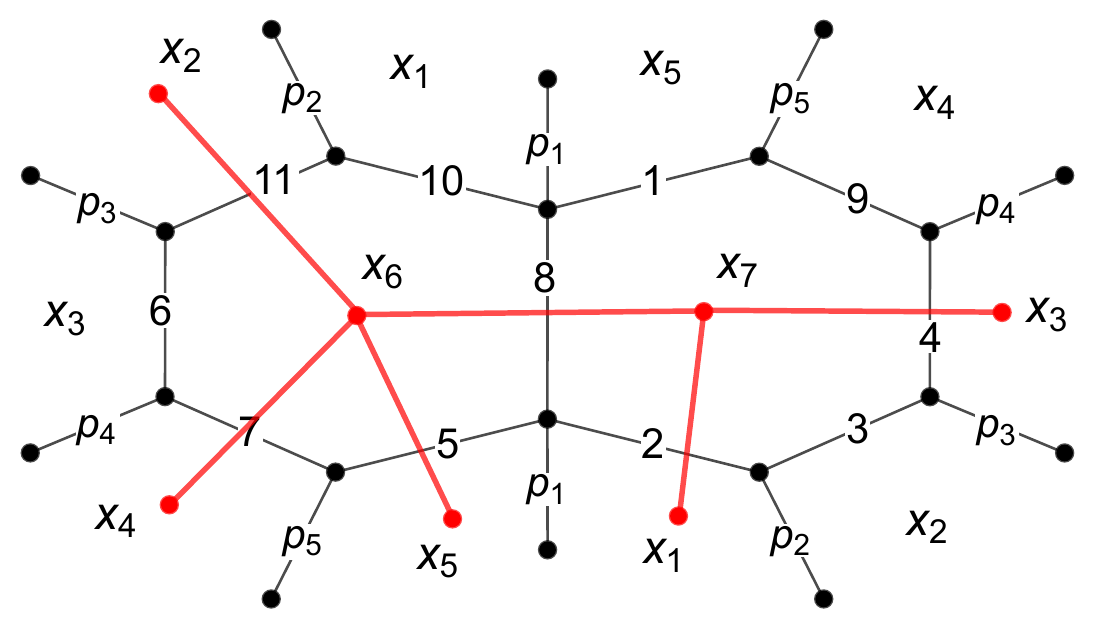}
 \label{fig:neg_geom_int_b}
}
\caption{
The diagrams show a non-planar six-propagator Feynman integral drawn in the (a) penta-box 8-propagator family (b) 11-propagator family. 
The red lines denote the propagators in the dual-momentum space.
Note the non-planarity means the crossing of $x^2_{26}$ and $x^2_{37}$ in diagram (a). However, the \textit{same} Feynman integral drawn in diagram (b) avoids this crossing by introducing double covering of external $x_1,\ldots,x_5$.
The diagram (b) allows us to translate back from dual-momentum coordinates and draw a momentum space box-triangle diagram $\mathrm{BT}_2$ in Fig.\ref{fig:I_np_prop_all}
}
\label{fig:neg_geom_int}
\end{figure}

\begin{figure}[h]
 \centerline{
   \subfloat[${\mathrm{KT}}$ (4 MIs)]{\includegraphics[scale=0.3]{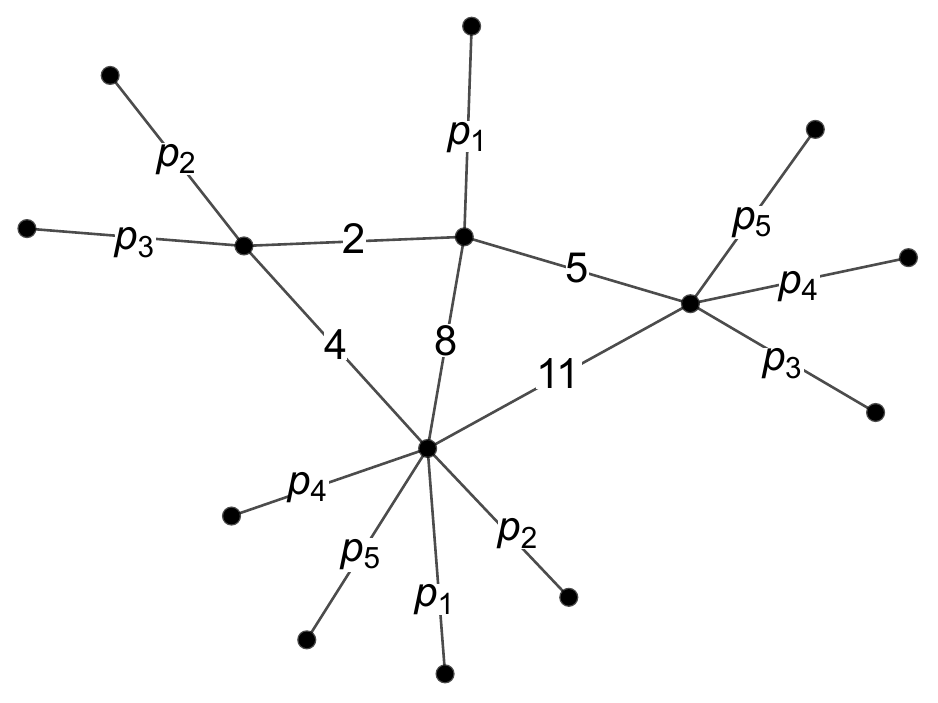}}
    \subfloat[${\mathrm{BT}_1}$ (1 MI)]{\includegraphics[scale=0.28]{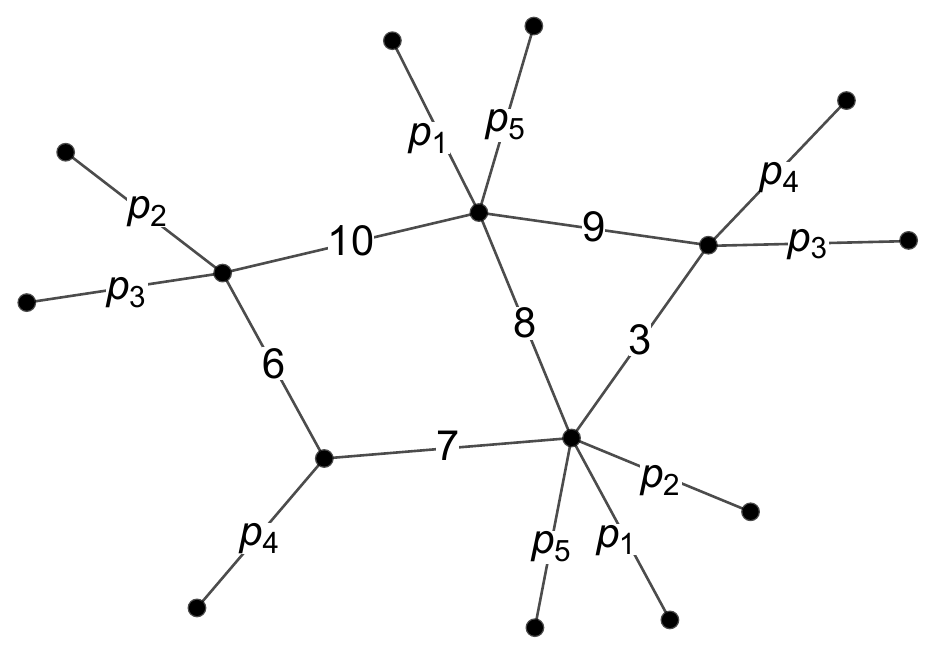}}
    \subfloat[${\mathrm{BT}_2}$ (6 MIs)]{\includegraphics[scale=0.29]{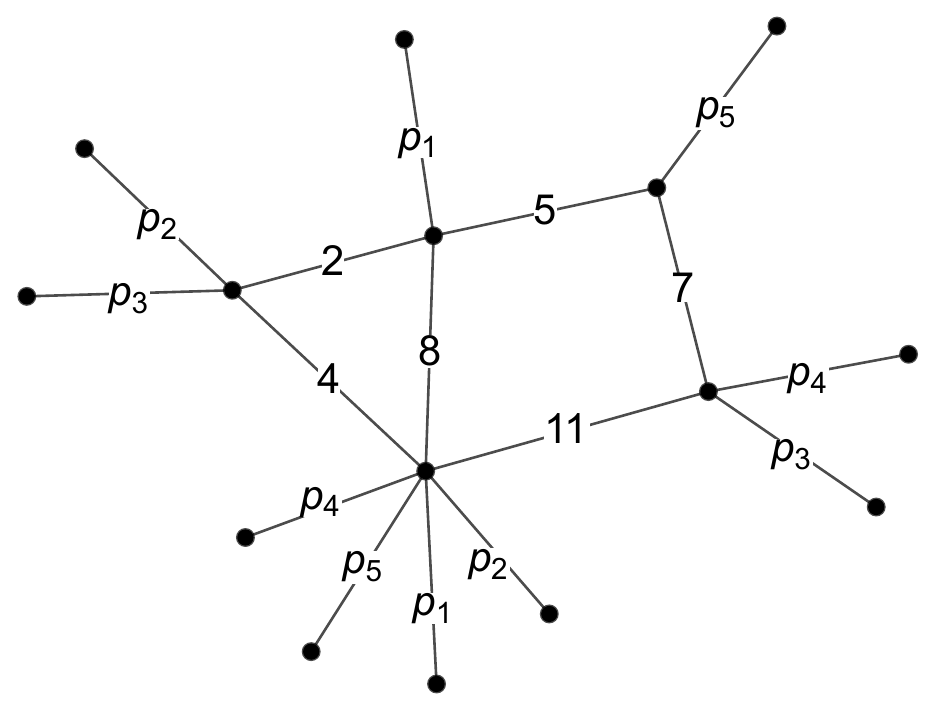}}      
 }
  \centerline{
   \subfloat[${\mathrm{PT}}$ (1 MI)]{\includegraphics[scale=0.275]{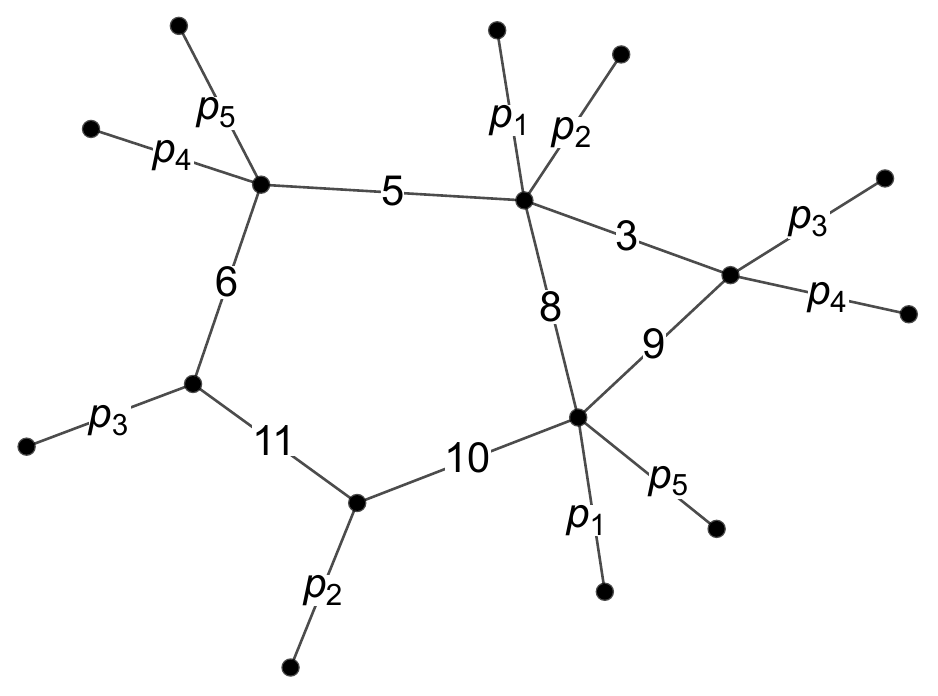}}
    \subfloat[${\mathrm{DB}_1}$  (2 MIs)]{\includegraphics[scale=0.275]{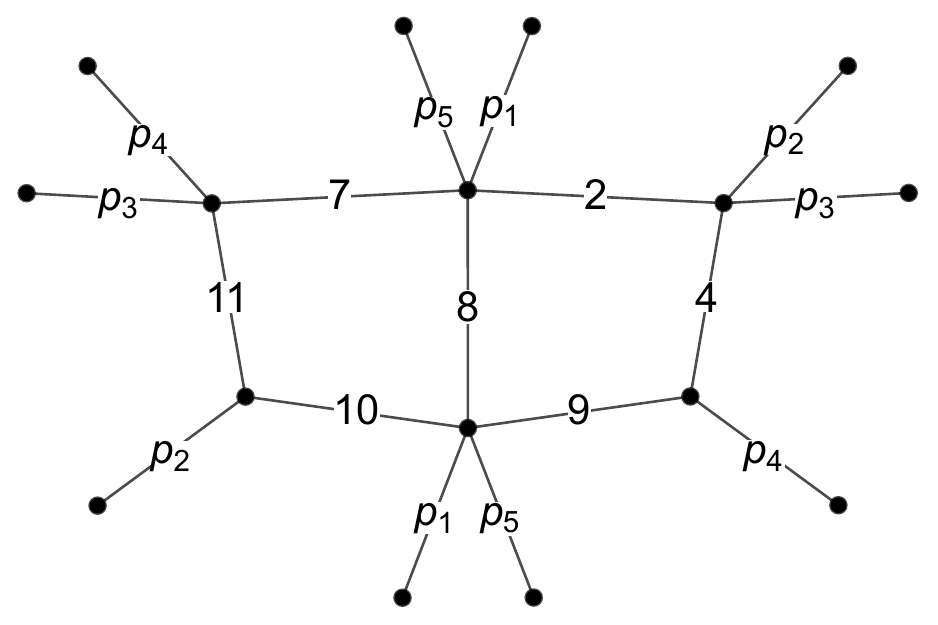}}
    \subfloat[${\mathrm{DB}_2}$ (3 MIs)]{\includegraphics[scale=0.25]{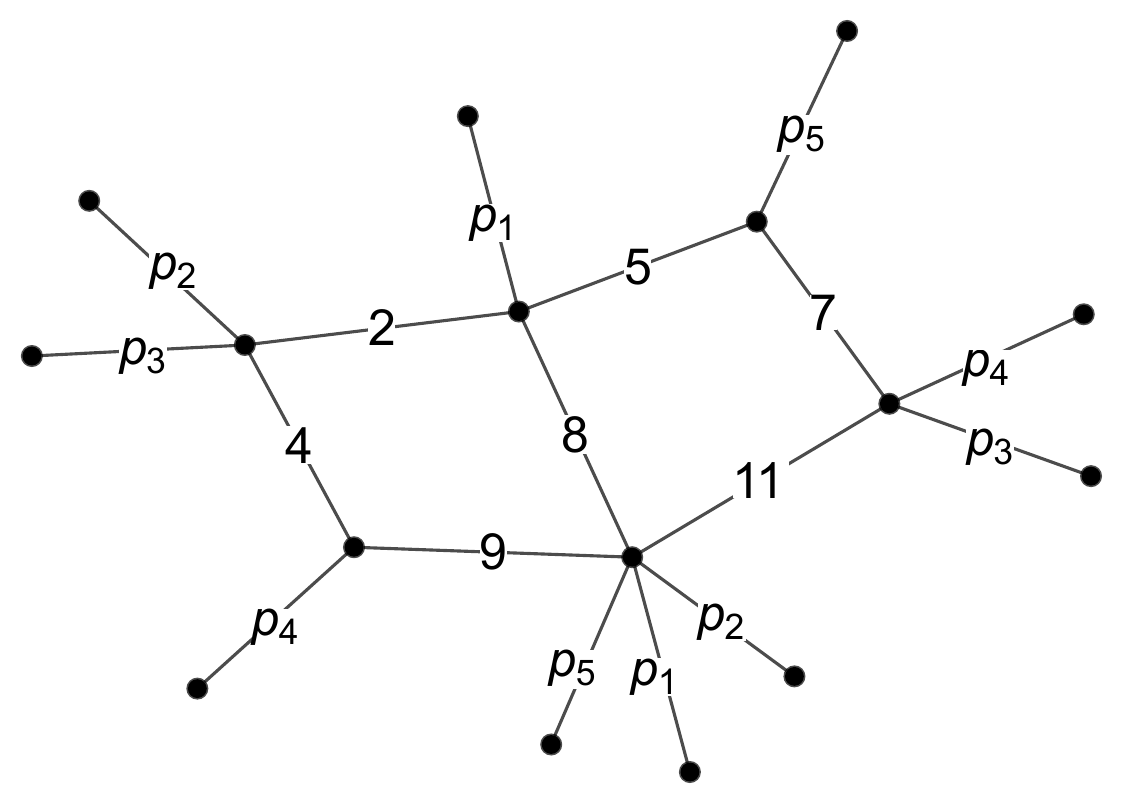}}    
 }

    \centerline{
   \subfloat[${\mathrm{PB}_{\mathrm{np}}}$ (3 MIs)]{\includegraphics[scale=0.25]{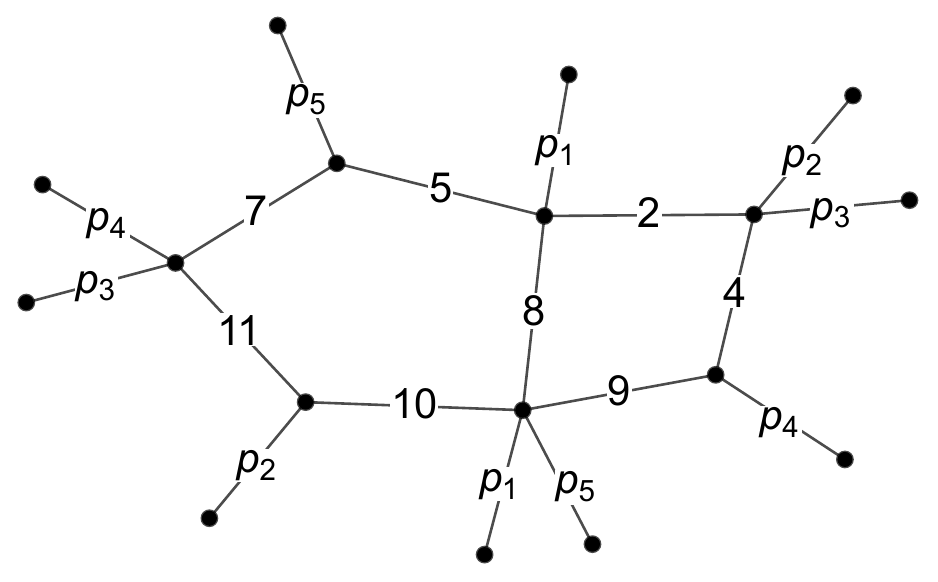}}
    \subfloat[${\mathrm{DP}_1}$  (1 MIs)]{\includegraphics[scale=0.25]{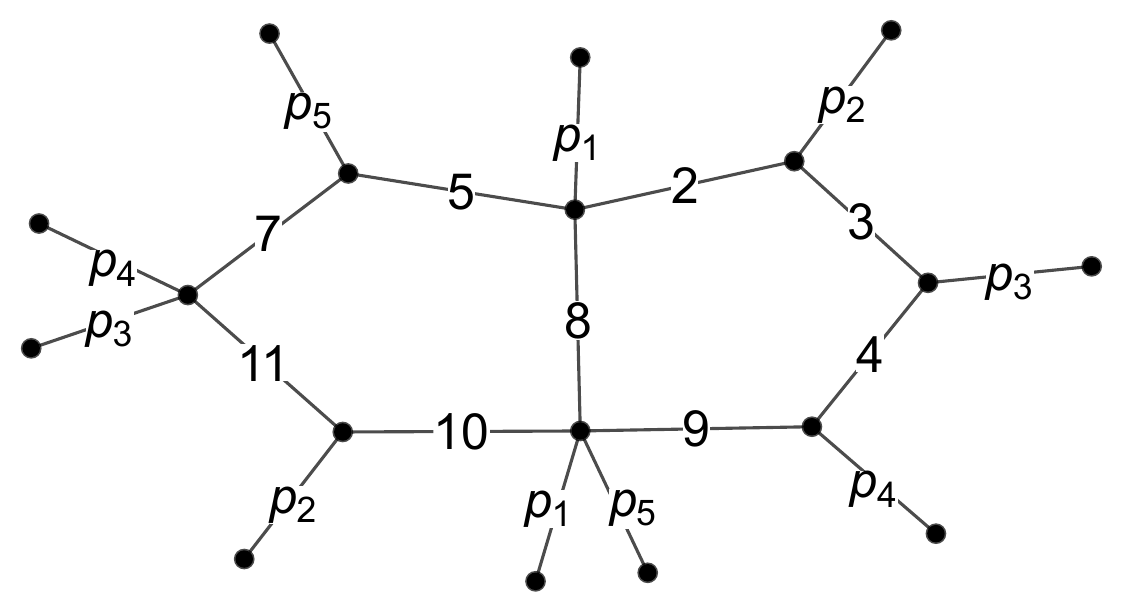}}
    \subfloat[${\mathrm{DP}_2}$ (1 MI)]{\includegraphics[scale=0.235]{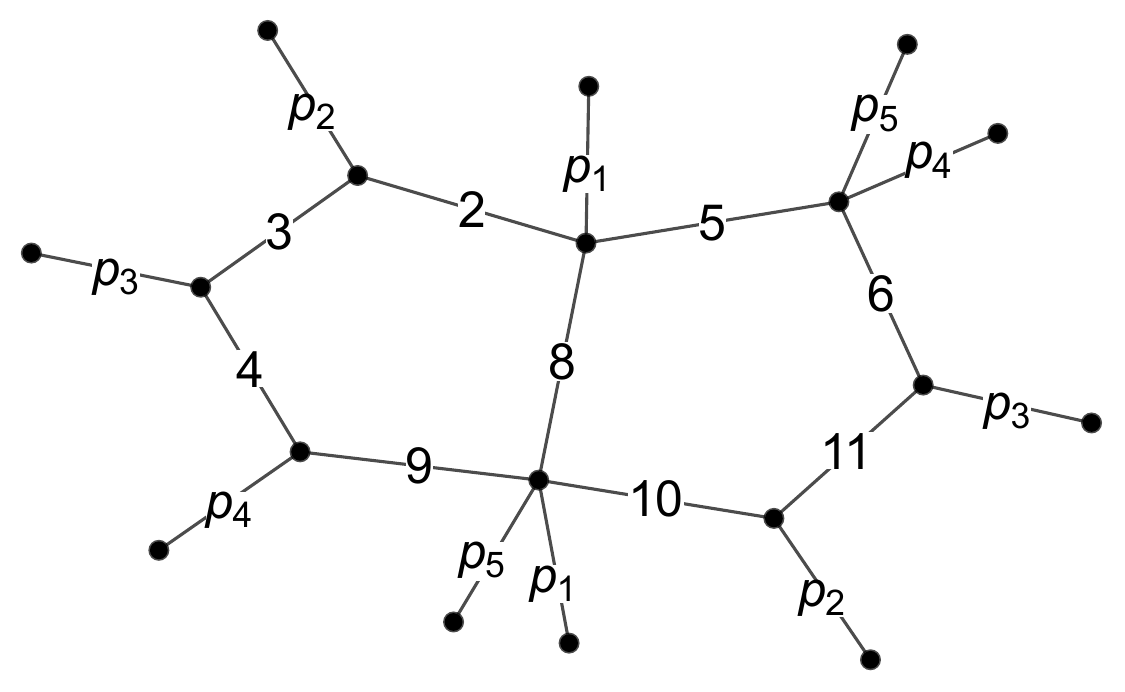}}    
 }
   \caption{The independent nonplanar sectors with up to 9 propagators, and the counting of corresponding MIs on the maximal cuts.}
 \label{fig:I_np_prop_all}
\end{figure}

As compared to the 8-propagator penta-box, the 11-propagator family \p{eq:integral_def} is {\em nonplanar} in the dual-momentum space. In general, the Feynman integrals \p{eq:integral_def} cannot be drawn as planar graphs on the penta-box 8-propagator family. We find it convenient to avoid drawing nonplanar graphs by introducing a double covering of external coordinates $x_1,\ldots,x_5$, see \cref{fig:I_np_prop11_b}. For example, let us consider a six-propagator Feynman integral \p{eq:integral_def} specified by
\begin{align}
\vec{a} = \{ 0, 1, 0, 1, 1, 0, 1, 1, 0, 0, 1 \} \,. \label{eq:exFI}
\end{align}
We depict it in \cref{fig:neg_geom_int_a} as a nonplanar graph, and as a planar graph using the double covering in \cref{fig:neg_geom_int_b}. Let us stress that non-planarity refers to the dual-momentum space, but not the momentum space! In general, the Feynman integrals \p{eq:integral_def} are not the usual amplitude Feynman integrals (planar or nonplanar) with five massless legs. Drawn in momentum space, they have up to 10 massless legs carrying momenta $p_1,\ldots,p_5,p_1,\ldots,p_5$, see \cref{fig:I_np_prop11_b}. For example, the six-propagator Feynman integral \p{eq:exFI} depicted in momentum space is $\mathrm{BT}_2$ in \cref{fig:I_np_prop_all}, which is obtained from \cref{fig:neg_geom_int_b} by pinching 5 propagators, and which carries external momenta $p_1,\,p_2+p_3,\,-p_3,\,p_3+p_4,\,p_5$.

\begin{table}[t]
\begin{center}
\begin{tabular}{ccc|c}
\toprule
Non-planar & Planar penta-box & One-loop factorized & Total MI \\ \midrule
135 & 140 & 66 & 341 \\
\bottomrule
\end{tabular}
\end{center}
\caption{Counting of the MIs in the family \p{eq:integral_def}.}
\label{tab:misall}
\end{table}

There are $\mathbb{Q}(X)$-linear dependences among Feynman integrals of the family \p{eq:integral_def} that follow from IBP relations \cite{Tkachov:1981wb,Chetyrkin:1981qh},. Using standard terminology, we refer to a basis of linearly independent Feynman integrals as {\em master integrals} (MIs). Further, we perform IBP reductions of the family \p{eq:integral_def} relying on the computer codes \cite{Lee:2012cn,Lee:2013mka,Smirnov:2019qkx} to identify MIs in various sectors. As usual, we say that $\{ a_i \}_{i=1}^{11}$ belongs to the {\em sector} $\{ \theta(a_i \geq 1 ) \}_{i=1}^{11}$, which is a list of $0$'s and $1$'s. We say that $\{ \mu_i \}_{i=1}^{11}$ is a subsector of $\{ \lambda_i \}_{i=1}^{11}$ iff $\mu_i \leq \lambda_i$ for $i=1,\ldots,11$ and $\sum_i \mu_i < \sum_i \lambda_i$. A {\em family} $\{ \mu_i \}_{i=1}^{11}$ of Feynman integrals comprises integrals from the sector $\{ \mu_i \}_{i=1}^{11}$ and all its subsectors. 

We find 341 master integrals (MIs) for the 11-propagator family \p{eq:integral_def}. Among them, we identify those which are well-known in the literature. Indeed, 66 MIs factorize into products of one-loop pentagon integrals, i.e. they have $a_8 = 0$. 140 MIs belong to the planar penta-box family \cite{Gehrmann:2015bfy,Gehrmann:2018yef}, i.e. they can be mapped by dihedral transformations \p{eq:sigmaG} onto the sector $\{ 1,1,1,1,1,1,1,1,0,0,0\}$, depicted in \cref{fig:I_np_prop11_a}, and its subsectors. We refer to the remaining 135 MIs as {\em non-planar}, see \cref{tab:misall}.

\begin{table}[t]
\centering
\begin{tabular}{|c|c|c|c|c|c|c|}
\hline 
\# propagators  & 5 & 6 & 7 & 8 & 9  & $\leq$ 9\tabularnewline
\hline 
\hline
\# sectors   & 5 & 15 & 20 & 10 & 10  & 60 \tabularnewline
\hline
\# sectors (mod dihedral)  & 1 & 2 & 3 & 1 & 2  & 9\tabularnewline
\hline
\end{tabular}
\caption{Number of non-planar sectors in the family \p{eq:integral_def} for a given number (up to 9) of propagators.}
\label{table:mis_np}
\end{table}

\begin{table}[t]
\centering
\begin{tabular}{|c|c|c|c|c|c|c|c|c|c|}
\hline 
sector & $\mathrm{KT}$ & $\mathrm{BT}_1$ & $\mathrm{BT}_2$ & $\mathrm{PT}$ & $\mathrm{DB}_1$ & $\mathrm{DB}_2$ & $\mathrm{PB}_{\mathrm{np}}$ & $\mathrm{DP}_1$ & $\mathrm{DP}_2$\tabularnewline
\hline 
\hline
\# propagators & 5 & 6 & 6 & 7 & 7 & 7 & 8 & 9 & 9 \tabularnewline
\hline
\# MIs & 10 & 16 & 29 & 42 & 26 & 55 & 82 & 147 & 155 \tabularnewline
\hline 
\# MIs on cuts & 4 & 1 & 6 & 1 & 2 & 3 & 3 & 1 & 1\tabularnewline
\hline 
\end{tabular}
\caption{Number of MIs, and MIs on maximal cuts, for each non-planar sector.}
\label{table:mis_top}
\end{table}

Let us discuss the 135 non-planar MIs in detail. The Feynman integrals \p{eq:integral_def} with $10$ and $11$ propagators are IBP-reduced to Feynman integrals with nine or fewer propagators. Thus, the highest nonreducible sectors of the family \p{eq:integral_def} contain 9 propagators. The non-planar MIs are categorized into 60
sectors, which are 9 \textit{independent} sectors after modding out dihedral permutations \p{eq:sigmaG}.
The counting of sectors for a given number of propagators is provided in \cref{table:mis_np}. We depict the nine independent non-planar sectors in \cref{fig:I_np_prop_all} and give them the following names: kite ($\mathrm{KT}$), two box-triangles (${\mathrm{BT}_1}$, ${\mathrm{BT}_2}$), penta-triangle (PT), two double-boxes (${\mathrm{DB}_1}$, ${\mathrm{DB}_2}$), non-planar penta-box ($\mathrm{PB}_{\mathrm{np}}$), and two double-pentagons (${\mathrm{DP}_1}$, ${\mathrm{DP}_2}$). In \cref{table:mis_top}, we summarize the counting of MIs in these nonplanar families. The number of MIs on the maximal cut is the number of MIs in the given family modulo its lower subsectors.
The 9-propagator non-planar sectors ${\mathrm{DP}_1}$ and ${\mathrm{DP}_2}$,
\begin{align}
&\vec{a}_{\mathrm{DP}_1} = \{ 0, 1, 1, 1, 1, 0, 1, 1, 1, 1, 1 \}\,, \notag\\
&\vec{a}_{\mathrm{DP}_2} = \{0, 1, 1, 1, 1, 1, 0, 1, 1, 1, 1\} \,,\label{eq:DPa}
\end{align}
are the highest independent IBP-irreducible sectors. All nonplanar sectors are contained in ${\mathrm{DP}_1}$ and ${\mathrm{DP}_2}$ and their dihedral permutations \p{eq:sigmaG}.

\subsection{Constructing the pure basis}

We calculate the Feynman integrals \p{eq:integral_def} relying on the method of differential equations (DE) \cite{Henn:2013pwa,Henn:2014qga}, in their canonical form \cite{Henn:2013pwa}.
Let $T$ be a family of Feynman integrals, which is contained in  \p{eq:integral_def}, and let $\vec{\bf u}_T$ denote (a particular choice) for its set of MIs. By definition, the MIs are closed under taking derivatives in kinematic variables. We say that $\vec{\bf u}_T$ is pure if it satisfies the following system of DEs \cite{Henn:2013pwa}, 
\begin{align}
\frac{\pa}{\pa v}\vec{\bf u}_T(X,\ep) = \ep \,  A_{T,v}(X) \,\vec{\bf u}_T (X,\ep) \;,\qquad v \in X \label{eq:DEv}
\end{align}
where we take derivatives in the Mandelstam variables $X$ of eq. \p{eq:s}. The entries of connection matrices $ A_{T,v}$ are algebraic functions of the kinematics.

The pure bases for the planar penta-box family and the product of one-loop pentagons, are known \cite{Gehrmann:2015bfy,Gehrmann:2018yef}.
Furthermore, some of the subsectors of the remaining integral families can be identified with Feynman integrals for which a pure basis is known in the literature.  
\begin{figure}[t]
\centering
\subfloat[]{
\raisebox{0\height}{
\includegraphics[scale=0.35]{figs/graph_KT.pdf}
}
\label{fig:KT_a}
}
\quad\quad
\subfloat[]{ 
\includegraphics[scale=0.35]{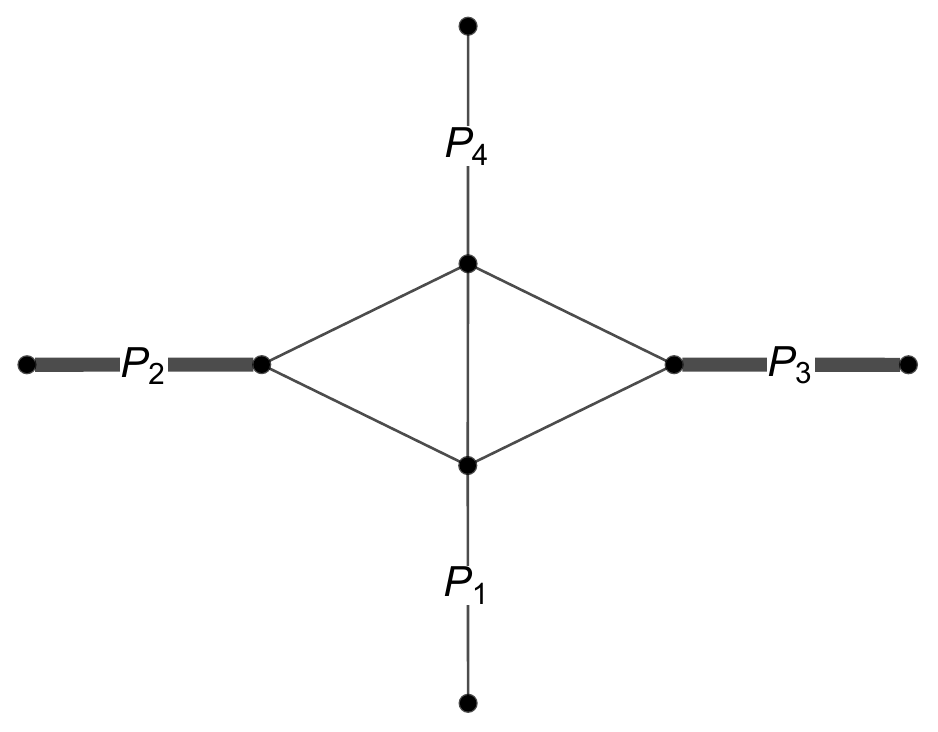}
\label{fig:KT_b}
}
\caption{
Diagram (a) is a kite integral in the non-planar space with 
5-point kinematics, which can viewed as a (planar) 4-point kite integral with 2 masses in diagram (b). The kinematic map between two diagrams is given in \eqref{eq:kin_map}.
}
\end{figure}

For example, consider the five-propagator kite integral ($\mathrm{KT}$) shown in \cref{fig:KT_a}. This sector coincides with a four-point two-mass two-loop family of Feynman integrals calculated in \cite{Henn:2014lfa}, see  \cref{fig:KT_a}, if we identify the kinematics
\begin{equation} \label{eq:kin_map}
   P_1=-p_3\;,\quad P_2=-p_1-p_2\;, \quad P_3 =p_2+p_3\;,\quad P_4=p_1\,,
\end{equation}
so $P_1^2 = P_4^2 = 0$ and $P_2^2, P_3^2 \neq 0$. This is a planar four-particle kinematics of the two-mass-easy type, and the pure MIs require the corresponding square root in their normalization, see \cref{eq:del2},
\begin{equation}
\sqrt{(S+T)^2-4 P_2^2\,P_3^2}=\sqrt{s_{45}^2-4\,s_{12}s_{23}} \equiv \sqrt{\Delta_2^{(4)}} \,,
\end{equation}
where  $S=(P_1+P_2)^2$ and $T=(P_1+P_3)^2$. According to \cref{table:mis_top}, there are 4 MIs on the maximal cut of the sector ${\rm KT}$
which we choose as in \cite{Henn:2014lfa}.

\begin{figure}
    \centering
    \includegraphics[width=0.5\linewidth]{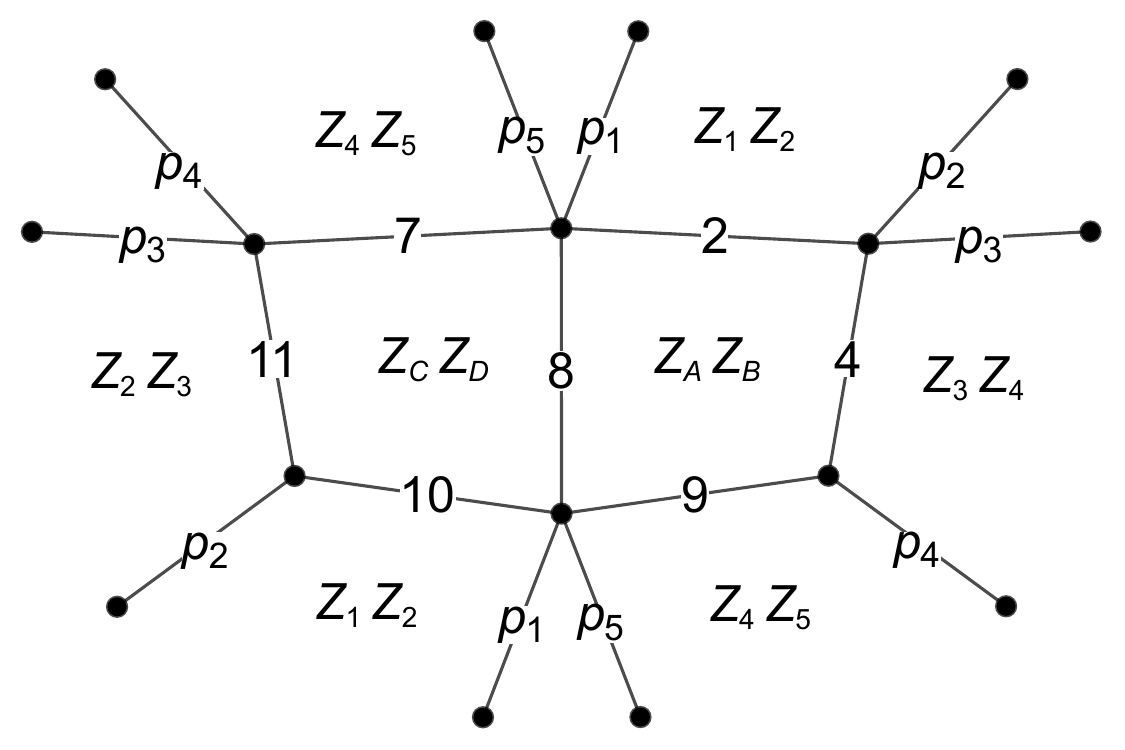}
    \caption{The double-box sector $\mathrm{DB}_1$ with 7 propagators in momentum twistor notation.}
    \label{fig:DB_1_Z}
\end{figure}

For the remaining non-planar sectors, we use the idea that loop integrands without double poles, and with constant leading singularities are expect to give pure Feynman integrals \cite{Arkani-Hamed:2010pyv,Henn:2013pwa}. These integrands are called $\dlog$ integrands, as their integrand can be written as (sums of) products of $d \log x = dx/x$ terms. 
One option is to classify all such integrands, for a given propagator structure and kinematics, cf. for example \cite{Henn:2020lye}. However, here we only need to provide a suitable basis for the differential equations. We therefore proceed in a simpler way and use a four-dimensional loop-by-loop approach.
This analysis of the integrands enables us to find candidates to form a pure basis of MIs on the maximal cut. 
We found the program {\tt{DLOGBASIS}}  \cite{Henn:2020lye} useful to verify the expected integrand properties.
Once a candidate basis is found, we calculate its derivatives, and explicitly verify that they satisfy the canonical ($\ep$-factorized) DE of eq. \p{eq:precanon}.

We find it convenient to employ the momentum twistor parametrization of the integrands. We assign the twistor lines $AB \sim x_7$ and $CD \sim x_6$ to the loop integrations in \p{eq:integral_def}, and intersecting twistors lines $Z_i Z_{i+1} \sim x_i$ with cyclic $i=1,\ldots,5$ to the external dual momenta. We also introduce the infinity bi-twistor $I$ since we have to deal with non-dual-conformal Feynman integrals. 

For example, the MIs of the double-box sector $\mathrm{DB}_1$ have the following form, see \cref{fig:DB_1_Z},
\begin{align} \label{eq:DB_1_UT}
I^{(i)}_{\mathrm{DB}_1} := \int\limits_{AB,CD}    \frac{N^{(i)}_{\mathrm{DB}_1}}{\vev{AB12}\vev{AB34}\vev{AB45}\vev{ABCD}\vev{CD12}\vev{CD23}\vev{CD45}\vev{ABI}\vev{CDI}}\,.
\end{align}
According to \cref{table:mis_top}, there are 2 MIs on the maximal cut. In order to find these MIs in pure form, we proceed loop-by-loop working out the $\dlog$ form of the integrands. We take into account that the following $CD$-subintegral of \p{eq:DB_1_UT} is a $\dlog$ four-form $\Omega^{(4)}$ accompanied by the three-mass box leading singularity factor, see (2.41) in \cite{Arkani-Hamed:2010pyv},
\begin{align}
\frac{1}{\vev{ABCD}\vev{CD12}\vev{CD23}\vev{CD45}} \frac{d^4 Z_C d^4 Z_D}{{\rm vol}(GL(2))}= \frac{1}{\vev{AB (123) \cap (245)}} \, \Omega^{(4)}  (CD)\,. \label{eq:dlog3mh}
\end{align}
Choosing the following numerator
\begin{align}
    N^{(1)}_{\mathrm{DB}_1}=\vev{1234}\vev{1245}\vev{2345}\vev{ABI}\vev{CDI}\,
\end{align}
and substituting the $\dlog$ form \p{eq:dlog3mh} in \cref{eq:DB_1_UT}, we end up with the one-mass box $AB$-subintegral with the unit leading singularity. Translating \cref{eq:DB_1_UT} with numerator \p{eq:dlog3mh} in momentum notations, we obtain 
\begin{align}
I^{(1)}_{\mathrm{DB}_1} = s_{23} s_{34} s_{45} \, G_{0,1,0,1,0,0,1,1,1,1,1}\,.
\end{align}
 
A second MI on the maximal cut of $\mathrm{DB}_1$ can be obtained e.g. by canceling the three-mass leading singularity in \p{eq:dlog3mh}. Namely, choosing the numerator
\begin{align} \label{eq:N_DB_1}
    N^{(2)}_{\mathrm{DB}_1}=\vev{AB(123)\cap(245)} \vev{I (124)\cap (345) } \vev{CDI}\,
\end{align}
and substituting the $\dlog$ form \p{eq:dlog3mh} in \cref{eq:DB_1_UT}, we end up with the three-mass box $AB$-subintegral with the unit leading singularity. Rewriting the numerator \p{eq:N_DB_1} in momentum notations with the help of the Schouten identity, we obtain 
\begin{align}
I^{(2)}_{\mathrm{DB}_1}  = (s_{23}-s_{15}) \left(s_{34}\,G_{0,0,0,1,0,0,1,1,1,1,1}-s_{15}\,G_{0,1,-1,1,0,0,1,1,1,1,1}\right) .
\end{align}

\begin{figure}
    \centering
    \includegraphics[width=0.5\linewidth]{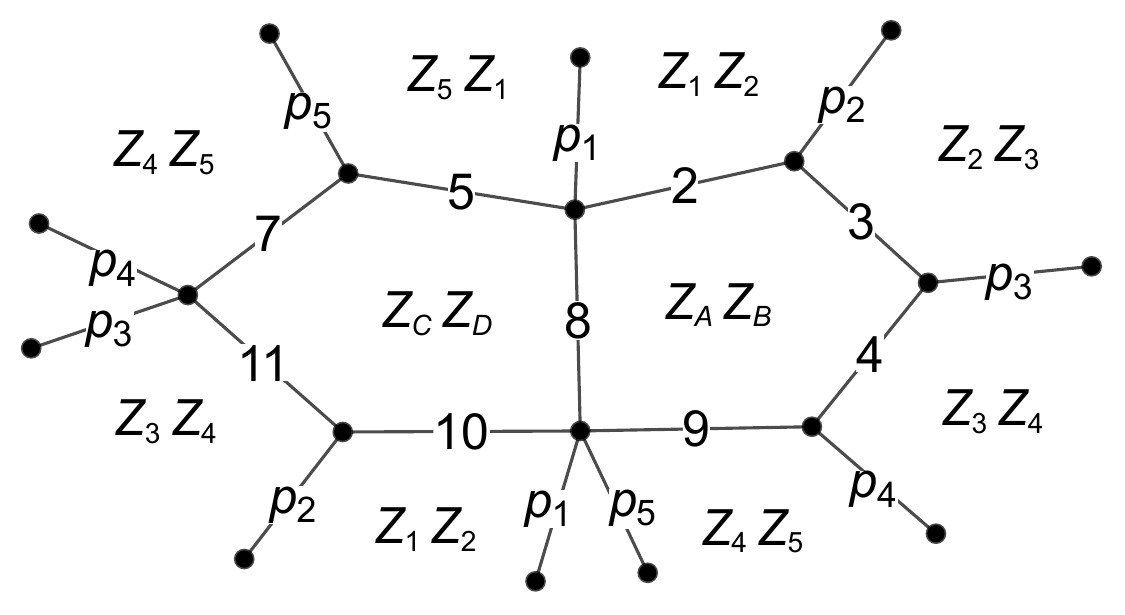}
    \caption{The double-pentagon, $\mathrm{DP}_1$, in the momentum twistor variables.}
    \label{fig:DP_1}
\end{figure}

Similarly, we perform a $\dlog$ analysis of the remaining nonplanar sectors. 
Let us explain here how it works for the sectors with highest number of propagators.
There are two such sectors, each containing one MI, see \cref{fig:I_np_prop_all}. In \cref{fig:DP_1} we present $\mathrm{DP}_1$ in momentum twistor notations, 
\begin{align}
I_{\mathrm{DP}_1} := \int\limits_{AB,CD} \frac{ N_{\mathrm{DP}_1}}{\vev{AB12}\vev{AB23}\vev{AB34}\vev{AB45}\vev{ABCD}\vev{CD12}\vev{CD23}\vev{CD45}\vev{CD51}}  \,. \label{eq:DP1mt}
\end{align}
The one-loop pentagon $CD$-subintegral is put in $\dlog$ from upon choosing the magic numerator \cite{Arkani-Hamed:2010pyv},
\begin{align}
\vev{CD \overline{24}} \equiv \vev{CD(123)\cap(345)}\,.
\end{align} 
The leading singularity $1/\vev{AB25}$ of the $CD$-subintegral complements the $AB$-subintegral to the pentagon one-loop integral, which takes the $\dlog$ form provided it has the magic numerator $\vev{AB24}$. Summarizing, the integral $\mathrm{DP}_1$ \p{eq:DP1mt} with complex numerator
\begin{align}
N_{\mathrm{DP}_1} = \vev{AB24}\vev{CD \overline{24}}\vev{1235}\vev{1245}
\end{align}
is in $\dlog$ form with unit leading singularities. 
Calculating derivatives of $I_{\mathrm{DP}_1}$, we verify that the DE is $\ep$-factorized on the maximal cut. 

Similarly, the magic numerator of the pentagon one-loop sub-integral and the loop-by-loop calculation of the $\dlog$ form, enable us to identify the pure MI in the $\mathrm{DP}_2$ sector
\begin{align}
I_{\mathrm{DP}_2} := \int\limits_{AB,CD} \frac{\vev{AB24}\vev{CD \overline{24}}\vev{1235}\vev{1245}}{\vev{AB12}\vev{AB23}\vev{AB34}\vev{AB45}\vev{ABCD}\vev{CD12}\vev{CD23}\vev{CD34}\vev{CD51}}  \,. 
\end{align} 

Ideally, the above analysis leads directly to differential equations in canonical form. 
However, it is sometimes convenient to find a pure basis for a sector $T$, we identify a pure basis on the maximal cut first. The derivatives of MIs on the maximal cut are coupled which corresponds to the diagonal block of the connection matrix. For a pure basis on the maximal cut, the diagonal block is in $\ep$-factorized form. We then proceed to the off-diagonal part. We supplement the MIs from the maximal cut with MIs belonging to all lower subsectors of $T$ and denote all of them $\vec{I}_{T}$. We choose the MIs from the lower subsectors to be pure by induction. Usually, one can easily choose a pure basis on the maximal cut such that the DE takes the pre-canonical form,
\begin{align}
\frac{\pa}{\pa v}\, \vec{I}_{T}(X,\ep)  = \left( B_{T,v}(X)+ \ep A_{T,v} (X)\right) \vec{I}_{T}(X,\ep) \label{eq:precanon}
\end{align}
where $v \in X$ \p{eq:s}. The nonzero entries of the off-diagonal block matrix $B_{T,v}$ correspond to an admixture of the lower subsector MIs to the maximal cut MIs. A redefinition of the maximal cut MIs eliminates $ B_{T,v}$ and puts the DE into the $\ep$-factorized form \p{eq:DEv}. 
We collect the full basis of MIs in an accompanying ancillary file.

\subsection{Canonical differential equations}
\label{sec:canon}

In the previous section, we have constructed a pure basis of MIs, which satisfy DE \p{eq:DEv}. As we explained at the end of \cref{sect:11propfamily}, any nonplanar Feynman integral of the family \p{eq:integral_def} can be expanded in the MIs of two independent $9$-propagator families ${\mathrm{DP}_1}$ and ${\mathrm{DP}_2}$ \p{eq:DPa} and their dihedral permutations \p{eq:sigmaG}. The pure bases for the planar subtopologies and the one-loop products are known. They are not required in our calculation of the ladder negative geometry. Thus, it will be enough to consider the pure bases $\vec{\bf u}_T$ for families $T={\mathrm{DP}_1},\,{\mathrm{DP}_2}$.

We combine DEs \p{eq:DEv} into the canonical DE \cite{Henn:2013pwa},
\begin{align}
d\vec{\bf u}_T(X,\ep) = \ep \,  dA_{T}(X) \,\vec{\bf u}_T (X,\ep) \label{eq:DEu}
\end{align}
with the total differential $d$  in all kinematic variables $X$, 
\begin{align}
d = \sum_{v \in X} d v \frac{\pa}{\pa v} \,.
\end{align}
The connection matrix $d A_T(X) $ is a $\mathbb{Q}$-linear combinations of the $\dlog$ forms which we refer to as the {\em alphabet letters},
\begin{align} \label{eq:dA_T}
d A_T(X) :=  \sum_{v \in X} A_{T,v} \, dv = \sum_{i=1}^{111} d\log\left(w_i(X)\right) A_{i,T} \,,
\end{align}
$A_{i,T}$ are matrices of rational numbers. In what follows, we find that the family \p{eq:integral_def} requires 111 alphabet letters $\{ w_i(X) \}_{i=1}^{111}$, which are algebraic functions of the kinematics $X$ \p{eq:s}. We present them in \cref{sec:2L_even_letters} and \cref{sect:AppLetters}. We also provide canonical DE \p{eq:DEu} for nonplanar families $T={\mathrm{DP}_1},\,{\mathrm{DP}_2}$ in ancillary files.

In order to solve the canonical DE \p{eq:DEu}, we series expand the pure MIs in $\ep$ and normalize them such that their expansion starts at finite order,
\begin{align}
\vec{\bf u}_T (X,\ep) = \sum_{k \geq 0} \ep^{k} \, \vec{\bf u}^{(k)}_T (X) \,.
\label{eq:uexpandeps}
\end{align}
Using values of the pure basis MIs at $X=X_0$ as initial values, we solve the DE analytically in terms of Chen iterated integrals \cite{Chen:1977oja},
\begin{align}
\vec{\bf u}^{(k)}_T (X) =  \sum_{m=0}^{k} \sum_{i_1,\ldots,i_m} \left( A_{i_m,T} \ldots  A_{i_1,T} \vec{\bf u}_T^{(k-m)}(X_0) \right) \left. [w_{i_1},\ldots,w_{i_m}]\right._{X_0}(X)  \,,\label{eq:itintDE}
\end{align}
where summations $i_1,\ldots,i_m$ run over the alphabet letters. The iterated integrals with the reference point $X_0$ are defined starting with $[]_{X_0}=1$,
\begin{align} 
\left.[ w_{i_1}, \ldots, w_{i_m} ]\right._{X_0}(X) & = \int_0^1 dy \, \frac{\partial \log\left[w_{i_m}(\gamma(y)) \right]}{\partial y} 
  \,\left. [ w_{i_1}, \ldots, w_{i_{m-1}} ]\right._{X_0}\left(\gamma(y)\right),
  \label{eq:defitint}
\end{align}
and the integration path $\gamma$ in kinematic space connects $\gamma(0)=X_0$ and $\gamma(1)=X$. Using analytic representation \p{eq:itintDE}, we can immediately calculate derivatives of the pure MIs and verify that they satisfy \p{eq:DEu}, since
\begin{align}
d \left. [w_{i_1},\ldots,w_{i_m}]\right._{X_0}(X)  = \left. [w_{i_1},\ldots,w_{i_{m-1}}]\right._{X_0}(X)\; d\log \left( w_{i_m}(X)\right). \label{eq:diffitint}
\end{align} 
 
We choose a reference point $X_0$ in the Euclidean region as follows,
\begin{align}
        X_0 = (s_{12}=-1, s_{23} = -1, s_{34} = -1 , s_{45} = -1 , s_{15} = -1)\,, \label{eq:X0}
\end{align}
and we choose the positive branch of the square root $\ep_5(X_0) = \sqrt{5}$ \p{eq:del5}.
Note that $X_0$ is invariant under dihedral transformations \p{eq:tau}.
The analytic solution \p{eq:itintDE} of the canonical DE involves initial values $\left\{ \vec{\bf u}_T^{(k)}(X_0)\right\}_{k\geq 0}$. In order to calculate the negative geometries, the $\ep$-expansion of the pure MIs is required up to order $k=4$. Then the initial values are required up to the same order. 
In principle, these initial values, up to a trivial overall normalization, can be obtained by requiring absence of spurious singularities, see e.g. section 7 of \cite{Henn:2020lye}.
Here, we evaluate them numerically with 70-digit precision using \texttt{AMFlow} \cite{Liu:2022chg}. The initial values $\vec{\bf u}_T^{(0)}(X_0)$ are rational numbers. We assign the transcendental weight $k$ to the initial values $\vec{\bf u}_T^{(k)}(X_0)$ to make \cref{eq:itintDE} of uniform transcendental weight.

Omitting the initial values of weight $k \geq 1$ in \cref{eq:itintDE} is equivalent to not specifying the reference point of the iterated integrals that results in the {\em symbol} expression \cite{Goncharov:2010jf} for the pure MIs, 
\begin{align}
{\cal S}\left( \vec{\bf u}^{(k)}_T (X) \right) =  \sum_{i_1,\ldots,i_k} \left( A_{i_k,T} \ldots  A_{i_1,T} \vec{\bf u}_T^{(0)}(X_0) \right) \left. [w_{i_1},\ldots,w_{i_k}]\right.(X)\,.
\end{align}

\subsection{Nonplanar extension of the pentagon alphabet}
\label{sec:2L_even_letters}

The analytic structure of the planar penta-box family in \cref{fig:I_np_prop11_a} is described by the 26-letter planar pentagon alphabet~\cite{Gehrmann:2015bfy},
\begin{align}
\mathbb{A}^\text{2-loop}_{\rm pl} := \{ W_{i}\}_{i=1}^{20} \cup \{ W_{i}\}_{i=26}^{31}\,, \label{eq:26pl}
\end{align}
we follow the standard convention of \cite{Chicherin:2017dob} labeling the planar letters $W_i$. It is contained in the family \p{eq:integral_def} for example as an 8-propagator sector $\{ 1,1,1,1,1,1,1,1,0,0,0 \}$ along with its subsectors.

The nonplanar sector of the 11-propagator family \p{eq:integral_def} has a more intricate analytic structure. Constructing the pure basis of MIs and canonical differential equations \p{eq:DEu}, we find that the nonplanar sector requires the 26 planar pentagon letters \p{eq:26pl} to be supplemented with new 85 letters. In total, a 111-letter alphabet is required to solve analytically \p{eq:itintDE} the 11-propagator family \p{eq:integral_def}. We denote them uniformly $\{w_i\}_{i=1}^{111}$ and separate them into planar $W_i$ and nonplanar $\tW_i$,
\begin{equation}
\{w_i\}_{i=1}^{111}=\{ W_{i}\}_{i=1}^{20} \cup \{ W_{i}\}_{i=26}^{31} \cup \{\tW_i\}_{i=1}^{85}\,. \label{eq:w111}
\end{equation}   

The alphabet is closed upon dihedral permutations of the kinematics, namely $\dlog$ forms of the alphabet letters linearly transform among themselves, 
\begin{align}
\sigma (d\log(w_i)) \in \vev{ d\log(w_1),\ldots, d\log(w_{111})}_{\mathbb{Q}} \label{eq:sigmaw}
\end{align}
where $\sigma = \tau,\,\rho$ is a dihedral transformation~\p{eq:tau}. Thus, the alphabet letters are naturally organized into orbits of the cyclic shift $\tau$.

The planar alphabet \p{eq:26pl} is also closed under dihedral permutations. Among 26 planar letters, 20 letters $\{ W_i \}_{i=1}^{20}$ are linear in Mandelstam variables \p{eq:s}, 5 letters $\{ W_i \}_{i=26}^{30}$ are algebraic with the square root $\sqrt{\Delta_5}$ \p{eq:del5}, and $W_{31} = \sqrt{\Delta_5}$. We recall their definitions in \cref{sect:planarletters}. We recall that $\Delta_5$ is dihedral invariant.

Along with $\sqrt{\Delta_5}$, the nonplanar sector requires 10 additional square roots. They are square roots of quadratic and quartic polynomials in the Mandelstam variables,
\begin{align}
\Delta_2^{(1)} =& s_{12}^2 - 4 s_{34} s_{45}\,, \label{eq:del2} \\
\Delta_4^{(1)}=& s_{15}^2 s_{12}^2+s_{23}^2 s_{12}^2-2 s_{15} s_{23} s_{12}^2-2 s_{23}^2 s_{34} s_{12}+2
   s_{15} s_{23} s_{34} s_{12} \notag\\ 
   & +2 s_{15} s_{34} s_{45} s_{12}+2 s_{23} s_{34} s_{45}
   s_{12}+s_{23}^2 s_{34}^2+s_{34}^2 s_{45}^2-2 s_{23} s_{34}^2 s_{45}\,, \label{eq:del4}
\end{align}   
each appearing in 5 cyclic permutations, $\Delta_2^{(i)}=\tau^{i-1}\,\left(\Delta_2^{(1)}\right)$
and $\Delta_4^{(i)}=\tau^{i-1}\,\left(\Delta_4^{(1)}\right)$ for $i=1,\cdots,5$. We identify them as among the normalization prefactors of the pure integrals. For example, $\mathrm{KT}$ sector requires $\sqrt{
\Delta_2^{(4)}}$ and $\mathrm{BT}_2$ sectors requires $\sqrt{
\Delta_4^{(1)}}$, see \cref{fig:I_np_prop_all}.
 
Among 85 nonplanar letters $\tW_i$ \p{eq:w111}, there are 20 letters, which are polynomial in the Mandelsatm variables. We organize them in cyclic orbits as follows,
\begin{align}
\label{eq:tildeW}
& \tW_{i} := \tau^{i-1}\,\left(\Delta_2^{(1)}\right) \;,\quad \tW_{5+i} := \tau^{i-1}\,\left(\tW_{6} \right) , \notag\\
& \tW_{10+i} := \tau^{i-1}\,\left(\tW_{11} \right) \;,\quad \tW_{15+i} := \tau^{i-1}\,\left(\Delta_4^{(1)}\right)  \;,\quad i=1,\ldots,5 
\end{align}
and $\tW_{6}$ and $\tW_{11}$ are cubic in the Mandelstam variables,
\begin{align} \label{eq:tW_even}
&\tW_6  :=s_{12} s_{15}^2 - s_{12} s_{15} s_{23} - s_{12} s_{23} s_{34} + s_{23} s_{34}^2 + s_{15} s_{34} s_{45} \,, \\ 
&\tW_{11} := s_{12}^2 s_{15} + s_{12} s_{15} s_{23} + s_{12} s_{23} s_{34} + s_{23}^2 s_{34} + s_{15}^2 s_{45} - 
 2 s_{15} s_{34} s_{45} + s_{34}^2 s_{45}\,.
\end{align}

The remaining 65 non-planar letters $\{ \tW_{i} \}_{i=21}^{85}$ are algebraic. They involve one or two square roots and have the following form
\begin{align}
\frac{P_1-P_2\,\sqrt{R}}{P_1+P_2\,\sqrt{R}} \, ,\quad
\frac{P-\sqrt{R_1}\sqrt{R_2}}{P+\sqrt{R_1}\sqrt{R_2}} \,,\label{eq:alglett}
\end{align}
with $P_1, P_2, P$ being homogeneous polynomials in Mandelstam variables, and $R,R_1,R_2 = \Delta_5,\,\Delta_2^{(i)},\, \Delta_4^{(i)}$. We count them in \cref{tab:letters}, and provide their explicit form in \cref{sec:2L_odd_letters}.

\begin{table}[t]
\begin{center}
\begin{tabular}{c|cccccc} 
\toprule
roots & ${\Delta_2^{(i)}}$ & ${\Delta_4^{(i)}}$ & $\Delta_5$ & $\Delta_2^{(i)},\Delta_5$ & $\Delta_4^{(i)},\Delta_5$ & $\Delta_2^{(i)},\Delta_2^{(i+1)}$ \\[0.1cm]
\midrule
\# letters & $20=4\times 5$ & $15=3\times 5$ & $10 = 2 \times 5$ & $10=2\times 5$ & $5=1\times 5$ & $5=1\times 5$ \\
\bottomrule
\end{tabular}
\end{center}
 \caption{Couning of the non-planar algebraic letters $\{ \tW_{i} \}_{i=21}^{85}$, which contain one or two square roots.}
    \label{tab:letters}
\end{table}

The alphabet letters appear in the canonical DE \p{eq:DEu} as linear combinations of derivatives $\{ d\log(w_i) \}_{i=1}^{111}$. We employed several complementary approaches to identify explicit expressions for the letters presented in this section. Some nonplanar letters $\tW_i$ appear in the alphabets of subtopologies known in literature. The normalization of the pure MIs suggests some of the letters $\tW_i$. We also rely on computer codes \cite{Fevola:2023kaw,Fevola:2023fzn,Jiang:2024eaj}, which implement the Landau analysis of the branch cut singularities. 
Let us note that using these codes we were able to identify all but the five quartic letters $\tW_{16},\ldots,\tW_{20}$.

\section{Results for integrated negative geometries and positivity properties}
\label{sect:integratednegativegeometries}

In the previous section, we calculated a family of nonplanar two-loop five-particle Feynman integrals \p{eq:integral_def}. Using these analytic expressions for the Feynman integrals, we perform loop integrations of the two-loop five-cusp ladder negative geometry integrand constructed in \cref{sect:mtintegrand}. As compared to Feynman integrals of \cref{sect:2LNPFI}, we find that the two-loop ladder $\laddtwopic$ has a simpler analytic structure. We show that the two-loop ladder belongs to a class of the planar pentagon functions \cite{Gehrmann:2018yef,Chicherin:2020oor}, and we also recall definitions and properties of these transcendental functions. Using the two-loop negative geometry decomposition \p{eq:Fdecomp2}, we also obtain a planar pentagon function expression for the ``loop'' negative geometry $\looppic$.  We discuss analytic properties and numerical evaluation of the integrated two-loop negative geometries.

\subsection{Integrating the two-loop ladder}
\label{sect:integr2LL}

We would like to express the two-loop ladder $\laddtwopic$ in terms of the Feynman integrals \p{eq:integral_def}. Thus, we have to rewrite the momentum twistor expression for the ladder integrand \p{eq:h2} in space-time coordinates (dual momenta). Let us introduce short-hand notations 
\begin{align}
\delta(0) = 0 \,,\qquad
\delta(1) = 6 \,,\qquad
\delta(2) = 7  \,.
\end{align}
The loop variables $x_6$ and $x_7$ from \p{eq:integral_def} and the Lagrangian coordinate $x_0$ are the moduli parameters of the momentum twistor lines, $AB_j \sim x_{\delta(j)}$, $j=0,1,2$. We integrate over the loop variables as follows,
\begin{align}
\frac{d^4 Z_{A_j} d^4 Z_{B_j}}{{\rm vol}(GL(2))} = \vev{AB_j}^4 \, d^4 x_{\delta(j)} \,, \qquad j=1,2 \,.  \label{eq:intZAZB}
\end{align}
We rewrite the twistor four-brackets as follows,
\begin{align}
&\vev{AB_j AB_k} = \vev{AB_j}\vev{AB_k} \, x^2_{\delta(j)\, \delta(k)} \,, \label{eq:brABAB} \\
&\vev{AB_j \,i \,i+1} = \vev{AB_j} \vev{i\,i+1} \, x_{\delta(j)\,i}^2 \,,  
\label{eq:brABii+1} \\
&\vev{AB_j \,i \,i+2} =  \frac{\vev{AB_j}\vev{i+2\,i}}{s_{i\,i+2}} \, \tr_{-} \left( \widehat{p}_i \, \widehat{x}_{i\,\delta(j)} \, \widehat{x}_{\delta(j)\,i+2}\, \widehat{p}_{i+2} \right) , \label{eq:traceABii+2}
\end{align}
where $j,k=0,1,2$; an index $i$ takes cyclic values from $\{1,2,\ldots,5 \}$; the duality relations between momenta and coordinates are given in \p{eq:dual_momentum}; the chiral trace is defined in \p{eq:trm}; and we recall definitions of the spinor-helicity brackets $\vev{mn} = \lambda^{\alpha}_m \lambda_{n\, \alpha}$.

The chiral trace in \p{eq:traceABii+2} contains the parity-even and parity-odd parts,   
\begin{align}
2 \,\tr_{-} \left( \widehat{p}_i \, \widehat{x}_{i\,\delta(j)} \, \widehat{x}_{\delta(j)\,i+2}\, \widehat{p}_{i+2} \right) =& -s_{i\,i+1} s_{i+1\,i+2} + x^2_{\delta(j)\,i-1} s_{i+1\,i+2} + x^2_{\delta(j)\,i} \left( s_{i\,i+1} - s_{i+3\,i+4} \right) \notag \\ & +x^2_{\delta(j)\,i+1} \left( s_{i+1\,i+2} - s_{i+3\,i+4} \right)+x^2_{\delta(j)\,i+2} s_{i\,i+1} \notag\\
& - \frac{16}{\ep_5} \, {\rm Gr}\left(
\{ p_{i} , x_{\delta(j)\,i}, x_{\delta(j)\, i+2} , p_{i+2} \} \, ; \,
\{ p_1 , p_2 , p_3 , p_4 \}
\right) .
\end{align}
The parity-even part is given by squared distances among the cusp coordinates $x_1,\ldots,x_5$ and $x_{\delta(j)}$. The parity-odd part is proportional to $\ep_5$, see \p{eq:ep5}. We express it in terms of the Gram determinant ${\rm Gr}$, which is a quartic polynomial in the squared distances. 

Substituting \crefrange{eq:intZAZB}{eq:traceABii+2} in the two-loop ladder integrand \p{eq:h2}, we verify that spinor helicities cancel out. The resulting expression is written in terms of $\ep_5$  and the squared distances among $x_0,x_1,\ldots,x_5,x_6,x_7$. In order to simplify the loop integrations, we choose $x_0 \to \infty$ by doing a dual-conformal transformation, see \cref{eq:Fx0inf}. Then, we find that the two-loop ladder \p{eq:h2} in the frame $x_0 \to \infty$ is expanded over the Feynman integrals \p{eq:integral_def}, 
\begin{align}
h_{13}^{(2)}(X) = \lim_{\ep \to 0} \, \sum_{\vec a} \left( A_{\vec a}(X) + \ep_5\, B_{\vec a}(X) \right) \, G_{\vec a}(X) \label{eq:ABG}
\end{align}
where $A$ and $B$ are rational functions in the Mandelstam variables $X$ \p{eq:s}. The sum contains $283$ Feynman integrals $G_{\vec a}$. We also introduced the dimensional regularization $D=4-2\ep$ in order to render the loop integrations in each term of \p{eq:ABG} well-defined.

We observe that the Feynman integrals appearing in the sum \p{eq:ABG} contain at most nine propagators. We find that each $G_{\vec a}$ contributing in \p{eq:ABG} belongs to one of the four 9-propagator families, which we denote $T_1,\ldots,T_4$,
\begin{align}
& \vec{a}_{T_1} = \{1, 1, 0, 1, 1, 1, 0, 1, 1, 1, 1\} \,, \qquad
 \vec{a}_{T_2} = \{ 1, 0, 1, 1, 1, 1, 0, 1, 1, 1, 1 \} \,, \\
& \vec{a}_{T_3} = \{0, 1, 1, 1, 1, 1, 0, 1, 1, 1, 1\}\,,  \qquad 
 \vec{a}_{T_4} =\{1, 1, 1, 1, 1, 0, 1, 1, 0, 1, 1\} \,.
\end{align}
These four families are dihedral permutations $\sigma$ (see \cref{eq:sigmaG}) of the families ${\rm DP}_1$ and ${\rm DP}_2$ \p{eq:DPa},
\begin{align}
& \vec{a}_{T_1} = \sigma_{\{ 1, 5, 4, 3, 2 \}}\left( {\vec{a}_{{\rm DP}_1}} \right) \,,\qquad \vec{a}_{T_2} = \sigma_{\{ 2, 3, 4, 5, 1 \} }\left( {\vec{a}_{{\rm DP}_1}} \right) ,\notag\\
&  \vec{a}_{T_3} = \sigma_{\{ 1, 2, 3, 4, 5 \}}\left( {\vec{a}_{{\rm DP}_2}} \right) \,,\qquad
\vec{a}_{T_4} = \sigma_{\{ 5, 1, 2, 3, 4 \}}\left( {\vec{a}_{{\rm DP}_2}} \right) . \label{eq:T1T4}
\end{align}
Thus, we need to calculate Feynman integrals $G_{\vec a}$ from the families ${\rm DP}_1$ and ${\rm DP}_2$ and to apply the dihedral transformations \p{eq:T1T4} to map them into the families $T_1,\ldots,T_4$. We use \texttt{FiniteFlow} \cite{Peraro:2019svx} to construct the IBP reduction rules for the Feynman integrals $G_{\vec a}$ from the families ${\rm DP}_1$ and ${\rm DP}_2$, and we expand them in the bases of pure MIs $\vec{\bf u}_{{\rm DP}_1}$ and $\vec{\bf u}_{{\rm DP}_2}$. The dihedral mappings \p{eq:T1T4} act on the UT Feynman integrals as follows, see \cref{eq:sigmaG}, 
\begin{align}
& \vec{\bf u}_{T_1} = (\tau \rho)\left( \vec{\bf u}_{{\rm DP}_1} \right) \,,\quad \vec{\bf u}_{T_2} = \tau\left( \vec{\bf u}_{{\rm DP}_1} \right) ,\quad \vec{\bf u}_{T_3} = \vec{\bf u}_{{\rm DP}_2}  \,,\quad
 \vec{\bf u}_{T_4} = {\tau}^{-1}\left( \vec{\bf u}_{{\rm DP}_2} \right) . \label{eq:utT1T4}
\end{align}
We calculated pure MIs bases of the families ${\rm DP}_1$ and ${\rm DP}_2$ solving the canonical DE \p{eq:DEu}, and we represented them as iterated integrals \p{eq:itintDE}. We choose the base point of the iterated integrals to be $X_0$ \p{eq:X0}, which is invariant under dihedral transformations of the kinematics. Thus, a dihedral transformation $\sigma$ acts only on the alphabet letters, but it does not change the initial values in the iterated integral solution \p{eq:itintDE},
\begin{align}
\sigma\left( \vec{\bf u}_{T} \right) (X) = \sum_{k \geq 0} \ep^{k} \sum_{m=0}^{k} A_{i_m,T} \ldots  A_{i_1,T} \, \vec{\bf u}_T^{(k-m)}(X_0) \, \left.[\sigma \left( w_{i_1} \right),\ldots, \sigma \left( w_{i_m} \right) ]\right._{X_0}(X)  \label{eq:sigmaut}
\end{align}
where $T={\rm DP}_1,\,{\rm DP}_2$. Let us recall that the alphabet is closed under dihedral transformations, see \cref{eq:sigmaw}. 
Eq.~\p{eq:sigmaut} immediately provides the iterated integral expressions for all pure MIs \p{eq:utT1T4}, which are required in our calculation of the two-loop ladder.

Let us note that the set \p{eq:utT1T4} of $2\times147 + 2\times155$ pure Feynman integrals (see the counting of MIs in \cref{table:mis_top}) is not linearly independent since there are overlaps among the sectors $T_1,\ldots,T_4$. In order to resolve the linear relations among them, we find identical MIs $G_{\vec a}$ belonging to the sectors $T_1,\ldots,T_4$, and then we IBP-reduce them to the pure MIs \p{eq:utT1T4}. In this way we find 345 $\mathbb{Q}$-linear relations among pure Feynman integrals $\{ \vec{\bf u}_{T_1}, \ldots, \vec{\bf u}_{T_4} \}$.

Substituting the IBP reduction rules and their dihedral transformations \p{eq:T1T4} in \cref{eq:ABG}, we rewrite it as follows, 
\begin{align}
h_{13}^{(2)}(X) = \lim_{\ep \to 0}\, \sum_{T \in \{ T_1,\ldots, T_4 \}} \left( {\vec A}_{ T}(X,\ep) + \ep_5\, {\vec B}_{T}(X,\ep) \right) \cdot \vec{\bf u}_{T}(X) \label{eq:htout}
\end{align}
where coefficients $A$ and $B$ are rational functions in Mandelstam variables $X$, dimensional regularization $\ep$, and also the square-roots \p{eq:del2}, \p{eq:del4}. The square roots in the IBP reduction rules come from the normalization of the pure MIs. For our choice of the pure MIs, they are finite $\ep = 0$, see \cref{eq:uexpandeps}, but the coefficients $A$ and $B$ do contain $\ep$-poles at $\ep = 0$. 

After substituting the iterated integral expressions for the pure MIs \p{eq:sigmaut} in \cref{eq:htout}, we find that 
\begin{itemize}
\item $\ep$-poles cancel out;
\item the finite part is of uniform transcendental weight four;
\item it has unit leading singularity;
\item only planar pentagon alphabet letters \p{eq:26pl} contribute to the iterated integrals.
\end{itemize}
In other words, \cref{eq:htout} takes the form,
\begin{align}
h_{13}^{(2)}(X) = \sum_{m=0}^{4} \sum_{j_1,\ldots,j_m} c^{(4-m)}_{j_1,\ldots,j_m} \, \left.\left[W_{j_1},\ldots,W_{j_m}\right]\right._{X_0}(X) \label{eq:h13itint}
\end{align}
where $c^{(k)}$ are transcendental weight-$k$ constants. Namely, $c^{(k)}$ are $\mathbb{Q}$-linear combination of weight-$k$ initial values of the pure MIs $\left\{ \vec{\bf u}_{{\rm DP}_i}^{(k)}(X_0) \right\}_{i=1,2}$. Obviously, \cref{eq:h13itint} could not hold for arbitrary initial values. Consequently, \cref{eq:h13itint} is equivalent to $\mathbb{Q}$-linear relations among the initial values of weight $k=0,1,2,3$. We obtain these linear relations. We verify that the initial values of the pure MIs, which we evaluated numerically in \cref{sec:canon}, do satisfy the exact linear relations with the expected numerical accuracy. 

The exact $\mathbb{Q}$-linear relations among the boundary constants also reduce the number of $\mathbb{Q}$-linear independent $c$'s in \cref{eq:h13itint},
\begin{align}
c^{(0)}_{j_1,\ldots,j_4} \in \mathbb{Q} \,,\quad
c^{(1)}_{j_1,j_2,j_3} = 0 \,,\quad
c^{(2)}_{j_1,j_2} \in \vev{ \pi^2}_{\mathbb{Q}} \,,\quad
c^{(3)}_{j_1}  \in \vev{ \zeta_3, \boldsymbol{c}_3 }_{\mathbb{Q}} \,,\quad
c^{(4)} =  \boldsymbol{c}_4
\end{align}
where $\boldsymbol{c}_3$ and $\boldsymbol{c}_4$ are transcendental constants of weights 3,4, respectively. 

We have chosen one of the pure functions, i.e. $h_{13}^{(2)} \equiv h_2^{(2)}$, in the decomposition of the two-loop ladder \p{eq:F2Ll}. The remaining pure functions are obtained by cyclic shifts $\tau$, which act only on the alphabet letters due to the dihedral invariance of the base point $X_0$ \p{eq:X0}, 
\begin{align}
h_{i\,i+2}^{(2)}(X) = \tau^{i-1} h_{13}^{(2)}(X) = \sum_{m=0}^{4} \sum_{j_1,\ldots,j_m} c^{(4-m)}_{j_1,\ldots,j_m} \, \left.\left[ \tau^{i-1} \left(W_{j_1}\right) ,\ldots,  \tau^{i-1} \left( W_{j_m} \right) \right]\right._{X_0}(X) .
\label{eq:ladderitint}
\end{align}
In what follows, we rewrite the iterated integrals from the previous equation as the planar pentagon functions \cite{Gehrmann:2018yef,Chicherin:2020oor}. 

\subsection{Planar pentagon function expressions for the integrated negative geometries}
\label{sect:pf}

To express all negative geometries up to the two-loop order,
we will use the planar pentagon functions, first introduced in \cite{Gehrmann:2018yef,Chicherin:2020oor} as a basis of the transcendental functions expressing all massless planar two-loop Feynman pentabox family, see \cref{fig:I_np_prop11_a}. 

The integrand of the two-loop correction $F^{(2)}$ involves only planar Feynman integrals belonging to the pentabox family, so $F^{(2)}$ is expressible in the planar pentagon functions of \cite{Gehrmann:2018yef,Chicherin:2020oor}, as was shown in \cite{Chicherin:2022zxo}. However, the integrands of the two-loop ladder $\laddtwopic$ and the ``loop'' topology $\looppic$ from the negative geometry decomposition of $F^{(2)}$ require a larger set of Feynman integrals, which we calculated in \cref{sec:canon} as iterated integrals over the $111$-letter nonplanar alphabet \p{eq:w111}. Thus, one may think that the planar pentagon functions are not sufficient to express these negative geometries. Despite the presence of the nonplanar letters in the expression of the individual Feynman integrals, they cancel out in the iterated integral expression for the two-loop ladder \p{eq:ladderitint} such that only 25 letters of the planar pentagon alphabet contribute. Then, these iterated integrals can be reduced to the planar pentagon functions of \cite{Gehrmann:2018yef,Chicherin:2020oor}.

\begin{table}
\begin{center}
\begingroup
\setlength{\tabcolsep}{10pt}
\renewcommand{\arraystretch}{1.5}
\begin{tabular}{c|ccccc}
weight  $w$ & 0 & 1 & 2 & 3 & 4 \\
\hline
number $l(w)$ of $f_{w,a}$ & 1 & 5 & 5 & $3*5+1$ & $11*5+1$
\end{tabular}
\endgroup
\end{center}
\caption{Counting of the planar pentagon functions $\{ f_{w,a}\}$ of the transcendental weight $w$ split into cyclic orbits each containing 1 or 5 functions.}\label{tab:pf}
\end{table}

The planar pentagon functions are defined as weight-$w$ iterated integrals \p{eq:defitint} over the planar pentagon alphabet with respect to the Euclidean reference point $X_0$ \p{eq:X0} for $w=0,1,\ldots,4$. We denote them as $\{ f_{w,a}\}_{w=0,\ldots,4}$ where the label $a$ discerns  $l(w)$ pentagon functions of weight $w$. This counting is summarized in \cref{tab:pf}. Their definitions respect the discrete dihedral symmetry. Namely, they are split into cyclic orbits of length one or five. The label $a$ specifies the orbit and position on the orbit. If the $n$-th orbit is of length one, then we put $a=n$ and the pentagon function is invariant under the cyclic shift $\tau( f_{w,n}) =  f_{w,n}$, see \cref{eq:tau}. If the $n$-th orbit is of length five, then the corresponding five pentagon functions carry labels $a=(n,1),\ldots,(n,5)$, and they are obtained from each other by the cyclic shifts,
\begin{align}
f_{w,(n,p)} = \tau^{p-1}\left(f_{w,(n,1)}\right) \,,\quad p=1,\ldots,5 \,.
\end{align}  
Obviously, at weight zero, there is only one pentagon function which is just a rational constant, which we choose $f_0 =1$. Then, according to \cref{tab:pf}, there is one length-five orbit at weights one and two, which are denoted as $\{ f_{1,(1,p)} \}_{p=1}^{5}$ and $\{ f_{2,(1,p)} \}_{p=1}^{5}$, respectively. At weights three and four, there are three and eleven length-five orbits, i.e.
$\{ f_{3,(n,p)}\}^{n=1,\ldots,3}_{p=1,\ldots,5}$ and $\{ f_{4,(n,p)}\}^{n=1,\ldots,11}_{p=1,\ldots,5}$. Also, at weights three and four, there are length-one orbits, $\{ f_{3,4}\}$ and $\{ f_{4,12}\}$.

\begin{table}
\begin{center}
\begingroup
\setlength{\tabcolsep}{10pt}
\renewcommand{\arraystretch}{1.5}
\begin{tabular}{c|cccc}
weight $w$ & 1 & 2 & 3 & 4 \\
\hline
$F^{(1)} \sim \laddonepic$ & 5 & 5 & 0 & 0 \\
$F^{(2)}$  & 5 & 5 & 16 & 56 \\
$\laddtwopic$ & 5 & 5 & 16 & 41 \\
$\laddladdpic$ & 5 & 5 & 0 & 0 \\
$\looppic$ & 5 & 5 & 16 & 56 \\
\end{tabular}
\endgroup
\end{center}
\caption{Counting of the weight-$w$ pentagon functions contributing to the one-loop and two-loop negative geometries.}\label{tab:pfng}
\end{table}

In \cref{sect:integr2LL}, we calculated the two-loop ladder $\laddtwopic$ by solving the contributing nonplanar Feynman integrals and expressed it as weight-four UT linear combinations of iterated integrals, see \cref{eq:ladderitint}. The latter involves only 25 planar pentagon letters. Now we are going to expand it in the basis of algebraically independent planar pentagon functions \cite{Gehrmann:2018yef,Chicherin:2020oor}.

Let us start with $F^{(1)}$, which is the one-loop ladder $\laddonepic$\,, see \p{eq:F1Ll}. The polylogarithmic expressions for its pure functions $\{ g_i^{(1)} \}_{i=1}^{5}$, see \p{eq:g1}, are the following UT polynomials of weight two in the pentagon functions, 
\begin{align}
g_i^{(1)} = f_{2,(1,i)} - f_{2,(1,i+2)} + \frac{1}{2}\left(2f_{1,(1,i)}-f_{1,(1,i+2)}-f_{1,(1,i+4)}\right)\left(f_{1,(1,i+2)}-f_{1,(1,i+4)}\right) + \frac{\pi^2}{6} \, \label{eq:g1topf}
\end{align}
where we imply that index $i$ takes cyclic values, i.e. $i+5\equiv i$. Substituting this expression in \cref{eq:F2Lfact}, we rewrite the factorizable two-loop negative geometry $\laddladdpic$ as a weight-four UT polynomial in the pentagon functions. In \cref{tab:pfng}, we summarize how many pentagon functions of various weights appear in the expression of the integrated negative geometries. 
Both $\laddonepic$ and $\laddladdpic$ involve pentagon functions only of weights one and two.

The pure functions of the two-loop ladder $\laddtwopic$ \p{eq:F2Ll} and the non-decomposed two-loop $F^{(2)}$ \p{eq:F2L} are more complicated. They are weight-four UT polynomials in the pentagon functions of the following form
\begin{align}
h_i^{(2)}\; , \; g_i^{(2)} \quad : \quad & \sum_a \alpha_{a} f_{4,a} + \sum_{a_1,a_2} \alpha_{a_1,a_2} f_{1,a_1} f_{3,a_2} + \sum_{a_1,a_2} \beta_{a_1,a_2} f_{2,a_1} f_{2,a_2} 
\notag\\
& 
+\sum_{a_1,a_2,a_3} \alpha_{a_1,a_2,a_3} f_{1,a_1} f_{1,a_2} f_{2,a_3} +\sum_{a_1,a_2,a_3,a_4} \alpha_{a_1,a_2,a_3,a_4} f_{1,a_1}f_{1,a_2}f_{1,a_3}f_{1,a_4} \notag\\
& 
+ \pi^2 \left( \sum_a \beta_{a} f_{2,a}
+ \sum_{a_1,a_2} \gamma_{a_1,a_2} f_{1,a_1} f_{1,a_2} \right) + \zeta_3 \sum_a \gamma_a f_{1,a} + \boldsymbol{c}_4 \label{eq:gtopf4}
\end{align}
where $\alpha,\beta,\gamma$ are rational numbers, summation indices $a,a_1,a_2,a_3,a_4$ run over the labels of the pentagon functions, and $\boldsymbol{c}_4$ are transcendental weight-4 constants. As we can see, they involve pentagon functions of weights up to four. We also notice that according to \cref{tab:pfng} all pentagon functions, enumerated in \cref{tab:pf}, appear in $F^{(2)}$, but 15 weight-four pentagon functions are absent from the two-loop ladder $\laddtwopic$\,. Finally, substituting the pentagon function expressions for the pure functions $h_i^{(2)}$ and $ g_i^{(2)}$ in \cref{eq:F2Lloop} we rewrite the ``loop'' negative geometry $\looppic$ in the pentagon functions. 

Let us summarize which of the 25 planar letters (see \cref{sect:planarletters}) are present in the iterated integral expressions for the pure functions of the negative geometries:
\begin{itemize}
\item The one-loop ladder $\laddonepic$ (i.e. $F^{(1)}$) involves ten letters $W_1,\ldots,W_5,W_{11},\ldots,W_{15}$. Whereas each $g_i^{(1)}$, $i=1,\ldots,5$, involves only 5 letters, e.g. $W_1,W_3,W_5,W_{13},W_{15}$ are present in $g_1^{(1)}$, and the letter content of the remaining four pure functions is obtained by cyclic shifts $\tau$ \p{eq:tau}.
\item
The two-loop ladder $\laddtwopic$ involves 20 letters 
\begin{align}
W_1,\ldots,W_5,W_{11},\ldots,W_{20},W_{26},\ldots,W_{30} \,.
\end{align}
The same letters appear in all its five pure functions $\{ h_i^{(2)} \}_{i=1}^5$.
\item The nondecomposed two-loop $F^{(2)}$, as well as its pure function $g_0^{(2)}$, involve 25 planar letters (i.e. all planar letters except for $W_{31}$), but the letter content of $\{ g_i^{(1)} \}_{i=1}^{5}$ is more restricted. Each of them contains only 22 letters. For example, $W_{6},W_{8},W_{10}$ are absent from $g^{(2)}_{1}$.
\item There are no bonus cancellations of the letters in the ``loop'' $\looppic$, see \p{eq:F2Lloop}. Namely, the pure function accompanying $r_i$ contains the same 22 letters as $g_i^{(2)}$, $i=1,\ldots,5$, and and the pure function accompanying $r_0$ contains 25 letters.  
\end{itemize}
In \cref{sect:boxing}, we derive a d'Alembertian differential equation for the ladder-type negative geometries. We explain in \cref{sect:lastentry,sect:NTlastentry} how it restricts their letter content.

We provide both iterated integral and pentagon function expressions for the negative geometries in the ancillary files.

\subsection{Numerical evaluation of the pentagon functions}
\label{sect:numpf}


In \cref{sect:positive}, we will study numerical values of the negative geometries in the Euclidean region and evaluate them in $O(10^5)$ kinematic points. Since the integrated negative geometries are polynomials in the planar pentagon functions, we recall the numerical evaluation of the pentagon functions.

We rely on two complementary approaches to evaluate the pentagon functions and their derivatives, see details in \cref{sect:PFDE}. Firstly, we use the $\texttt{C++}$ code of \cite{Chicherin:2022zxo}, which relies on a rewriting of the iterated integrals as univariate integrations of logarithms and dilogarithms and performs the quadrature numerically. Evaluations are easily parallelizable, the resulting precision is $\sim 11$ digits, and evaluation time is $\sim 5$ min per kinematic point per CPU.  

Secondly, having at our disposal canonical DE, summarised in \p{eq:DEpf}, and the boundary condition $\vec{\bf F}(X_0)$, we apply \texttt{DiffExp} \cite{Hidding:2020ytt} to integrate the DE numerically using the generalized series expansions. Since the initial values are known analytically, we can achieve arbitrarily high precision of evaluations. Also, using this approach, we can evaluate the pentagon functions very close to singularities. The evaluation time is comparable with the first approach, but it could vary significantly depending on the location of the kinematic point $X$ and the choice of the integration path connecting $X_0$ and $X$.

\subsection{Final result and checks}

In this work, we constructed the integrand of the two-loop ladder negative geometry in \cref{sect:mtintegrand} and performed two-loop integration expressing the result in terms of the planar pentagon functions, see \cref{eq:gtopf4}. In the ancillary files, we provide both Chen's iterated integral expressions and pentagon function expressions for all two-loop negative geometries. In this section, we would like to summarize the checks we performed on the resulting expression.

The obtained analytic expression possesses all expected properties. Although the negative geometries are well-defined in four spacetime dimensions and finite, we had to introduce the dimensional regulator $D=4-2\ep$ to apply the usual calculation procedure (based on IBP-reductions) for the two-loop Feynman integrals \p{eq:integral_def}. Because of the regulator, the intermediate expressions in our calculation contain poles $1/\ep^p$ with $p=1,\ldots,4$. The poles do cancel out among each other which is a nontrivial check on the result. Then, we expect that $L$-loop negative geometries evaluate to weight-$2L$ UT functions. It does hold for our explicit results. In \cref{sect:limits}, we calculate the soft/collinear limits of the five-cusp negative geometries and verify that they do reproduce the known four-cusp expressions, see \cref{sect:App4cusp}. Finally, based on previous four-cusp and five-cusp calculations \cite{Arkani-Hamed:2021iya,Chicherin:2022zxo}, we could expect that the negative geometries are positive if evaluated inside the Amplituhedron region. In \cref{sect:positive}, we provide evidence that the obtained pentagon function expressions verify this property. 

We also performed a direct numerical check of the obtained results. We evaluate them at an Euclidean point $X_1$,
\begin{align}
X_1 = \left( s_{12} = -1\,,\; s_{23} = -3 \,,\; s_{34} = -11 \,,\; s_{45} = -17\,,\; s_{15} = -13 \right). \label{eq:X1}
\end{align}
On the one hand side, we have a representation for the two-loop ladder pure function $h^{(2)}_2$ as a linear combination of the two-loop Feynman integrals (before performing IBP-reductions), see \cref{eq:ABG}. We evaluate them numerically at $X_1$ with 70-digit precision using \texttt{AMFlow} \cite{Liu:2022chg}. On the other hand side, we evaluate the pentagon functions at $X_1$, as explained in \cref{sect:numpf}, and find the value of $h^{(2)}_2$. Both evaluations match at the expected precision level. We also use \texttt{AMFlow} to check numerically the IBP reduction rules, which we apply in our calculation, and reductions of the scalar integrals to the pure bases of MIs \p{eq:utT1T4}.

We provide reference values for $F^{(L)}$ with $L=0,1,2$ as well as for the negative geometries at $X=X_1$, see~\p{eq:X1}, with 12-digit precision which we obtained evaluating the pentagon functions,
\begin{align}
& F^{(0)} = -269.770449477 \,,\qquad
F^{(1)} = 3089.22986379\,, \notag\\
& F^{(2)} = -43647.6529114\,,\qquad
F^{(\laddtwopic)} = 31920.5063313 \,, \notag\\
& F^{(\laddladdpic)} =35083.3435008 \,,\qquad
F^{\left(\looppic\right)} = 11629.0503406\,.
\end{align}    
One can easily see that the provided values agree with the two-loop decomposition \p{eq:Fdecomp2}.
 
Note that the values for each $F^{(\laddtwopic)}$, $F^{(\laddladdpic)}$, $F^{\left(\looppic\right)}$ are defined up to a sign, as the corresponding canonical form has a sign ambiguity. However, all these signs are fixed once we consider these negative geometries in the context of the expansion (\ref{eq:Omega_L}) where it needs to agree with the expansion in terms of products of amplitudes (represented by specific positive geometries). 
    

\subsection{Positivity properties in the Amplituhedron region}
\label{sect:positive}

One motivation for studying negative geometry is to understand the positivity of the integrated quantities, from the Lagrangian insertion in the Wilson loop, down to individual geometric objects. 
The results of observables expressed in terms of pentagon function, with numerical evaluation implemented in \cref{sect:numpf}, provide us 
considerable amount of data to investigate this subject. 
We first recall that the positivity of the four-point observable as well as negative geometries have been fully explored in \cite{Arkani-Hamed:2021iya}.
Later, a five-point positivity hypothesis for the Lagrangian insertion in the Wilson loop was proposed in \cite{Chicherin:2022zxo}, which states the observable is positive/negative definite (depending on even/odd loop order) within the Amplituhedron region, i.e.
\begin{equation}
\label{eq:post_WL}
F^{(L)}_{5}|_{{\rm Eucl}^+} < 0 \; \text{ at even} \; L\,,\qquad
F^{(L)}_{5}|_{{\rm Eucl}^+} > 0 \; \text{ at odd} \; L\,.
\end{equation}
The Amplituhedron region ${{\rm Eucl}^+}$ is the five-particle one-loop MHV amplituhedron \cite{Arkani-Hamed:2013jha}. In the frame $x_0 \to \infty$, see \cref{sect:5part}, the momenta twistor constraints specifying this region are equivalent to, see \cite{Chicherin:2022zxo} for more details, 
\begin{align}
{{\rm Eucl}^+} \;: \qquad
\epsilon_5>0 \,,\qquad s_{i\, i+1}> 0 \,,\qquad i=1,\ldots,5\,. \label{eq:eucl+}
\end{align}
Thus, the Amplituhedron region is a subregion of the Euclidean region, where all adjacent Mandelstam variables are positive. It is further divided into 11 subregions according to the sign of the non-adjacent Mandelstam variables, $s_{i\, i+2}$ in \eqref{eq:non-adj-s}, which are summarized in \cref{table:subregions}.

\begin{table}[t]
\begin{center}
\begin{tabular}{c|c|c}
\toprule
Region & configuration of  $s_{i,i+2}$  & \# subregions  \tabularnewline
\midrule
$\mathrm{(A)}$ & {\rm all} $s_{i,i+2}<0$  & 1  \tabularnewline
\midrule
$\mathrm{(B)}$ & {\rm one} $s_{i,i+2}$ {\rm positive}, \text{others negative}   & 5   \tabularnewline
\midrule
$\mathrm{(C)}$ & $s_{i,i+2},s_{i+1,i+3}$ {\rm positive}, \text{others negative}  & 5   \tabularnewline
\bottomrule
\end{tabular}
\end{center}
\caption{The Amplituhedron subregions.}
\label{table:subregions}
\end{table}

Apart from the positivity property of $F^{(L)}_{5}$ in
\eqref{eq:post_WL},
we further argue the positivity property holds for individual negative geometries that decompose the full observable. At one-loop, only the ladder $\laddonepic$ contributes to the negative geometry decomposition, so its positivity is equivalent to \cref{eq:post_WL} at $L=1$, 
\begin{align}
F^{(\laddonepic)}_5 \biggr|_{{\rm Eucl}^+}  >0\,.
\end{align}
The two-loop negative geometry decomposition \p{eq:Fdecomp2} involves three geometries,
\begin{align} 
F^{(2)}_5 = - F_5^{(\laddtwopic)} - \frac{1}{2} F_5^{(\laddladdpic)} +\frac{1}{2} F_5^{\left(\looppic\right)} \,. \label{eq:ngdF2}
\end{align}
We conjecture that they take positive values,
\begin{align} \label{eq:positivity_5pt}
F^{(\laddtwopic)}_5 \,,\; F^{(\laddladdpic)}_5 \,,\; F^{\left(\looppic\right)}_5 \biggr|_{{\rm Eucl}^+}  >0\,,\quad {\rm and} \quad F^{(2)}_5\biggr|_{{\rm Eucl}^+} <0\,. 
\end{align}

\begin{table}[t]
\centering
\begin{tabular}{c|cc c c}
\toprule
\rule{0pt}{5ex}
Kinematics & $F^{(2)}$  & $F^{(\laddtwopic)}$ & $F^{(\laddladdpic)}$ & $F^{\left(\looppic\right)}$  \tabularnewline
\midrule
$X_\mathrm{(A)}$ & \small{$-166839.228227$}  & \small{$95179.9740231$} & \small{$215781.615620$} & \small{$72463.1072129$}  \tabularnewline
\midrule
$X_\mathrm{(B)}$ & \small{$-457.130590830$}  & \small{$316.229580873$} & \small{$448.738525078$} & \small{$166.936505165$} \tabularnewline
\midrule
$X_\mathrm{(C)}$ & \small{$-18408369.7482$}  & \small{$34472138.5938$} & \small{$24656713.7133$} & \small{$56784251.4046$}  \tabularnewline
\bottomrule
\end{tabular}
\caption{Numerical values of the two-loop Lagrangian insertion in the Wilson loop $F^{(2)}$ and the negative geometries at the kinematic points \p{eq:ABC} from the three different Amplituhedron subregions.}
\label{table:data_subregions}
\end{table}

With the explicit expressions for the integrated two-loop negative geometries in terms of the pentagon functions, see \cref{eq:F2Ll,eq:F2Lfact,eq:F2Lloop,eq:g1topf,eq:gtopf4}, we examine $10^5$ random kinematic points from ${{\rm Eucl}^+} $. We evaluate the pentagon functions using the {\tt{C++}} code~\cite{Gehrmann:2018yef}.\footnote{Usually, the Euclidean region is defined such that all adjacent Mandelstam variables are negative, which is opposite to our conventions in \cref{eq:eucl+}. This is irrelevant for us, since the pure functions are dimensionless. They depend on four dual-conformal cross-ratios \p{eq:u} which are ratios of the Mandelstam variables in the conformal frame $x_0 \to \infty$. In particular, they are invariant under the sign flip of all Mandelstam invariants $X \to -X$ \p{eq:s}.} This gives strong evidence that each negative geometry is positive \p{eq:positivity_5pt}. 
For illustrating, 
in \cref{table:data_subregions}, we list numeric values of two-loop Lagrangian insertion in the Wilson loop and the negative geometries at the following kinematic points, which reside in the three subregions of ${{\rm Eucl}^+}$, see \cref{{table:subregions}},   
\begin{align}
    X_{\mathrm{(A)}}=\biggl( s_{12}&=1, &s_{23}&=\frac{102}{31},& s_{34}&=\frac{153}{4}, & s_{45}&=\frac{13}{31}, & s_{15}&=\frac{159}{4}\biggr)\,,
    \notag\\
    X_{\mathrm{(B)}}=  
    \biggl(s_{12}&=1, &s_{23}&=\frac{23}{33}, & s_{34}&=\frac{4}{11}, & s_{45}&=\frac{19}{11},  &s_{15}&=\frac{4}{33}\biggr)\,,\notag\\
     X_{\mathrm{(C)}}=  
    \biggl( s_{12}&=1, &s_{23}&= \frac{27}{137}, & s_{34}&= \frac{9}{37}, & s_{45}&= 108, & s_{15}&=135\biggr)\,. \label{eq:ABC}
\end{align}

On top of the positivity of negative geometries \eqref{eq:positivity_5pt},  we discovered the ladder geometries can be further \emph{triangulated} within specific kinematic region.
We found that the individual terms of one-loop and two-loop ladders, i.e. $h^{(1)}_i$ and $h^{(2)}_i$ in \eqref{eq:F1Ll} and \eqref{eq:F2Ll}, display a particularly nice positivity property within the Amplituhedron sub-region (A) as follows
\begin{align}
{\rm Region\, (A):}\qquad
(r_i-r_0)>0\,, \quad h^{(1)}_i>0\,,\quad h^{(2)}_i>0\,,\quad {\rm for} \quad i=1,...,5\, .
\end{align}
Therefore, the postitvity of $F^{(\laddonepic)}$ and $F^{(\laddtwopic)}$ in sub-region (A) naturally follows as each cyclic term, as well as the accompanying leading singularity, are already positive definite in the region. However, in region (B) and (C), the positivity of ladders $F^{(\laddtwopic)}$ requires summing all 5 cyclic terms in \eqref{eq:F2Ll}.

Lastly, the negativity of $F^{(2)}$, see \cref{eq:post_WL}, does not follow from the positivity of the individual geometries owing to the presence of both plus and minus signs in the two-loop negative geometry decomposition \p{eq:ngdF2}. In the soft and collinear limit, summarised in \cref{sect:limits}, the five-point positivity \p{eq:positivity_5pt} reduces to the four-point positivity~\cite{Arkani-Hamed:2021iya} at the level of individual negative geometries.
In the multi-Regge limit, see detailed discussion in \cref{sect:MR}, the dominating leading logarithmic term also exhibits the positivity property, which once again is manifest at the individual geometries.
\begin{align}
\frac{F_5^{(L)}}{F_5^{(0)}}=   q^{(L)}_{2L} \left(\log \delta \right)^{2L} + {\cal O}\left(\left(\log \delta\right)^{2L-1}\right)\,,
\end{align}
where $q^{(1)}_2=-4$ and $q^{(2)}_4=8$ are consistent with \eqref{eq:post_WL} (note the ratio $F_5^{(0)}$ is negative definite within Euclidean region). The leading two-loop term is further divided into
\begin{equation}
\label{eq:h_decomp}
 q^{(2)}_4=-q^{(\laddtwopic)}_4 -\frac{1}{2} q^{(\laddladdpic)}_4+\frac{1}{2} q^{(\looppic)}_4\,, 
\end{equation}
where the values of $q^{({\bf g})}_4$ are of uniform sign
\begin{equation}
    q^{(\laddtwopic)}_4=-\frac{8}{3}\,, \quad q^{(\laddladdpic)}_4=-16\,,\quad q^{(\looppic)}_4=-\frac{16}{3}\,,
\end{equation}
and the relative minus signs on the RHS of \eqref{eq:h_decomp} do not spoil the positivity of $q^{(2)}_4$.

\section{d'Alembertian differential equation for the ladder-type geometries}
\label{sect:boxing}


The negative geometries of the ladder type have an especially simple form, at least at the integrand level. Indeed, their integrands are constructed recursively at any loop order in \cref{sect:mtintegrand}. The form of the integrand suggests that the loop integrations could be performed much simpler than going through the Feynman graph calculation with dimensional regularization, as we did in \cref{sect:integr2LL}. Indeed, in the four-cusp case, the ladders (as well as all negative geometries of the tree topology) have been solved in a closed form at any loop order \cite{Arkani-Hamed:2021iya}. That was possible due to a second-order differential equation of the d'Alembertian type which relates relates $L$-loop and $(L+1)$-loop ladders. Starting with $L = 0$ and solving the DE recursively, one finds ladders at any $L$. 

In this section, we derive an analogous d'Alembertian DE for the five-cusp ladders. We explicitly verify that the one-loop and two-loop ladders, $\laddonepic$ and $\laddtwopic$\,, which are expressed in terms of the pentagon functions, see \cref{eq:g1topf,eq:gtopf4}, satisfy the d'Alembertian DE. 

Then, instead of using the d'Alembertian DE as an extra check of our Feynman graph calculation, we find the integrated ladders by solving the DE. In this way, we bypass loop integrations in dimensional regularization. Namely, we rely on the symbol bootstrap and pin down the symbol solution of the DE. In \cref{sect:bootstrap}, we work out constraints that the DE imposes on the last and next-to-last entries of the symbols and perform the symbol bootstrap analysis.

\subsection{d'Alembertian differential equation in momentum twistor variables}

In contrast to several previous sections where we work in the frame $x_0 \to \infty$, we restore the Lagrangian coordinate $x_0$. Let us consider the integrand for one of the pure functions $h^{(L)}_{ij}$ of the $L$-loop ladder \p{eq:inthL}. It depends on the Lagrangian coordinate $AB_0 \sim x_0$ in a very special way, 
\begin{align}
h^{(L)}_{ij}(AB_0) = \int\limits_{AB_1,\ldots,AB_{L}} \frac{\vev{A B_0 ij}}{\vev{AB_0 A B_1}} I_{L}(AB_1,\ldots,A B_{L})\,, \label{eq:gI}
\end{align}
where $(ij) = (13),(24),(35),(14),(25)$. The explicit expression for the rational function $I_{L}$ follows from \cref{genladder}, which is the 
product of ${\cal C}_{ab,cd}$'s factors \p{eq:Cabcd}. We aim to find a differential operator acting on $AB_0 \sim x_0$ which simplifies the right-hand side of \p{eq:gI}. We relate the integration over the twistor line $AB_1 \sim y_1$ with the space-time integration,
\begin{align}
\frac{d^4 Z_{A_1} d^4 Z_{B_1}}{{\rm vol}(GL(2))} = \vev{AB_1}^4  d^4 y_1 \,,
\end{align} 
and rewrite the momentum twistor four-brackets from \p{eq:gI} in terms of the space-time coordinates (dual momenta)
\begin{align}
\vev{AB_0 ij} = \vev{AB_0}\vev{ij} (x_0 - x_*)^2 \;,\qquad
\vev{AB_0 AB_1} = \vev{AB_0}\vev{AB_1} (x_0 - y_1)^2 
\end{align}
where $x_{*} := \ket{j} \bra{i} x_i + \ket{i}\bra{j} x_j$ is the space-time coordinate corresponding to the twistor line $Z_i Z_j \sim x_*$. Then we notice that the second-order differential operator 
\begin{align}
{\cal D}_{ij} := (x_{0}-x_{*})^2 \Box_{x_0} \frac{1}{(x_{0}-x_{*})^2} \label{eq:Dop}
\end{align}
freezes the loop integration over $y_1$ in \p{eq:gI}, 
\begin{align}
{\cal D}_{ij} \, h^{(L)}_{ij}(AB_0) = - 4 \vev{AB_0 ij} \vev{AB_0}^2 \int \limits_{AB_2,\ldots,AB_{L}} I_{L}(AB_0,AB_2,\ldots,AB_{L}) \label{eq:DgI}
\end{align}
since the propagator is a Green's function of the d'Alembertian operator $\Box_{x_0}$,
\begin{align}
\Box_{x_0}  \frac{1}{(x_0-y_1)^2} = -4\i \pi^2 \delta^4(x_0-y_1) \,.
\end{align} 
Moreover, taking into account the explicit expression for $I_{L}$, we rewrite the right hand side of \p{eq:DgI} as a linear combination of the pure functions of $({L}-1)$-loop ladder,
\begin{align}
& {\cal D}_{ij} \, h^{(L)}_{ij}(AB_0) = - 4 \sum_{kl} \overline{{\cal C}}_{kl,ij}(AB_0) \, h^{(L-1)}_{kl}(AB_0)\; ,\qquad {L}>1 \label{eq:Dg}\\
& {\cal D}_{ij} \, h^{(1)}_{ij}(AB_0) = 4 \, \overline{{\cal C}}_{i+1\,j+1,ij}(AB_0) \,, \label{eq:Dg1}
\end{align}
where $\overline{{\cal C}}_{ij,kl}(AB_0)$
are obtained from ${\cal C}_{ij,kl}(CD,AB_0)$ \p{eq:Cabcd} by taking residue at $CD =AB_0$,
\begin{align}
\overline{{\cal C}}_{ij,kl}(AB_0) := \vev{AB_0}^2 \left. \left[ \vev{AB_0 CD} {\cal C}_{ij,kl}(CD,AB_0) \right] \right|_{CD=AB_0}\,.
\end{align}

\subsection{Differentiating the planar pentagon functions}

In order to use d'Alembertian DE \p{eq:Dg1}, we need to rewrite the differential operator ${\cal D}_{ij} $ \p{eq:Dop} in convenient variables. The operator acts on a dimensionless dual-conformal invariant function $h^{(L)}_{ij}$, which depends on $x_0$ only via dual-conformal cross-ratios \p{eq:u}. We apply the chain rule and rewrite the derivatives in $x_0$ as derivatives in the dual-conformal cross-ratios. A choice of four independent cross-ratios inevitably breaks the dihedral symmetry, so we would like to find a more symmetric form of ${\cal D}_{ij} $. In the present work, we mainly work in the frame $x_0 \to \infty$, see \cref{sect:5part}. Once ${\cal D}_{ij} $ is written as an operator in the dual-conformal cross-ratios, we can easily choose the frame $x_0 \to \infty$ replacing the cross-ratios with ratios of Mandelstam variables $X$ \p{eq:s}. Finally, we can rewrite ${\cal D}_{ij} $ as a differential operator in five Mandelstam variables instead of their ratios. At this point, it is helpful to take into account that ${\cal D}_{ij}$ acts on dimensionless functions, which are invariant under rescaling of the Mandelstam variables,
\begin{align}
\sum_{v \in X} v \frac{\pa}{\pa v} \, h^{(L)}_{ij}(X) = 0 \,.
\end{align}
Using this freedom, we obtain for example for ${\cal D}_{13}$,
\begin{align}
-\frac{1}{4}{\cal D}_{13} = \sum_{i=1}^{5} s_{i+1\,i} s_{i\,i+2} s_{i+2\,i+1}\, \pa_{s_{i\,i+1}}\pa_{s_{i+1\,i+2}} - \frac{1}{s_{13}} \tr_{-} \left(\widehat{p}_3 \widehat{p}_4\widehat{p}_5 \widehat{p}_1 \right) \left( s_{34} \pa_{s_{34}} + s_{15} \pa_{s_{15}}\right) \label{eq:D13s}
\end{align}
where we assume cyclicity of the indices, i.e. $6 \equiv 1$ and so on, and the chiral traces are defined in \p{eq:trm} and evaluated in \p{eq:trmeval} in terms of the Mandelstam variables and parity-odd $\ep_5$ \p{eq:ep5}. 
The corresponding DE \p{eq:Dg} and \p{eq:Dg1} take the following form in the frame $x_0 \to \infty$,
\begin{align}
-\frac{1}{4}{\cal D}_{13} \, h^{(1)}_{13} = & s_{13} \,, \label{eq:D13g1s}\\
-\frac{1}{4}{\cal D}_{13} \, h^{(L)}_{13} = & (s_{12} + s_{23}) \, h^{(L-1)}_{13} - s_{13} \left( h^{(L-1)}_{24} + h^{(L-1)}_{25} \right) \notag\\
&  + \left( s_{23} - s_{45} + \frac{1}{s_{35}} \tr_{-} \left(\widehat{p}_{5}\, \widehat{p}_{1}\, \widehat{p}_{2}\, \widehat{p}_{3}\right)  \right) h^{(L-1)}_{35} \notag \\ 
&  + \left( s_{12} - s_{45} + \frac{1}{s_{14}} \tr_{-} \left(\widehat{p}_{1}\, \widehat{p}_{2}\, \widehat{p}_{3}\, \widehat{p}_{4}\right)  \right) h^{(L-1)}_{14} \;,\qquad L >1 \,. \label{eq:D13gs}
\end{align}
The differential equations and ${\cal D}_{ij}$ in the frame $x_0 \to \infty$ for the remaining four pairs $(ij)$ are obtained by cyclic shifts $\tau$ \p{eq:tau} of the indices in \p{eq:D13s}, \p{eq:D13g1s}, \p{eq:D13gs}. The cyclic shift alters the first-order derivatives in the differential operator \p{eq:D13s}, but the second-order derivatives are invariant. Let us stress that the d'Alembertian DE in the frame $x_0 \to \infty$, see \cref{eq:D13gs}, and the d'Alembertian DE at arbitrary $x_0$, see \cref{eq:Dg}, are equivalent.
 
Let us notice that the right-hand side of  \cref{eq:D13gs} contains four linear independent rational factors. Namely, among five rational functions in the right-hand side of d'Alembertian DE \p{eq:Dg}, 
\begin{align}
\overline{{\cal C}}_{12,ij} \,,\; \overline{{\cal C}}_{23,ij} \,,\; \overline{{\cal C}}_{34,ij} \,,\; \overline{{\cal C}}_{45,ij} \,,\; \overline{{\cal C}}_{15,ij} \label{eq:5C}
\end{align}
with $j=i+2$, only four of them are linear independent, since $\overline{{\cal C}}_{i+1\,i+3,i\,i+2} = \overline{{\cal C}}_{i-1\,i+4,i\,i+2}$.
 
The one-loop ladder $\laddonepic$ and the two-loop ladder $\laddtwopic$, which we calculated in the previous sections, have to satisfy the d'Alembertian DE. We choose $(ij)=(13)$, and recall various notations for the corresponding pure function of the one-loop ladder $h^{(1)}_{13} \equiv g^{(1)}_2 \equiv h^{(2)}_2$ and of the two-loop ladder $h^{(2)}_{13} \equiv h^{(2)}_2 $, see \cref{eq:hab-hi,eq:h1-g1}. The one-loop DE \p{eq:D13g1s} is immediate to verify by acting with the second-order differential operator \p{eq:D13s} on the dilogarithmic expressions for the pure functions of the one-loop ladder \p{eq:g1}.

Then, we would like to verify DE \p{eq:D13gs} at $L=2$, which relates the one-loop and two-loop ladders. Both the one-loop and two-loop ladders are expressed as UT polynomials in the pentagon functions of the transcendental weights two and four, respectively, see \cref{eq:g1topf,eq:gtopf4}. As reviewed in \cref{sect:PFDE}, taking derivatives of the pentagon functions, we stay in the space of the pentagon functions. Thus, we calculate the second-order derivatives of $h^{(2)}_{13}$ and express them in the pentagon functions. In general, the resulting expression is a polynomial in pentagon functions (with rational coefficients in Mandesltam variables and $\ep_5$), which has transcendental weight-two and weight-three components. We verify that differentiating $h^{(2)}_{13}$ by ${\cal D}_{13} $ \p{eq:D13s}, the weight-three component vanishes and the weight-two component reproduces the right-hand side of \cref{eq:D13gs} at $L=2$.

In principle, the DE \p{eq:D13g1s} and \p{eq:D13gs} supplemented with boundary conditions uniquely fix the ladders at any loop order. Instead of trying to solve the second-order partial differential equations, we work out the constraints they impose on pure functions $h^{(L)}_{ij}$ and use them in the symbol bootstrap analysis in the next section.

\section{Symbol bootstrap of the ladder-type negative geometries} 
\label{sect:bootstrap}

We performed loop integrations of the one-loop and two-loop ladder negative geometries relying on a conventional Feynman diagram calculation and IBP-reductions, see \cref{sect:integr2LL}. In this approach, we had to calculate a family of the two-loop Feynman integrals \p{eq:integral_def}. Their analytic structure is governed by the 111-letter alphabet, \p{eq:w111}. However, a much smaller sub-alphabet -- the planar pentagon alphabet -- is required for the negative geometries up to the two-loop order. In this section, we would like to show that the loop integrations of ladders $\laddonepic$ and $\laddtwopic$ can be performed in a much simpler way relying on the symbol bootstrap.
 
Due to the leading singularity analysis of the ladder integrand, we know that $h_{ij}^{(L)}$ is a pure function. The absence of 
double poles
in the integrand suggests that $h_{ij}^{(L)}$ is UT of transcendental weight $2L$. Let us assume that 
this function is
expressible in terms of iterated integrals with $\dlog$ kernels, and that the relevant symbol alphabet is the planar pentagon alphabet \p{eq:26pl}, \cref{sect:planarletters}. The assumption about the alphabet is crucial for the bootstrap. 

We recall that the pure functions of the ladder are related by cyclic shifts $\tau$. For example, once $h^{(L)}_{13}$ is known, other $h^{(L)}_{ij}$ are obtained by cyclic shifts, $\{ \tau^k h^{(L)}_{13} \}_{k=0}^{4}$.
In the following, for the sake of simplicity, we prefer to work at the symbol level omitting all transcendental constants in the iterated integral representation, see \cref{sec:canon}. Each symbol term of $h_{ij}^{(L)}$ contains $2L$ entries,
\begin{align}
{\cal S}\left( h_{ij}^{(L)} \right) = \sum_{i_1,\ldots,i_{2L}} c_{i_1,\ldots,i_{2L}} [W_{i_1},\ldots,W_{i_{2L}}] \label{eq:symb2L}
\end{align} 
where summation indices run over labels of the planar pentagon alphabet letters \p{eq:26pl}, and $c$'s are indeterminate rational numbers, which we would like to pin down using some constraints on $h_{ij}^{(L)}$. A symbol can be lifted to a function if and only if the symbol satisfies the integrability condition for each pair of adjacent entries,
\begin{align}
\sum_{i_1,\ldots,i_L} c_{i_1,\ldots,i_{2L}} [W_{i_1},\ldots,W_{i_{k-1}},W_{i_{k+2}},\ldots,W_{i_{2L}}] \, d\log W_{i_k} \wedge d\log W_{i_{k+1}} = 0 \,\label{eq:integrable}
\end{align}
where $k=1,\ldots,2L-1$.

According to our assumption, the symbol entries are letters of the planar pentagon alphabet \p{eq:26pl}. The pure function $h_{ij}^{(L)}$ is dimensionless, but some of the alphabet letters have nonzero dimensions. So, we 
draw symbol entries from a set of 25 dimensionless combinations of the letters, see \cref{sect:planarletters},
\begin{align}
\overline{\mathbb{A}}^\text{2-loop}_{\rm pl} := \{ [W_1]-[W_i] \}_{i=2}^{20} \cup \{ [W_i] \}_{i=26}^{30} \cup \{ 2[W_1] - [W_{31}] \} \,.
\label{eq:dimlessW}
\end{align}
This reduces the number of indeterminates $c$'s in \p{eq:symb2L}.

The first entries of the symbol specify the location of the discontinuities. If $h_{ij}^{(L)}$ has a nonzero discontinuity, then one of the adjacent Mandlestam invariants vanishes, $s_{i \,i+1} = 0$. For example, inspecting the two-loop Feynman diagrams \p{eq:integral_def}, we conclude that their unitarity cuts are located at $s_{i \,i+1} = 0$. Thus, there are four dimensionless combinations of the letters which are allowed first entries, 
\begin{align}
\text{First entries :} \qquad
[W_1]-[W_2],\, [W_1]-[W_3] ,\, [W_1]-[W_4] ,\, [W_1]-[W_5] \,. \label{eq:first}
\end{align}

The counting of integrable weight-$2L$ symbols with first entries drawn from \cref{eq:first} and other entries drawn from \p{eq:dimlessW} is provided in the first column of \cref{tab:bootstrap}.

\subsection{The last entry condition}
\label{sect:lastentry}

Differentiation acts on the last entry of the symbol. We use d'Alembertian differential equation \p{eq:Dg} to obtain some simple constraints on the last entries of the symbol \p{eq:symb2L}. Let us split weight-$2L$ symbol \p{eq:symb2L} into weight-$(2L-2)$ and weight-2 symbols,
\begin{align}
{\cal S}\left( h_{ij}^{(L)} \right) = \sum_a S^{(2L-2)}_a \otimes S^{(2)}_a  \label{eq:symbgsplit}
\end{align}
where we arrange the terms such that both sets $\{ S_a^{(2L-2)}\}$ and $\{ S^{(2)}_a \}$ are linearly independent. Acting with the second-order differential operator on an integrable weight-$2L$ symbol results in a linear combination of weight-$(2L-1)$ and weight-$(2L-2)$ symbols. That means that ${\cal D}_{ij} S^{(2)}_a$ is a linear combination of weight-1 and weight-0 symbols, see \cref{eq:diffitint},
\begin{align}
{\cal D}_{ij}\, {\cal S}\left( h_{ij}^{(L)} \right) = \sum_a S^{(2L-2)}_a \otimes {\cal D}_{ij} S^{(2)}_a \,. \label{eq:symbDgsplit}
\end{align}
On the other hand, the right-hand side of eq. \p{eq:Dg} has weight $(2L-2)$, so it does not contain a weight-$(2L-1)$ symbol component. Thus, because of linear independence, each ${\cal D}_{ij} S^{(2)}_a $ has to be a rational function. In other words, the last entries of the symbol ${\cal S}\left( g_{ij}^{(L)} \right)$ are annihilated by ${\cal D}_{ij}$. 

For the planar pentagon alphabet, we find 11 dimensionless weight-1 symbols that are annihilated by ${\cal D}_{ij}$. For example, for ${\cal D}_{13}$ we have,
\begin{align}
{\rm ker}({\cal D}_{13}) \cap \overline{\mathbb{A}}^\text{2-loop}_{\rm pl} =  \{ &  [W_1] - [W_2],\, [W_1] - [W_4] ,\, [W_3] - [W_5], \, [W_1] - [W_{11}] ,\,  \notag \\ 
& [W_4] - [W_{14}] ,\, [W_5] - [W_{15}] ,\, [W_3] - [W_{17}] - [W_{26}] ,\, \notag \\  & [W_3] - [W_{18}] + [W_{27}], \, 
[W_3] - [W_{19}] + [W_{28}], \, \notag \\  & [W_3] - [W_{20}] - [W_{29}] ,
\, [W_1] - [W_{16}] + [W_{30}] \}\,, \label{eq:lastlett}
\end{align}
as admissible
last entries of ${\cal S}\left( h_{13}^{(L)} \right)$. The last entries of ${\cal S} \left( h^{(L)}_{ij} \right)$ are obtained by cyclic permutations $\tau$ \p{eq:tauW} in \cref{eq:lastlett}.

\subsection{The next-to-last entry condition}
\label{sect:NTlastentry}

\begin{table}[t]
\begin{center}
\begin{tabular}{cccc}
\toprule
weight-2 symbols &
integrable & last entry & annihilated by ${\cal D}_{ij}$ \\ \midrule
$25*25$ & 394 & 91 & 65 \\
\bottomrule
\end{tabular}
\end{center}
\caption{The counting of the planar pentagon weight-2 symbols $\{ S^{(2)}_a \}$, which are the two rightmost entries in \cref{eq:symbgsplit}, and are annihilated by the second-order differential operator ${\cal D}_{ij}$ \p{eq:Dop}.} \label{tab:lasttwo}
\end{table} 
 
We can obtain more constraints on $h_{ij}^{(L)}$, even
without knowing the explicit expressions of the pure functions on the right-hand side of DE \p{eq:Dg}.
Indeed, each ${\cal D}_{ij} S^{(2)}_a $ \p{eq:symbDgsplit} has to be a linear combination of the rational functions appearing on the right-hand side of eq. \p{eq:Dg}. This fact imposes constraints on the next-to-last entries of the symbol ${\cal S}\left( h_{ij}^{(L)} \right)$. Let us work out these constraints more explicitly. 

The weight-2 symbols $\{S_a^{(2)} \}$, which are the two rightmost entries in \p{eq:symbgsplit}, have to be integrable in order to correspond to a pure function, i.e. \cref{eq:integrable} at $k=2L-1$ has to be satisfied. For the planar pentagon alphabet, we find 394 linearly independent integrable symbols built from dimensionless combinations of the letters \p{eq:dimlessW}. Imposing the last letter condition \p{eq:lastlett} brings their number down to 91. Among these 91 weight-2 symbols, 65 are annihilated by ${\cal D}_{ij}$, see \cref{tab:lasttwo}. As we noted previously, among the five rational functions \p{eq:5C}, only four are linearly independent. For each of the four rational functions $\overline{{\cal C}}_{kl,ij}$, we find a solution $S^{(2)}_a$ in the linear space of the 91 integrable weight-2 symbols,
\begin{align}
{\cal D}_{ij} S^{(2)}_a = \overline{{\cal C}}_{kl,ij} \,. \label{eq:DS2}
\end{align}
Thus, among the 91 weight-2 symbols $\{ S^{(2)}_a \}$ in \cref{eq:symbgsplit}, only $65+4$ symbols are compatible with the rational factors on the right-hand side of \cref{eq:Dg}.

\subsection{Bootstrap constraints}

Let us combine together the constraints on the symbol outlined above to pin down the indeterminates $c$'s in \cref{eq:symb2L}. The counting at loop order $L=1,2,3$ is summarized in \cref{tab:bootstrap}. In the first column of \cref{tab:bootstrap}, we count the number of integrable symbols, see \cref{eq:integrable}, with dimensionless entries \p{eq:dimlessW} and the first entries drawn from \p{eq:first}. In the second column, we impose the last entry condition \p{eq:lastlett}. In the third column, we also constrain the next-to-last entries as explained in \cref{sect:NTlastentry}. 
The d'Alembertian DE provides more constraints on $h_{ij}^{(L)}$ provided $h_{ij}^{(L-1)}$ is known. Namely, we demand that inhomogeneous d'Alembertian DE \p{eq:D13gs} with the known right-hand side is satisfied. In the fourth column, we count the number of indeterminates provided the DE is satisfied. 

In order to fix a unique solution of the DE we need to supplement it with the boundary conditions. We fix the remaining indeterminates in the symbol expression \p{eq:symb2L} at $L=1,2$ demanding that the ladder has correct singularities. Indeed, the rational prefactor $r_2-r_0$, which accompanies $h_{13}^{(L)}$ in expression \p{eq:F2Ll} of the ladder, has a pole at $s_{13} = 0$, see \p{eq:r1}. This spurious pole should be absent from the ladder, so it has to be suppressed by the zero of $h_{13}^{(L)}$ at $s_{13} = 0$. In \cref{sect:spur}, we checked explicitly that spurious poles are absent using explicit pentagon function expressions for the one-loop and two-loop ladders. Now, we use this property as a bootstrap constraint for the symbol \p{eq:symb2L}, 
\begin{align}
\left. {\cal S}\left( h_{i\,i+2}^{(L)} \right) \right|_{s_{i\,i+2} = 0} = 0 \,. \label{eq:nospurSymb}
\end{align}

\begin{table}[t]
\begin{center}
\begin{tabular}{c|ccccc}
\toprule
$L$ & integrability & last entry & next-to-last entry & DE & physical singularities \\
\midrule
1 & 20 & 9 & 6 & 6 & 0
\\ 
2 & 525 & 84 & 49 & 24 & 0
\\
3 & 14990 & 1012 & 354 & $-$ & $-$ \\
\bottomrule
\end{tabular}
\end{center}
\caption{The counting of indeterminates in the weight-$2L$ symbol ansatz for the pure function $h_{ij}^{(L)}$ of the $L$-loop ladder negative geometry after consecutively imposing constraints: integrability of the symbol \p{eq:integrable} built from dimensionless letters \p{eq:dimlessW} of the planar pentagon alphabet and having correct first entries \p{eq:first}, the last entry condition \p{eq:lastlett}, the next-to-last entry condition, d'Alembertian DE \p{eq:D13gs}, and absence of spurious singularities \p{eq:nospurSymb}.}
\label{tab:bootstrap}
\end{table}

In this way, we find all indeterminates $c$ in the symbol expression \p{eq:symb2L} of the one-loop and two-loop ladders. They agree with the result of the Feynman graph calculation in \cref{sect:integr2LL}. 
However, we find a contradiction in the three-loop bootstrap imposing the d'Alembertian DE, see \cref{tab:bootstrap}. This implies that the main bootstrap assumption -- the planar pentagon alphabet \p{eq:26pl} -- fails at $L \geq 3$. In other words, the three-loop ladder $\laddthreepic$ requires a larger symbol alphabet.

\subsection{Looking into the three-loop planar pentagon alphabet}

The symbol bootstrap analysis performed above suggests that the 26-letter planar pentagon alphabet \p{eq:26pl} is not sufficient to express the three-loop ladder. Besides the three-loop ladder $\laddthreepic$\,, we could ask about the symbol alphabet for other three-loop negative geometries and for the three-loop Lagrangian insertion in the Wilson loop $F^{(3)}$.

In contrast with negative geometries, the Lagrangian insertion in the Wilson loop is planar in the large color limit. The integrand of $F^{(L)}$ \p{eq:Fl} is built from the four-dimensional loop integrands of the planar MHV amplitudes, as explained in \cite{Alday:2012hy,Henn:2019swt,Chicherin:2022zxo}. Thus, the integrand of $F^{(L)}$ involves only planar families of Feynman integrals. More precisely, after switching to the frame $x_0 \to \infty$ and introducing momentum variables \p{eq:dual_momentum}, the integrand of $F^{(L)}$ is decomposed in a basis of the five-particle $L$-loop planar families of Feynman integrals, as well as products of lower-loop planar families. For example, in the two-loop case, $L=2$, the planar pentabox family, \cref{fig:I_np_prop11_a}, and the product of the one-loop pentagon families are required. In the three-loop case, $L=3$, there are four five-particle three-loop planar families.

\begin{figure}[h]
\centering
\subfloat[]{
\includegraphics[scale=0.29]{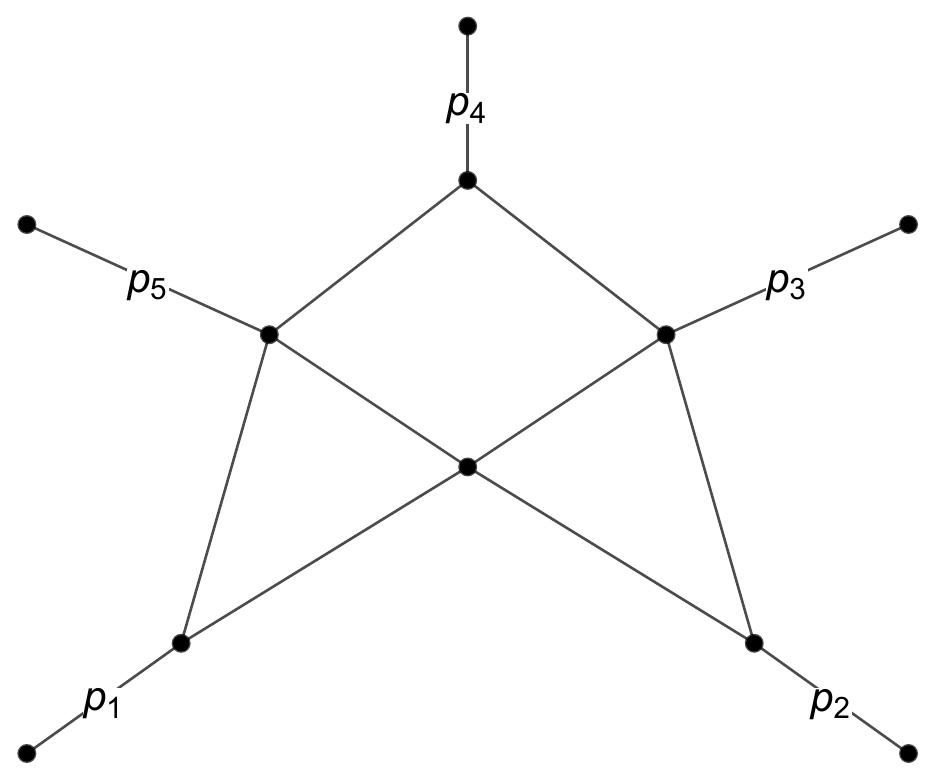}
\label{fig:3L_Z}
}
\;\qquad
\subfloat[]{ 
\includegraphics[scale=0.31]{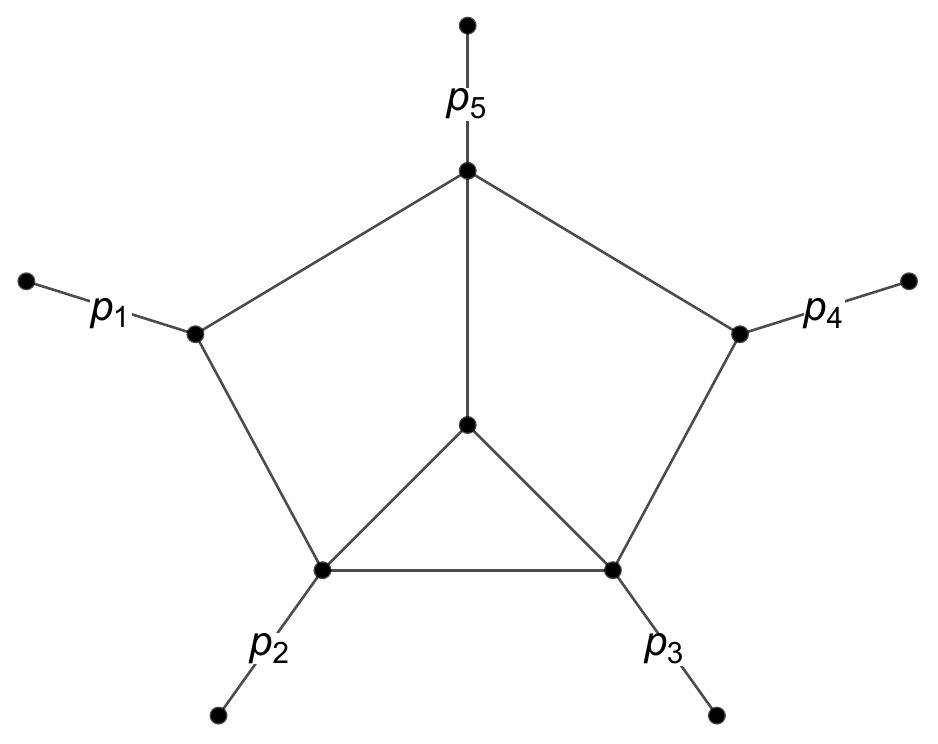}
\label{fig:3L_Z_2}
}
\caption{Diagram (a) The 8-propagator sub-sector in a five-particle three-loop planar family, which requires $\widehat{W}_1$ \p{eq:What1} in its iterated integral expression. 
Diagram (b) Another 8-propagator sub-sector in a five-particle three-loop planar family, which requires the square-root letter $\widetilde{W}_{16}$ \eqref{eq:tildeW} and algebraic letters $\tW_{41},\tW_{46},\tW_{51},\tW_{76}$ in its iterated integral expression.}
\end{figure}

We refrain from a detailed discussion of the planar three-loop families here. We only note that they do require the 26-letter planar pentagon alphabet \p{eq:26pl} to be extended by at least 30 new letters. We consider certain subsectors of the three-loop planar topologies, construct canonical DE on their maximal cuts, and identify new letters in them. 

Firstly, we find a cyclic orbit of 5 new letters $\{\widehat{W}_i\}_{i=1}^{5}$, which are quadratic in the Mandelstam variables,
\begin{align}
\widehat{W}_{i}=\tau^{i-1} \left( \widehat{W}_1 \right) \;,\qquad 
\widehat{W}_1=s_{23} s_{34} - s_{34} s_{45} + s_{45} s_{15} \,,\qquad i=1,\ldots,5\,.\label{eq:What1}
\end{align}
The letter $\widehat{W}_1$ is present in the iterated integral expressions for the five-particle three-loop planar Feynman integrals depicted in \cref{fig:3L_Z}. We constructed a pure basis of 2 MIs on the maximal cut of this 8-propagator sector and derived the canonical DE. The connection matrix of the DE does contain the letter $\widehat{W}_1$. 

Secondly, we consider the 8-propagator planar three-loop sector in \cref{fig:3L_Z_2}, and we construct a pure basis of 8 MIs of its maximal cut. Surprisingly, we find that the corresponding canonical DE involves some letters of the two-loop non-planar 11-propagator topologies discussed in \cref{sec:2L_even_letters}. Namely, we observe that letter $\tW_{16} =\Delta_4^{(1)}$ \p{eq:tildeW} is present, as well as the algebraic letters $\tW_{41},\tW_{46},\tW_{51}$, see \cref{eq:tW41,eq:tW46,eq:tW51}, which involve the square root $\sqrt{\Delta_4^{(1)}}$, and the algebraic letter $\tW_{76}$ \p{eq:tW76}, which involves simultaneously $\sqrt{\Delta_4^{(1)}}$ and $\sqrt{\Delta_5}$. In other words, this three-loop sector requires 5 new letters, which are absent in the two-loop planar topologies. Supplementing them with their cyclic permutations results in 25 new letters, i.e. the square-roots of $\{ \Delta_4^{(i)} \}_{i=1}^{5}$ \p{eq:del4} and all algebraic letters from \cref{sec:2L_odd_letters} involving these square roots. 

The 8-propagator sectors in \cref{fig:3L_Z,fig:3L_Z_2} are subsectors of the same 11-propagator family of the three-loop planar five-point Feynman integrals. We also attempted a more systematic search for new letters in the planar three-loop five-point families and their subsectors using the computer codes \cite{Fevola:2023kaw,Fevola:2023fzn,Jiang:2024eaj}, which perform the Landau analysis. Using these codes, we were able to detect letters $\widehat{W}_1$ and $\tW_{16}$ in the sectors depicted in \cref{fig:3L_Z,fig:3L_Z_2}, and we found no new letters in other inequivalent sectors of the three-loop planar families. Relying on this evidence, we conjecture the {\em three-loop planar pentagon alphabet},
\begin{align}
\mathbb{A}^\text{3-loop}_{\rm pl} =  \mathbb{A}^\text{2-loop}_{\rm pl} \cup \{\widehat{W}_i\}_{i=1}^{5} \cup  \{ \tW_{i} \}_{i=16}^{20} \cup \{ \tW_{i} \}_{i=41}^{55} \cup \{ \tW_{i} \}_{i=76}^{80} \,,
\end{align}
where we recall expressions  for the letters a given in \cref{eq:26pl,eq:tildeW,eq:What1,eq:tWorb,eq:tW41,eq:tW46,eq:tW51,eq:tW76}. This 56-letter alphabet is necessary for calculating three-loop five-particle scattering amplitudes in a massless QFT. Indeed, all 56 letters do appear in the iterated integral expressions of individual three-loop Feynman integrals. However, only a subset of letters could be necessary for the three-loop Lagrangian insertion $F_5^{(3)}$ \p{eq:Fl}.

Compared to the planar integrand of $F^{(3)}_5$, the integrands of the three-loop ladder and other nonfactorizable negative geometries are nonplanar. They are expanded in a basis of more complicated three-loop families of Feynman integrals, which in general require an extension of the 111-letter nonplanar alphabet \p{eq:w111}. These nonplanar letters could in principle contribute to the three-loop ladder. In \cite{CHMTYZ}, we study this question and perform the symbol bootstrap of the three-loop ladder and other negative geometries.


\section{Summary and discussion}
\label{sect:summary}

Let us summarize the main results of this paper. 
In the ancillary files, we give
\begin{itemize}
\item The analytic formulae of full Wilson loop observable, and its negative geometry decomposition up to $L=2$, expressed as iterated integrals and pentagon functions
\cite{Gehrmann:2018yef}, e.g.
the two-loop full observable is provided in $\tt{2loop\_ full-WL\_ iter\_ integrals.m}$ and $\tt{2loop\_ full-WL\_ pent\_ functions.m}$, with the leading singularities and constants (numerical values) defined in $\tt{rCoefficients.m}$, and $\tt{constant\_ definitions.m}$ ($\tt{constant\_ values.m}$), respectively.

\item The pure basis for two-loop five-point family \eqref{eq:integral_def} for negative geometry integrals, $\tt{MIs \_ DP\_ 1.txt}$ and $\tt{MIs \_ DP\_ 2.txt}$,
and their differential equations, $\tt{Atilde \_ DP\_ 1.m}$ and $\tt{Atilde \_ DP\_ 2.m}$, in terms of the 111 letters in \eqref{eq:dA_T}, $\tt{letters.m}$, discussed in Sec.\ref{sect:2LNPFI}.

\item The multi-Regge limit behavior, see \cref{sect:MR}, of full Wilson loop observable and negative geometries up to $L=2$, with the weight-1 and weight-3 functions defined in $\tt{funct\_ definitions\_ regge.m}$.

\end{itemize}

Using these results, we analyzed the decomposition of the five-point Wilson loop with Lagrangian insertion into geometric building blocks, to see whether this makes the observed positivity properties \cite{Chicherin:2022zxo} of the integrated one- and two-loop expressions more manifest. The building blocks involve Feynman integrals that go beyond those computed in reference \cite{Gehrmann:2015bfy,Gehrmann:2018yef}, so we computed the additional integral families with the help of the differential equations method. 
This allowed us to determine the full negative geometry decomposition of the five-point two-loop Wilson loop with Lagrangian insertion.
We found that the new scalar master integrals that we computed involve additional alphabet letters, which however drop out in the two-loop negative geometry building blocks. As a consistency check of our result, we evaluated soft and collinear limits.

Having obtained the two-loop negative geometry expansion, we then investigated the positivity properties of the individual contributions. We found that each piece has a uniform sign when the kinematics is evaluated in the Amplituhedron region, cf. Table \ref{table:data_subregions}.

Finally, we studied a complementary approach that may be suitable for obtaining higher-loop results. To begin with, we derived the all-loop integrand for the five-point ladder-type geometries, with the final expression given in eq. (\ref{genladder}).
We organized the integrals in a way that makes their leading singularities manifest. Interestingly, for products of ladder geometries, new leading singularities appear, which however drop out for the full observable. The general structure of leading singularities will be discussed in more detail in upcoming work \cite{THMT}.

From our explicit integrand expressions it is easy to see that the ladder-type integrals satisfy a d`Alembertian differential equation, similar to the four-point case \cite{Arkani-Hamed:2021iya}.
Such an equation is known to be powerful, especially when combined with bootstrap ideas. We therefore rederived the expression for the two-loop ladder from such a bootstrap approach. We then attempted to obtain the answer for the three-loop ladder from the same bootstrap ansatz, but this was unsuccessful. We conclude that at this loop order, the ansatz needs to be extended by additional symbol letters. Performing a Landau analysis of planar three-loop integrals we identified several potential new letters. A dedicated bootstrap analysis involving these additional letters will be discussed elsewhere \cite{CHMTYZ}. 

There are several further directions for future investigations:
\begin{itemize}
\item Given the uniform sign properties we observed in the individual integrated negative geometries, it would be very interesting also look for possible monotonicity properties in derivatives, as suggested in reference \cite{Henn:2024qwe}.

\item Furthermore, it would be interesting to derive the negative geometry integrands for six and more particles. The six-particle case would be particularly interesting in view of recent progress on the planar two-loop function space \cite{Henn:2024ngj}, as this could serve as a first application of those hexagon functions.

\item Multiple Lagrangian insertions in the Wilson loop are also of interest. In particular, the double Lagrangian insertion in the four-cusp Wilson loop has been considered recently in \cite{Abreu:2024flk} and its positivity in the Amplituhedron region has been observed. It would be interesting to study this case from our geometric viewpoint.

\item Finally, another natural extension of this work could be to analyse similar properties for integrated negative geometries in the ABJM theory, cf. \cite{Henn:2023pkc,He:2023exb,Lagares:2024epo,Li:2024lbw}.
\end{itemize}

\section*{Acknowledgments}

It is a pleasure to thank Taro Brown, Antonela Matija\v{s}i\'{c}, Elia Mazzuchelli, Chenyu Wang, and Qinglin Yang for discussions. Funded by the European Union (ERC, UNIVERSE PLUS, 101118787). Views and opinions expressed are however those of the authors only and do not necessarily reflect those of the European Union or the European Research Council Executive Agency. Neither the European Union nor the granting authority can be held responsible for them. D.C. is supported by ANR-24-CE31-7996. J.T. is supported by the U.S. Department of Energy, grant No. SC0009999 and the funds of the University of California.

\newpage

\appendix
\section{Two-loop alphabet letters}
\label{sect:AppLetters}

In this Appendix, we complement definitions of the alphabet letters outlined in \cref{sec:2L_even_letters}.

\subsection{Planar pentagon letters}
\label{sect:planarletters}

The planar pentagon letters $\mathbb{A}^\text{2-loop}_{\rm pl}$ \p{eq:26pl}  are in accordance with the definition in the literature \cite{Gehrmann:2018yef}. They are organized in the cyclic orbits, see \p{eq:tau},
\begin{align}
W_{i+5k} & = \tau^{i-1}\,\left( W_{1+5k} \right)\,,\qquad i=1,\ldots,5, \qquad k = 0,1,2,3,5 \,,
\label{eq:tauW}
\end{align}
and $W_{31}$, which is the
square root of the Gram determinant \p{eq:del5}, is cyclic invariant,
\begin{align}
W_{31}=\sqrt{\Delta_5}\,.
\end{align}
20 letters $\{ W_i\}_{i=1}^{20}$ are linear in Mandelstam variables. They are cyclic shifts of
\begin{align}
&W_1=s_{12}\,,& 
&W_6=s_{34}+s_{45}\,,&\nonumber\\
&W_{11}=s_{12}-s_{45}\,,&
&W_{16}=s_{12}+s_{23}-s_{45}\,.&
\end{align}
5 letters $\{ W_i\}_{i=26}^{30}$ are algebraic. They are cyclic shifts of
\begin{align}
&W_{26}=\frac{-s_{12} s_{15} + s_{12} s_{23} - s_{23} s_{34} - s_{15} s_{45} + s_{34} s_{45}-\sqrt{\Delta_5}}{-s_{12} s_{15} + s_{12} s_{23} - s_{23} s_{34} - s_{15} s_{45} + s_{34} s_{45}+\sqrt{\Delta_5}} \,.
\end{align}
Let us also note that the algebraic letters can be rewritten in terms of the chiral traces \p{eq:trm} and in terms of the rational prefactors $\{ r_{i} \}_{i=1}^{5}$ \p{eq:r1},
\begin{align}
W_{26} = \frac{\tr_{-} \left(\widehat{p}_4 \widehat{p}_5 \widehat{p}_1 \widehat{p}_2 \right)}{\tr_{+} \left(\widehat{p}_4 \widehat{p}_5 \widehat{p}_1 \widehat{p}_2 \right)} = \frac{s_{24} \left(r_3\right)^2}{s_{12} s_{45} s_{15}^3}
\end{align}
provided we choose the branch of the square root as $\sqrt{\Delta_5} = \ep_5$.

%

\subsection{Nonplanar algebraic letters}
\label{sec:2L_odd_letters}

The 11-propagator family of Feynman integrals \p{eq:integral_def} requires the planar pentagon alphabet to be extended by 85 letters $\{ \tW_i \}_{i=1}^{85}$, see \cref{eq:w111}. We presented definitions of 20 polynomial letters $\{ \tW_i \}_{i=1}^{20}$ in \cref{sec:2L_even_letters}. Here we collect definitions of 65 algebraic letters $\{ \tW_i \}_{i=21}^{85}$, organizing them in cyclic orbits,
\begin{align}
\tW_{i+5k} = \tau^{i-1} \left( \tW_{1+5k} \right) \,,\qquad i = 1 , \ldots, 5 , \qquad k=4,\ldots,16\,. \label{eq:tWorb}
\end{align}
The algebraic letters involve the square roots of $\Delta_5$, $\Delta_2^{(i)}$, $\Delta_4^{(i)}$, see \cref{eq:del5,eq:del2,eq:del4}. They have the form \p{eq:alglett} and involve one or two square roots, see \cref{tab:letters}.

45 algebraic letters involve a single square root. Indeed, 20 letters $\{ \tW_{i} \}_{i=21}^{40}$ involve the square root of $\Delta_2^{(i)}$ \p{eq:del2}. They are cyclic shifts of 
\begin{align}
\tW_{21} & = \frac{s_{12}-\sqrt{\Delta^{(1)}_2}}{s_{12}+\sqrt{\Delta^{(1)}_2}}
\,,\\ 
\tW_{26} & = \frac{s_{12}-2\,s_{34}-\sqrt{\Delta^{(1)}_2}}{s_{12}-2\,s_{34}+\sqrt{\Delta^{(1)}_2}}
\,,\\ 
\tW_{31} & = \frac{s_{12} s_{15} + s_{12} s_{34} + 2 s_{23} s_{34}-(s_{15} - s_{34})\sqrt{\Delta^{(1)}_2}}{s_{12} s_{15} + s_{12} s_{34} + 2 s_{23} s_{34}+ (s_{15} - s_{34})\sqrt{\Delta^{(1)}_2}}
\,,\\ 
\tW_{36} & = \frac{s_{12} s_{23} + s_{12} s_{45} + 2 s_{15} s_{45}-(s_{23} - s_{45}) \sqrt{\Delta^{(1)}_2}}{s_{12} s_{23} + s_{12} s_{45} + 2 s_{15} s_{45}+ (s_{23} - s_{45}) \sqrt{\Delta^{(1)}_2}} \,.
\end{align}
15 letters $\{ \tW_{i} \}_{i=41}^{55}$ involve the square root of $\Delta_4^{(i)}$ \p{eq:del4}. They are cyclic shifts of 
\begin{align}
\tW_{41} & = 
\frac{-s_{12} s_{15} + s_{12} s_{23} + 2 s_{15} s_{34} - s_{23} s_{34} + s_{34} s_{45}-\sqrt{\Delta^{(1)}_4} }{-s_{12} s_{15} + s_{12} s_{23} + 2 s_{15} s_{34} - s_{23} s_{34} + s_{34} s_{45}+\sqrt{\Delta^{(1)}_4}}
\,, \label{eq:tW41}\\ 
\tW_{46} & = 
\frac{-s_{12} s_{15} - s_{12} s_{23} + s_{23} s_{34} - s_{34} s_{45}-\sqrt{\Delta^{(1)}_4} }{-s_{12} s_{15} - s_{12} s_{23} + s_{23} s_{34} - s_{34} s_{45}+\sqrt{\Delta^{(1)}_4}}
\,,\label{eq:tW46}\\ 
\tW_{51} & = 
\frac{-s_{12} s_{15} + s_{12} s_{23} - s_{23} s_{34} - s_{34} s_{45}-\sqrt{\Delta^{(1)}_4} }{-s_{12} s_{15} + s_{12} s_{23} - s_{23} s_{34} - s_{34} s_{45}+\sqrt{\Delta^{(1)}_4}}
\,.\label{eq:tW51}
\end{align}
10 letters $\{ \tW_{i} \}_{i=56}^{65}$ involve the square root of $\Delta_5$ \p{eq:del5}. They are cyclic shifts of 
\begin{align}
\tW_{56} & = 
\frac{-s_{12} s_{15} - s_{12} s_{23} - s_{23} s_{34} - s_{15} s_{45} + s_{34} s_{45}-\sqrt{\Delta_{5}} }{-s_{12} s_{15} - s_{12} s_{23} - s_{23} s_{34} - s_{15} s_{45} + s_{34} s_{45}+\sqrt{\Delta_{5}}}
\,,\\ 
\tW_{61} & = 
\frac{q_{61}-(s_{34}+s_{45})\sqrt{\Delta_{5}} }{q_{61}+(s_{34}+s_{45})\sqrt{\Delta_{5}}}
\,,
\end{align}
where we introduce a shorthand notation for the polynomial in the Mandelstam variables,
\begin{align}
  q_{61} := &-s_{12} s_{15} s_{34} + s_{12} s_{23} s_{34} - s_{23} s_{34}^2 + s_{12} s_{15} s_{45} - s_{12} s_{23} s_{45} - 
 s_{15} s_{34} s_{45} \nonumber\\
 &-s_{23} s_{34} s_{45} - s_{34}^2 s_{45} - s_{15} s_{45}^2 - s_{34} s_{45}^2 \,.
\end{align}
Similar to the planar case, these letters are related to the rational prefactors $\{ r_i \}_{i=0}^{5}$, see \cref{eq:r0,eq:r1},
\begin{align}
\tW_{61} = - \frac{s_{35}}{s_{34} s_{45}} \frac{\left( r_0 - r_4 \right)^2}{\tW_{13}} \,.
\end{align}

The remaining 20 algebraic letters involve a pair of square roots.
10 letters $\{ \tW_i \}_{i=66}^{75}$ involve the square roots of $\Delta_5$ and $\Delta_2^{(i)}$ simultaneously, and they are cyclic shifts of
\begin{align}
\tW_{66} = 
\frac{q_{66}-\sqrt{\Delta^{(1)}_2}\sqrt{\Delta_{5}} }{q_{66}+\sqrt{\Delta^{(1)}_2}\sqrt{\Delta_{5}}}\,, \qquad
\tW_{71} = 
\frac{q_{71}-\sqrt{\Delta^{(1)}_2}\sqrt{\Delta_{5}} }{q_{71}+\sqrt{\Delta^{(1)}_2}\sqrt{\Delta_{5}}}\,,
\end{align}
where
\begin{align}
  q_{66} := & s_{12}^2 s_{15} - s_{12}^2 s_{23} + s_{12} s_{23} s_{34} - s_{12} s_{15} s_{45} \notag\\
  &  - s_{12} s_{34} s_{45} - 
 2 s_{15} s_{34} s_{45} + 2 s_{23} s_{34} s_{45} + 2 s_{34}^2 s_{45} \,,\notag\\
 q_{71} :=& -s_{12}^2 s_{15} + s_{12}^2 s_{23} - s_{12} s_{23} s_{34} + s_{12} s_{15} s_{45} \notag\\ 
 & - s_{12} s_{34} s_{45} + 
 2 s_{15} s_{34} s_{45} - 2 s_{23} s_{34} s_{45} + 2 s_{34} s_{45}^2\,.
\end{align}
5 letters $\{ \tW_{i} \}_{i=76}^{80}$ involving square roots of $\Delta_5$ and $\Delta_4^{(i)}$ simultaneously are cyclic shifts of
\begin{align}
\tW_{76}  = 
\frac{q_{76}-\sqrt{\Delta^{(1)}_4}\sqrt{\Delta_{5}} }{q_{76}+\sqrt{\Delta^{(1)}_4}\sqrt{\Delta_{5}} }
\,, \label{eq:tW76}
\end{align}
where 
\begin{align}
q_{76} := & -s_{12}^2 s_{15}^2 + 2 s_{12}^2 s_{15} s_{23} - s_{12}^2 s_{23}^2 - 2 s_{12} s_{15} s_{23} s_{34} +  2 s_{12} s_{23}^2 s_{34} \notag\\ 
&  - s_{23}^2 s_{34}^2 + s_{12} s_{15}^2 s_{45} - s_{12} s_{15} s_{23} s_{45} - 
 2 s_{12} s_{15} s_{34} s_{45} - 2 s_{12} s_{23} s_{34} s_{45}  \notag\\ 
& - s_{15} s_{23} s_{34} s_{45} + 
 2 s_{23} s_{34}^2 s_{45}  + s_{15} s_{34} s_{45}^2 - s_{34}^2 s_{45}^2 \,.
\end{align}
 5 letters $\{ \tW_{i} \}_{i=81}^{85}$
involve the square roots of $\Delta^{(i)}_2$ and $\Delta^{(i+1)}_2$ simultaneously, and they are cyclic shifts of 
\begin{align}
\tW_{81}  = 
\frac{s_{12} s_{23} + 2 s_{15} s_{45} + 2 s_{34} s_{45}-\sqrt{\Delta^{(1)}_2}\sqrt{\Delta^{(2)}_2} }{s_{12} s_{23} + 2 s_{15} s_{45} + 2 s_{34} s_{45}+\sqrt{\Delta^{(1)}_2}\sqrt{\Delta^{(2)}_2} }
\,.
\end{align}


\section{Pentagon functions}
\label{sect:pf_review}
\subsection{Review of planar pentagon functions}
The pentagon functions \cite{Gehrmann:2018yef,Chicherin:2020oor} employed in \cref{sect:integratednegativegeometries} have the following schematic expressions as iterated integrals 
\begin{align}
f_{1,a} =& \sum_{i} \alpha_{a,i} \left.[W_i]\right._{X_0} \,, \label{eq:pf1}\\[0.2cm]
f_{2,a} =& \sum_{i,j} \alpha_{a,ij} \left.[W_i,W_j]\right._{X_0} \,, \label{eq:pf2}\\[0.2cm]
f_{3,a} =& \sum_{i,j,k} \alpha_{a,ijk} \left.[W_i,W_j,W_k]\right._{X_0} + \pi^2 \sum_{i} \beta_{a,i} \left.[W_i]\right._{X_0} + \beta_{a} \boldsymbol{c}_3 \,, \label{eq:pf3}\\[0.2cm]
f_{4,a} =& \sum_{i,j,k,l} \alpha_{a,ijkm} \left.[W_i,W_j,W_k,W_m]\right._{X_0} + \pi^2 \sum_{i,j} \beta_{a,ij} \left.[W_i,W_j]\right._{X_0} \notag\\
& \qquad + \sum_{i} (\gamma_{a,i} \boldsymbol{c}_3 + \delta_{a,i} \zeta_3)\left.[W_i]\right._{X_0} \label{eq:pf4}
\end{align}
where summation indices $i,j,k,m$ run over labels of the planar letters $\mathbb{A}^\text{2-loop}_{\rm pl}$ \p{eq:26pl}, $\alpha,\beta,\gamma,\delta$ are rational numbers, and we denote by $\boldsymbol{c}_3$ a transcendental weight-3 constant.\footnote{$\boldsymbol{c}_3$ is denoted as $d_{37,3} \approx -6.02201193$ in \cite{Gehrmann:2018yef}, and it is known analytically as a weight-3 combination of Goncharov polylogarithms.} Taking into account transcendental weight of $\pi^2$ and $\zeta_3$ constants, one can easily see that $f_{w,a}$ is UT of weight $w$.  

Since the iterated integrals vanish when evaluated at their reference point, we immediately obtain analytic values of the pentagon functions at $X=X_0$,    
\begin{align}
f_{1,a}(X_0) = f_{2,a}(X_0) = f_{4,a}(X_0) = 0 \,,\qquad f_{3,a}(X_0) = \beta_a \boldsymbol{c}_3 \,. 
\label{eq:pfX0}
\end{align}

The iterated integrals obey the shuffle algebra relations which imply that the product of two iterated integrals of weights $w_1$ and $w_2$ is an iterated integral of weight $w_1 + w_2$,
\begin{align}
\left.[W_{i_1},\ldots,W_{i_{w_1}}]\right._{X_0} 
\left.[ W_{j_1},\ldots,W_{j_{w_2}}]\right._{X_0} 
=\sum \left.[W_{k_1},\ldots,W_{k_{w_1+w_2}}]\right._{X_0} \label{eq:shuffle}
\end{align}
where summation $\{ k_1,\ldots,k_{w_1+w_2}\}$ runs over the shuffle product $\{ i_1,\ldots,i_{w_1} \} \shuffle \{ j_1,\ldots,j_{w_2} \}$.
Then, the product of pentagon functions respects the weight-grading, i.e. $f_{w_1,a_1} f_{w_2,a_2}$ is a weight-$(w_1+w_2)$ UT combination of the iterated integrals. The pentagon functions are defined such that they are algebraically independent, i.e. all monomials in the pentagon functions are linearly independent. Namely, for each weight $w=1,\ldots,4$, all weight-$w$ monomials 
\begin{align}
& f_{w,a} \,,\quad
 f_{w_1,a_1} f_{w_2,a_2} |^{w_1,w_2 > 0}_{w_1+w_2=w} \,,\quad
  f_{w_1,a_1} f_{w_2,a_2}f_{w_3,a_3} |^{w_1,w_2,w_3 > 0}_{w_1+w_2+w_3=w} \,,\notag \\[0.4cm] 
& f_{w_1,a_1} f_{w_2,a_2} f_{w_3,a_3} f_{w_4,a_4} |^{w_1,w_2,w_3,w_4 > 0}_{w_1+w_2+w_3+w_4=w}
\end{align}
are linearly independent.

Let us notice that the basis of the pentagon functions, as they are defined in \cite{Gehrmann:2018yef,Chicherin:2020oor}, can be not optimal for the negative geometries. Indeed, the iterated integral expressions of the negative geometries do not contain the planar letter $W_{31} = \log(\sqrt{\Delta_5})$, whereas $W_{31}$ does appear in some weight-four pentagon functions $f_{4,a}$, see \p{eq:pf4}, present in \p{eq:gtopf4}. From this point of view, the letter $W_{31}$ is spurious, and it cancels from \p{eq:gtopf4} upon substitution of \cref{eq:pf1,eq:pf2,eq:pf3,eq:pf4} and applying the shuffle algebra \p{eq:shuffle}. A similar observation on letter $W_{31}$ has been made for finite parts of five-particle amplitudes in ${\cal N} =4$ sYM \cite{Abreu:2018aqd,Chicherin:2018yne}, maximal super-gravity \cite{Chicherin:2019xeg,Abreu:2019rpt}, and massless QCD.

\subsection{Derivatives of planar pentagon functions}
\label{sect:PFDE}

In \cref{sect:boxing}, we calculate second derivatives of the negative geometries in the kinematic variables, and in \cref{sect:limits}, we study their singular limits. Both these tasks require differentiation of the pentagon functions. In this appendix, we calculate their first-order derivatives, reexpress them in the pentagon function basis, and derive a system of the first-order differential equations for them. The latter is also helpful for numerical evaluation of the pentagon functions.

The pentagon functions are defined as iterated integrals, \cref{eq:pf1,eq:pf2,eq:pf3,eq:pf4}, so their differentiation is straightforward \p{eq:diffitint} and decreases the transcendental weight by one,
\begin{align}
df_{w,a} = \sum_{i} h^{(i)}_{w-1,a} \, d\log W_i \label{eq:df}
\end{align}
where $w=0,\ldots,4$, summation index $i$ runs over the planar pentagon letters, $h_{-1} = 0$, and $h^{(i)}_{w-1,a}$ with $w=1,\ldots,4$ are weight-$(w-1)$ UT linear combinations of the iterated integrals. They are expandable in the pentagon function basis, i.e. they are polynomials in the pentagon functions and transcendental constants $\pi^2$ and $\zeta_3$, 
\begin{align}
& h^{(i)}_{0,a} = \alpha^{(i)}_{a}\,, \label{eq:diffpf1}\\[0.2cm]
& h^{(i)}_{1,a} = \sum_b \alpha^{(i)}_{a,b} \,f_{1,b}\,, \label{eq:diffpf2}\\[0.2cm]
& h^{(i)}_{2,a} = \sum_{b} \beta^{(i)}_{a,b} f_{2,b}  + \sum_{b,c} \alpha^{(i)}_{a,bc} f_{1,b} f_{1,c} +  \pi^2 \beta^{(i)}_{a}\,, \label{eq:diffpf3} \\[0.2cm]
& h^{(i)}_{3,a} = \sum_{b} \gamma^{(i)}_{a,b} f_{3,b}  + \sum_{b,c} \beta^{(i)}_{a,bc} f_{1,b} f_{2,c} + \sum_{b,c,d} \alpha^{(i)}_{a,bcd} f_{1,b} f_{1,c} f_{1,d} + \pi^2 \sum_b \delta^{(i)}_{a,b} f_{1,b} +  \zeta_3 \gamma^{(i)}_{a} \label{eq:diffpf4}
\end{align}
where $\alpha,\beta,\gamma, \delta$ are rational constants, and indices $a,b,c,d$ run over labels of the pentagon functions.

Thus, the derivatives of the pentagon functions are expressed in the pentagon functions. In order to derive a closed system of differential equations for the pentagon functions, we need to differentiate all $\{ h^{(i)}_{w,a} \}_{w=0}^{3}$ from \cref{eq:df},
\begin{align}
d h^{(i)}_{w,a} = \sum_{j} h^{(ij)}_{w-1,a} \, d\log W_j \label{eq:dh}
\end{align}
where $h^{(ij)}_{w-1,a}$ are weight-$(w-1)$ UT linear combinations of the iterated integrals for $w=1,\ldots,3$, which are again expressible in the basis of the pentagon functions. Continuing iterative differentiations of $\{ h^{(ij)}_{w,a} \}_{w=0}^{2}$, we eventually end up with rational constants $h^{(ijkl)}_{0,a}$. Their derivatives vanish. Thus, we complement the 83 pentagon functions (see \cref{tab:pf}) by their multiple derivatives
\begin{align}
\{ f_{w,a} \}^{4}_{w=0} \, ,\quad \{ h^{(i)}_{w,a} \}^{3}_{w=1} \, , \quad \{ h^{(ij)}_{w,a} \}^{2}_{w=1} \, , \quad \{ h^{(ijk)}_{1,a} \} \, ,
\end{align}
and choose a maximal $\mathbb{Q}$-linear independent set $\vec{\bf F}$ among them, which we find to contain 165 elements. We choose the first 83 entries of $\vec{\bf F}$ to be the pentagon functions, and the last 82 entries are UT polynomials in them of weights $2,3$. Then, the iterative derivatives \p{eq:df}, \p{eq:dh}, and so on, are combined in a system of canonical DE, 
\begin{align}
d\, \vec{\bf F}(X) = \sum_i A^{(i)}\vec{\bf F}(X) \, d\log\left(W_i(X)\right) \label{eq:DEpf}
\end{align} 
where $A^{(i)}$ are $165 \times 165$ nilpotent matrices of rational numbers, and summation $i$ runs over planar pentagon letters. As compared to canonical DE \p{eq:DEu} for the pure MIs, the canonical DE \p{eq:DEpf} does not involve $\ep$.

In order to be able to solve for $\vec{\bf F}$, we need to supplement DE \p{eq:DEpf} with the initial values. We know values $\vec{\bf F}(X_0)$ at the Euclidean $X_0$ \p{eq:X0}, which is the reference point of the iterated integrals. Indeed, we know analytic values $\{ f_{w,a}(X_0)\}$ of the pentagon functions, see \p{eq:pfX0}, and all entries of $\vec{\bf F}$ are polynomial in the pentagon functions.
\section{Soft, collinear, and multi-Regge limits}
\label{sect:limits}

In this Appendix, we calculate the asymptotics of the five-cusp negative geometries in various singular regimes. By singular regimes, we imply kinematics for which letters of the planar pentagon alphabet become infinite or vanish. In \cref{sect:soft}, we consider the soft limit when one of the pentagon contour edges shrinks to a point. In this limit, we expect the negative geometries to reduce smoothly to their four-cusp counterparts, see \cref{sect:App4cusp}. Similarly, in the collinear limit considered in \cref{sect:coll}, when two adjacent edges of the pentagon contour become parallel, the four-cusp expressions are recovered.  Let us denote with $\delta$ a small parameter that controls the approach to the soft/collinear limit. In the following, we verify  
\begin{align}
F^{(L)}_5 & \xrightarrow[\delta \to 0]{} F^{(L)}_4 \;, \;
& F^{(\laddladdpic)}_5 & \xrightarrow[\delta \to 0]{} F^{(\laddladdpic)}_4 \,, \\
F^{(\laddtwopic)}_5 & \xrightarrow[\delta \to 0]{} F^{(\laddtwopic)}_4 \;, \;
& F^{\left(\looppic\right)}_5 & \xrightarrow[\delta \to 0]{} F^{\left(\looppic\right)}_4 \,
\end{align}
where $L=1,2$, and we restored an index $n=4,5$ to distinguish between the four-cusp and five-cusp cases. 

In \cref{sect:DEasympt}, we explain how we calculate asymptotics of the pentagon functions and apply these results to various singular regimes. In \cref{sect:spur}, we show that the negative geometries are finite inside the Euclidean region despite some rational prefactors in their expression having poles. In \cref{sect:MR}, we consider the multi-Regge asymptotics inherent to the five-particle scattering amplitudes.

\subsection{Asymptotics of the pentagon functions}
\label{sect:DEasympt}


We rely on the canonical DE \p{eq:DEpf} for the pentagon functions, and apply the method of \cite{Caron-Huot:2020vlo,wasow1965asymptotic} to calculate its asymptotic solution. Let us briefly summarize the main steps. We assume that the kinematics is parameterized by four variables $Y=(y_1,y_2,y_3,y_4)$ and $\delta$ such that $\delta \to 0$ in the singular regime. The asymptotic solution of the DE is governed by the singular terms of its matrix. Thus, we rewrite the alphabet letters in $(\delta,Y)$ parametrization and series expand the matrix of the DE \p{eq:DEpf} at $\delta \to 0$,
\begin{align}
\sum_i A^{(i)} \, d\log(W_i) = B\, d\log(\delta) + d{\bf C}(Y) +  O(\delta) \label{eq:DEpfdelta}
\end{align}
where $B$ is a matrix with rational number entries and $d{\bf C}(Y)$ is a matrix of $\dlog$ forms in $Y$ variables. We recall that both $B$ and $d{\bf C}$ are nilpotent. Ignoring higher-order terms of the expansion, we omit the power corrections in the asymptotic solution. 

We are looking for an asymptotic solution of the DE in the form of iterated integrals, so we need to choose a reference point $Y_0$. Then we use \texttt{DiffExp} \cite{Hidding:2020ytt} to transport the known values $\vec{\bf F}(X_0)$ of the pentagon functions to the point $(\delta,Y_0)$ at $\delta \to 0$. Since some of the pentagon functions are singular at $\delta \to 0$, we end up with a logarithmic asymptotics,  
\begin{align}
\vec{\bf F}(\delta,Y_0) = \sum_{p=0}^{4}  \log^p(\delta)\, \vec{\bf f}^{(p)} + O\left(\delta \log^4(\delta)\right) \label{eq:Flogdelta}
\end{align}
where $\vec{\bf f}^{(p)}$ are numerical vectors, which we can evaluate with arbitrarily high precision. The leading term of $\delta$-expansion \p{eq:DEpfdelta} sums up the singular logarithms in \cref{eq:Flogdelta},
\begin{align}
\vec{\bf F}(\delta,Y_0)  = \exp\left(\log(\delta) B\right) \vec{\bf f}  +  O\left(\delta \log^4(\delta)\right)
\label{eq:Fexplogdelta}
\end{align}
that we use to find numerical vector $\vec{\bf f}$. Eqs. \p{eq:Flogdelta} and \p{eq:Fexplogdelta} agree, since $B$ is nilpotent. Then we use \p{eq:Fexplogdelta} as the initial condition of the DE and find the asymptotic solution
\begin{align}
\vec{\bf F}(\delta,Y) = \exp\left(\log(\delta)B\right) \cdot {\rm Pexp}\left( \int^{Y}_{Y_0} d{\bf C} \right) \vec{\bf f} + O\left(\delta \log^4(\delta)\right).
\label{eq:Fasympt}
\end{align}
One can easily see that it satisfies DE \p{eq:DEpf} with the right-hand side \p{eq:DEpfdelta}. 
The series expansion of the path-ordered exponent ${\rm Pexp}$ in eq.~\p{eq:Fasympt} is truncated since $d{\bf C}$ is nilpotent. Thus, ${\rm Pexp}$ in \cref{eq:Fasympt} is a linear combination of the iterated integrals \p{eq:defitint} defined with respect to the reference point $Y_0$. We recall that the first 83 entries of $\vec{\bf F}$ are the planar pentagon functions $\{ f_{w,a}\}$, so eq.~\p{eq:Fasympt} provides their logarithmic asymptotics at $\delta \to 0$. 

\subsection{Soft limit}
\label{sect:soft}

We provide some details on how we calculate the soft limit $p_5 \to 0$ of the pentagon functions and of the negative geometries. We introduce the following parametrization of the kinematics 
\begin{align}
& s_{12} = \frac{s}{1+\left(\frac{1}{x}+z_1\right)\delta} \,,\quad
s_{23} = s x \,,\quad
s_{34} = \frac{s}{1+z_2 \delta} \,,\quad \notag \\
& s_{45} = \frac{s \delta}{1+\left(\frac{1}{x}+z_1\right)\delta}\,,\quad
s_{15} = \frac{s x z_1 z_2 \delta}{\left(1+z_1+\frac{1}{x}\right)(1+z_2\delta)}\,.\label{eq:MandSoft}
\end{align}
Namely, instead of five Mandelstam variables, we specify the kinematic configuration by $\delta$ and $Y=(s,x,z_1,z_2)$. Let us note that the square root of the planar pentagon alphabet $\sqrt{\Delta_5}$ is rationalized in this parametrization. In the soft limit $\delta \to 0$, we reproduce the four-particle kinematics described by two Mandelstam invariants $s$ and $sx$, 
\begin{align}
s_{12} \to s \,,\quad
s_{23} \to s x \,,\quad
s_{34} \to s \,,\quad 
s_{45} \to 0 \,,\quad
s_{15} \to 0\,.
\end{align}
The parameters $z_1,z_2$ specify the directions in which we approach the limit.

The rational prefactors $\{ r_i \}_{i=0}^{5} \cup \{ \overline{r}_i \}_{i=1}^{5}$ of the negative geometries are finite in the soft limit $\delta \to  0$ and simplify as follows, 
\begin{align}
r_0,\, r_5 \to -s^2 x \,,\qquad
r_2,\, r_3,\, \overline{r}_1,\, \overline{r}_4,\, \overline{r}_5 \to 0 \,, \qquad \overline{r}_2 \to -r_4 \,, \qquad
\overline{r}_3 \to -r_1  \,, \label{eq:rsoft}
\end{align}
and $r_0,\,r_1,\,r_4$ remain linear independent in the limit.

In order to calculate soft asymptotics of the pentagon functions, we follow \cref{sect:DEasympt}. The 26-letter planar pentagon alphabet \p{eq:26pl} reduces to a 13-letter alphabet at $\delta \to 0$, which contains letter $\log(\delta)$ and 12 letters present in the connection $d{\bf C}(Y)$ \p{eq:DEpfdelta}. The latter are the four-cusp letters
\begin{align}
\log(s)\,,\, \log(x) \,,\, \log(1+x) \label{eq:lett4cusp}
\end{align}
as well as nine spurious letters depending on $z_1$ and $z_2$. These spurious letters do appear in the asymptotics of the pentagon functions but they have to cancel out from the negative geometries at $\delta \to 0$.

The reference point $X_0$ \p{eq:X0} of the pentagon functions takes the following form in parametrization \p{eq:MandSoft},
\begin{align}
X_0 \;: \; \left( \delta = 1\,,\, s = \frac{1}{2}(1-\sqrt{5})\,,\, x = \frac{1}{2}(\sqrt{5}+1) \,,\, z_1 = -1 \,,\, z_2 = \frac{1}{2}(\sqrt{5}-3)\right) \,. \label{eq:X0soft}
\end{align}
Relying on \texttt{DiffExp}, we numerically integrate canonical DE \p{eq:DEpf} written in variables $(\delta,Y)$ from $X_0$ \p{eq:X0soft} to a point $Y_0$ on the surface $\delta =0$.
This gives us the logarithmic asymptotics of the pentagon functions \p{eq:Flogdelta} at $Y=Y_0$. Then, we factor out powers of $\log(\delta)$ according to \p{eq:Fexplogdelta} and obtain the soft asymptotics \p{eq:Fasympt} of the pentagon functions where we neglect power corrections in $\delta$. Namely, the pentagon functions are polynomials in $\log(\delta)$ whose coefficients are the iterated integrals \p{eq:defitint} for the 12-letter alphabet and the reference point $Y_0$.

Substituting the asymptotics of the pentagon functions to the pure functions of the negative geometries, we find that some of them are $O(\delta)$, i.e.
\begin{align}
g^{(1)}_1 \,,\; g^{(1)}_4 \,,\; g^{(2)}_1 \,,\; g^{(2)}_4 \,,\;  h^{(2)}_1 \,,\; h^{(2)}_4 \to 0 \,. \label{eq:g1g4soft}
\end{align}
The pure functions $g_2^{(1)},\, g_3^{(1)}$ of the one-loop ladder and $h_2^{(2)},\, h_3^{(2)}$ of the two-loop ladder are finite and contain single letter $\log(x)$ in their iterated integral expression at $\delta \to 0$. Transforming the iterated integrals
into logarithmic functions, we confirm that the one-loop and two-loop five-cusp ladders \p{eq:F1L} and \p{eq:F2Ll} reduce to their four-cusp counterparts \p{eq:F4-1L} and \p{eq:F4-2L-ladd} in the soft limit,  
\begin{align}
\lim_{\delta \to 0} F^{(1)}_5 & = s^2 x \lim_{\delta \to 0}\left( g^{(1)}_2 + g^{(1)}_3 \right) = F^{(1)}_4 \,,\\
\lim_{\delta \to 0} F^{(\laddtwopic)}_5 & = s^2 x  \lim_{\delta \to 0}\left( h^{(2)}_2 + h^{(2)}_3 \right) = F^{(\laddtwopic)}_4  \,.
\end{align}

Then we notice that the last term \p{eq:F2Lfact} of the factorized two-loop negative geometry $\laddladdpic$ vanishes at $\delta \to 0$ in view of \p{eq:rsoft} and \p{eq:g1g4soft},
\begin{align}
\sum_{i=1}^{5} \overline{r}_{i}\, g_{i+2}^{(1)} g_{i+3}^{(1)} \to 0 \,,
\end{align}
and $g^{(1)}_5$ cancels out among the first and second term, so we obtain the expected four-cusp expression \p{eq:F4-2L-laddprod},
\begin{align}
\lim_{\delta \to 0} F^{(\laddladdpic)}_5 = s^2 x  \lim_{\delta \to 0} \left( g^{(1)}_2 + g^{(1)}_3 \right)^2 = F^{(\laddladdpic)}_4 \,.
\end{align}

We recall \cite{Chicherin:2022zxo} how the two-loop correction \p{eq:F2L} reduces to the four-cusp expression in the soft limit due to \p{eq:rsoft} and \p{eq:g1g4soft},
\begin{align}
\lim_{\delta \to 0} F^{(2)}_5 = -s^2 x  \lim_{\delta \to 0}\left( g^{(2)}_0 + g^{(2)}_5 \right) = F^{(2)}_4 \,. \label{eq:F2soft}
\end{align}
Let us note that both $g^{(2)}_0$ and $g^{(2)}_5$ contain divergent terms $\log^p(\delta)$ with $p=1,\ldots,4$ and spurious letters, but they do cancel out in the sum \p{eq:F2soft}. 

Finally, the soft limit of the ``loop'' negative geometry \p{eq:F2Lloop} also reproduces the four-cusp ``loop'' \p{eq:F4-2L-loop},
\begin{align}
\lim_{\delta \to 0} F^{\left(\looppic\right)}_5 = -2 s^2 x \lim_{\delta \to 0}\left( g^{(2)}_0 + g^{(2)}_5 - h^{(2)}_2 - h^{(2)}_3 - \frac{1}{2}\left( g^{(1)}_2 + g^{(1)}_3 \right)^2 \right) = F^{\left(\looppic\right)}_4 \,.
\end{align}

\subsection{Collinear limit}
\label{sect:coll}


In order to consider the collinear limit $p_4 || p_5$, we find convenient the following parametrization of the kinematics 
\begin{align}
& s_{12} = \frac{s}{1+\delta\left(1+\frac{1}{x}\right)\left(  \frac{1}{y} + \delta\right)} \,,\quad
s_{23} = s x \,,\quad
s_{34} = \frac{s z}{1 + y (1+x) (1-z)\delta}\,,\notag\\
& s_{45}  = \frac{s(1+x)\delta^2}{1+\delta\left(1+\frac{1}{x}\right)\left(  \frac{1}{y} + \delta\right)} \,,\quad s_{15} = \frac{s x (1-z)}{1 + y (1+x) (1-z)\delta} \,.
\label{eq:scoll}
\end{align} 
In the notations of \cref{sect:DEasympt}, $Y = (s,x,z,y)$ parametrizes the collier configuration $\delta = 0$. Here $s$ and $sx$ are Mandelstam variables of the four-particle kinematics, $z$ is the fraction of the momentum split between $p_4$ and $p_5$, i.e. $p_4 \to z P$ and $p_5 \to (1-z) P$, such that in the collinear limit $\delta \to 0$,
\begin{align}
s_{12} \to s \,,\quad
s_{23} \to s x \,,\quad
s_{34} \to s z \,,\quad 
s_{45} \to 0 \,,\quad
s_{15} \to sx(1-z)\,,
\end{align}
and $y$ specifies the direction in which we approach the collinear limit.

The rational prefactors $\{ r_i \}_{i=0}^{5} \cup \{ \overline{r}_i \}_{i=1}^{5}$ of the negative geometries either vanish or are proportional to the four-cusp rational prefactor $s^2 x$,  
\begin{align}
r_0,\, r_4,\, r_5 \to -s^2 x \,,\qquad
\overline{r}_{2} \to s^2 x\,,\qquad
r_1,\,r_2,\, r_3,\, \overline{r}_{1},\, \overline{r}_{3},\, \overline{r}_{4},\, \overline{r}_{5} \to 0 \,. \label{eq:rcoll}
\end{align}

In order to calculate the collinear asymptotics of the pentagon functions and pure functions of the negative geometries, we series-expand the planar pentagon alphabet \p{eq:26pl} at $\delta \to 0$. This results in letter $\log(\delta)$ and 10 other letters, which depend on $Y$ and include the four-cusp letters \p{eq:lett4cusp}, present in the connection $d{\bf C}(Y)$ \p{eq:DEpfdelta}. We expect that only letters $\log(x),\, \log(1+x)$ remain in the expressions for the negative geometries at $\delta \to 0$. Then we employ the boundary values $\vec{\bf F}(\delta,Y_0)$ in the collinear limit $\delta \to 0$ choosing somehow $Y_0$.
Namely, we use \texttt{DiffExp} to transport numerically the boundary values $\vec{\bf F}(X_0)$ of DE \p{eq:DEpf} at $X_0$ \p{eq:X0}, which has the following form in parametrization \p{eq:scoll},
\begin{align}
X_0 \;: \; \left( \delta = \frac{1}{2}(\sqrt{5}-1)\,,\, s = \frac{1}{2}(1-\sqrt{5})\,,\, x = \frac{1}{2}(\sqrt{5}+1) \,,\, y = -1 \,,\, z = \frac{1}{2}(\sqrt{5}-1)\right) ,\label{eq:X0coll}
\end{align}
to the kinematic point $(\delta \to 0 , Y_0)$.
In this way, eq.~\p{eq:Fasympt} provides the collinear asymptotics of the pentagon functions as iterated integrals with the reference point $Y_0$.

Taking into account the collinear limit of the rational prefactors \p{eq:rcoll}, we conclude that the pure functions $g^{(1)}_i$ and $h^{(2)}_i$ with $i=1,2,3$ contribute to the ladders, see \p{eq:F1L} and \p{eq:F2Ll}. Individually, these pure functions contain powers of the divergent logarithm $\log(\delta)$ and spurious letters, but their sum is finite and reproduces the four-cusp ladders \p{eq:F4-1L} and \p{eq:F4-2L-ladd},
\begin{align}
\lim_{\delta \to 0} F^{(1)}_5 & = s^2 x \lim_{\delta \to 0}\left( g^{(1)}_1 + g^{(1)}_2 + g^{(1)}_3 \right) = F^{(1)}_4 \,,\\
\lim_{\delta \to 0} F^{(\laddtwopic)}_5 & = s^2 x  \lim_{\delta \to 0}\left( h^{(2)}_1 + h^{(2)}_2 + h^{(2)}_3 \right) = F^{(\laddtwopic)}_4  \,.
\end{align}

Owing to \p{eq:rcoll}, the pure functions $g^{(1)}_4$ and $g^{(1)}_5$ drop out in the collinear limit from the factorized two-loop five-cusp negative geometry $\laddladdpic$ \p{eq:F2Lfact},
\begin{align}
\lim_{\delta \to 0}F^{(\laddladdpic)}_5 = s^2 x  \lim_{\delta \to 0}\left( g^{(1)}_1+ g^{(1)}_2 + g^{(1)}_3 \right)^2 = F^{(\laddladdpic)}_4 \,.
\end{align}
Finally, we obtain the collinear limit of the two-loop five-cusp correction $F^{(2)}_5$ \p{eq:F2L} and of the ``loop'' negative geometry \p{eq:F2Lloop},
\begin{align}
\lim_{\delta \to 0} F^{(2)}_5 & =  -s^2 x \lim_{\delta \to 0} \left( g^{(2)}_0 + g^{(2)}_4 + g^{(2)}_5 \right) = F^{(2)}_4  \,, \\
\lim_{\delta \to 0} F^{\left(\looppic\right)}_5 & = -2 s^2 x \lim_{\delta \to 0}\Bigl( g^{(2)}_0 + g^{(2)}_4 + g^{(2)}_5 - h^{(2)}_1 - h^{(2)}_2 - h^{(2)}_3 \notag\\ 
& \qquad\qquad\quad - \frac{1}{2}\left( g^{(1)}_1 + g^{(1)}_2 + g^{(1)}_3 \right)^2 \Bigr) = F^{\left(\looppic\right)}_4 \,,
\end{align}
where individual pure terms $g^{(2)}_0$, $g^{(2)}_4$, $g^{(2)}_5$ contain powers of the divergent logarithm $\log(\delta)$ and spurious letters, which cancel out in the sum.

\subsection{Absence of spurious singularities}
\label{sect:spur}

We expect the negative geometries to be smooth inside inside the Euclidean region. For example, the two-loop ladder integrand \p{eq:h2} does not have singularities in the Euclidean region. However, this property is not manifest in the representations \p{eq:F1Ll}, \p{eq:F2Ll}, \p{eq:F2Lfact}, \p{eq:F2Lloop} of the integrated negative geometries. The planar pentagon functions are smooth in the Euclidean region \cite{Gehrmann:2018yef}, but the prefactors $\{ r_i \}_{i=0}^{5} \cup \{ \overline{r}_i \}_{i=1}^{5} $ \p{eq:r0r11} become singular inside the Euclidean region. Indeed, the rational prefactor $r_i$ with $i=1,\ldots,5$ has a pole at $s_{i+1,i+4} = 0$, and $\overline{r}_{i}$ with $i=1,\ldots,5$ has poles at $s_{i+1\,i+3} = 0$ and $s_{i+2\,i+4} = 0$, see \cref{eq:r1,eq:r6}. In other words, the rational prefactors in the expression of the negative geometries have simple poles at zero loci of nonadjacent Mandelstam variables, $s_{ij} = 0$. 

We verify that the poles of the rational prefactors are suppressed by the accompanying pure functions. Using explicit polylogarithmic expression \p{eq:g1} for the one-loop pure functions $\{ g_i^{(1)}\}_{i=1}^{5}$, it is easy to see
\begin{align}
g^{(1)}_i \left( s_{i+1\, i+4} = 0 \right) = 0 \,,
\end{align}
so the term $r_i \, g^{(1)}_i$ is finite in the Euclidean region. In the same way, $g^{(1)}_{i+2}$ and  $g^{(1)}_{i+3}$ suppress poles of $\{ \overline{r}_{i} \}_{i=1}^{5}$ at $s_{i+1\,i+3} =0$ and $s_{i+2\,i+4} = 0$, respectively, and the term $\overline{r}_{i} \, g^{(1)}_{i+2}\, g^{(1)}_{i+3}$ of $\laddladdpic$ \p{eq:F2Lfact} is finite in the Euclidean region.

A more nontrivial calculation is required at the two-loop order in order to verify suppression of the poles of $\{ r_i \}_{i=1}^{5}$ by the pure functions in eqs.~\p{eq:F2Ll} and \p{eq:F2Lloop}. We rely on the approach of \cref{sect:DEasympt}. For example, for $i=1$, we choose $s_{25} = \delta$ and $Y$ a complementing set of four independent Mandelstam variables. We choose $Y_0$ inside the Euclidean region, so $\vec{\bf F}(\delta ,Y_0)$ is finite at $\delta \to 0$, see \p{eq:Flogdelta}, and $\vec{\bf F}(\delta = 0,Y_0) = \vec{\bf f}$, see \p{eq:Fexplogdelta}. Using the iterated integral expression \p{eq:Fasympt} for the pentagon functions at $s_{25} = 0$, we verify that 
\begin{align}
g^{(2)}_1 \left( s_{25} = 0 \right) = h^{(2)}_1 \left( s_{25} = 0 \right) = 0 \,.
\end{align}
Thus, we have explicitly checked that all two-loop iterated negative geometries are free from unphysical poles in the Euclidean region.

\subsection{Multi-Regge limit}
\label{sect:MR}


Another interesting asymptotic regime of the five-particle scattering is the multi-Regge kinematics. This is the high-energy scattering regime where the rapidities of the final state particles are strongly ordered and their transverse momenta are comparable. For the process $12 \to 345$, it is common to reach the multi-Regge regime using the following parametrization with $\delta \to 0$,
\begin{align}
s_{12} = \frac{s_1 s_2}{\kappa \delta^2} \,,\quad s_{23} = -z_1 z_2 \kappa \,, \quad s_{34} = \frac{s_1}{\delta}\,,\quad s_{45} = \frac{s_2}{\delta} \,,\quad s_{15} = - (1-z_1) (1-z_2) \kappa \,, \label{eq:MRkin}
\end{align}
However, instead of the three-particle production channel, we study the negative geometries in the Euclidean region with all adjacent Mandelstam invariants being positive. So we have to impose $z_1 >1$ and $z_2<0$ or $z_1 <0$ and $z_2>1$, and $s_1,\,s_2,\,\kappa,\, \delta>0$.

In the multi-Regge limit $\delta \to 0$, the rational prefactors $\{ r_i \}_{i=0}^{5} \cup \{ \overline{r}_i \}_{i=1}^{5} $  of the negative geometries are divergent. $r_0$ and $r_4$ are the most singular. We keep them and omit the remaining rational prefactors which would provide power corrections to the leading term,
\begin{align}
r_0,\, r_4 = - s_1 s_2 z_1 (1-z_2)\delta^{-2} +O(\delta^{-1})  \,,\quad
r_1,\, r_2,\, r_3,\, r_5,\overline{r}_1\ldots,\overline{r}_{5} = O(\delta^{-1}) \,.
\end{align}
Here and in the following we assume the Euclidean region with $z_1>1$ and $z_2 <0$.

Then we normalize the negative geometries by the Born-level $F^{(0)}$ \p{eq:F0} and obtain the following expressions at one-loop level, see \cref{eq:F1Ll},
\begin{align}
\frac{F^{(\laddonepic)}}{F^{(0)}} = -\sum_{i=1,2,3,5} g^{(1)}_i + O\left(\delta \log^2(\delta)\right)\,, \label{eq:FMR1Ll}
\end{align}
and at two-loop level, see \cref{eq:F2Ll,eq:F2Lfact,eq:F2Lloop},
\begin{align}
\frac{F^{(\laddtwopic)}}{F^{(0)}} & = - \sum_{i=1,2,3,5} h^{(2)}_i + O(\delta \log^4(\delta)) \,, \label{eq:FMR2Ll} \\
\frac{F^{(\laddladdpic)}}{F^{(0)}} & =- \left( \sum_{i=1,2,3,5} g^{(1)}_i \right)^2 + O(\delta \log^4(\delta)) \,,\label{eq:FMR2Lfact} \\
\frac{F^{\left(\looppic\right)}}{F^{(0)}} & = 2 \left( g_0^{(2)} + g_4^{(2)} \right) -2 \sum_{i=1,2,3,5} h^{(2)}_i - \left( \sum_{i=1,2,3,5} g^{(1)}_i \right)^2  + O(\delta \log^4(\delta)) \,. \label{eq:FMR2Lloop}
\end{align}

Then we apply the approach of \cref{sect:DEasympt} to calculate the multi-Regge asymptotics of the pentagon functions and of the pure functions from the previous equations. The planar pentagon alphabet \p{eq:26pl} simplifies to 11 letters in the multi-Regge limit $\delta \to 0$ which are split into four alphabets depending on the nonoverlapping set of variables. Namely, letter $\log(\delta)$ as well as ten letters,
\begin{align}
\{ \kappa\} \; , \; \{ s_1, s_2, s_1+s_2\}, \{z_1, 1-z_1, z_2, 1-z_2, z_1 - z_2, 1- z_1-z_2 \} \,. \label{eq:alphMR}
\end{align}
figuring in the connection $d{\bf C}(Y)$ \p{eq:DEpfdelta}. Let us mention that despite the set of five parameters $Y = (s_1,s_2,z_1,z_2,\kappa)$ is redundant, the calculation procedure outlined \cref{sect:DEasympt} stays the same. Then, we numerically integrate canonical DE \p{eq:DEpf} and transport its boundary values at $X_0$ in the parametrization \p{eq:MRkin},
\begin{align}
X_0 = \left( \delta=1 \, , \, s_1 = -1 \, , \, s_2=-1 \, ,\, \kappa=-1 \,,\, z_1 = \frac{1}{2}(1+\sqrt{5}) \,,\, z_2 = \frac{1}{2}(1-\sqrt{5})  \right)
\end{align}
to a point $(\delta \to 0, Y_0)$. 
Eq.~\p{eq:Fasympt} provides the logarithmic asymptotics of the pentagon functions in the multi-Regge limit. Substituting the asymptotics in \crefrange{eq:FMR1Ll}{eq:FMR2Lloop} we find
\begin{align}
\frac{F^{(\bf g)}}{F^{(0)}} = \sum_{k=0}^{2\ell(\bf g)} q^{({\bf g})}_k(z_1,z_2,s_1,s_2,\kappa) \log^k(\delta) + O\left(\delta \log^{2\ell(\bf g)}(\delta)\right) \, \label{eq:FMRasympt}
\end{align}
where 
\begin{equation}
{\bf g} \in \{ \laddonepic,\laddtwopic, \laddladdpic, \looppic \}
\end{equation}
labels the negative geometries and $\ell(\bf g)$ counts their loop order (number of black blobs). The pure functions $q^{({\bf g})}_k$ are iterated integrals over the alphabet \p{eq:alphMR} with the reference point $Y_0$.

The functions $q^{({\bf g})}_k$ \p{eq:FMRasympt} are expressible in terms of the classical polylogarithms. They are graded UT polynomials of the transcendental weight $(2\ell({\bf g})-k)$ in the logarithms
\begin{align}
\log\left( \frac{s_1}{\kappa}\right)\,,\; 
\log\left( \frac{s_2}{\kappa}\right) \,,\;  \log(-z_1 z_2) \,,\;   \log\left((z_1-1)(1-z_2)\right) , \label{eq:logMR}
\end{align}
transcendental constants $\pi^2,\,\zeta_3$, and a weight-3 UT combination of the dilogarithms and trilogarithms. We refrain from typing in here its explicit expression. The explicit expressions for $q^{({\bf g})}_k$ are provided in the ancillary files.

Let us note that this weight-3 polylogarithmic combination appears in the two-loop ladder $q^{(\laddtwopic)}_k$ and ``loop'' $q^{(\looppic)}_k$ negative geometries with $k=3,4$. However, it cancels out in the two-loop negative-geometry decomposition \p{eq:Fdecomp2},
\begin{align}
 - q^{(\laddtwopic)}_k - \frac{1}{2} q^{(\laddladdpic)}_k + \frac{1}{2} q^{\left(\looppic\right)}_k
\end{align} 
which involves only logarithms \p{eq:logMR} and constants $\pi^2,\,\zeta_3$. In other words, the negative geometries, including the two-loop ladder, have a more complicated form than the two-loop correction $F^{(2)}$ in the multi-Regge limit.
\section{Four-cusp negative geometries}
\label{sect:App4cusp}

In this Appendix, we summarise perturbative results up to the two-loop order for the Lagrangian insertion in the four-cusp Wilson loop, $F_4^{(L)}$ with $L=0,1,2$ in \cref{eq:Fl}, and for the corresponding negative geometries \cite{Arkani-Hamed:2021iya}.

The loop corrections
are harmonic polylogarithms of the dual conformal cross-ratio,
\begin{align}
x := \frac{x_{20}^2 x_{40}^2 x_{13}^2}{x_{10}^2 x_{30}^2 x_{24}^2} \,. 
\end{align}
They are proportional to the unique leading singularity given by the Born-level approximation,
\begin{align}
F_4^{(0)}(x_0;x_1,\ldots,x_4) = -\frac{x_{13}^2 x_{24}^2}{x_{10}^2 x_{20}^2 x_{30}^2 x_{40}^2}   \,,
\end{align}
and have the following form
\begin{align}
& F_4^{(1)}/F_4^{(0)} = F_4^{(\laddonepic)}/F_4^{(0)} = - \left[ \log^2(x) + \pi^2\right] \,, \label{eq:F4-1L}\\
& F_4^{(\laddtwopic)}/F_4^{(0)} = -\frac{1}{6} \left[\pi^2 + \log^2(x) \right]\left[5\pi^2 + \log^2(x) \right]\,,\label{eq:F4-2L-ladd}\\
& F_4^{(\laddladdpic)}/F_4^{(0)} = -\frac{1}{6} \left[\pi^2 + \log^2(x) \right]^2\,,\label{eq:F4-2L-laddprod}\\
& F_4^{\left(\looppic\right)}/F_4^{(0)} = 
 -8 H_{\text{0,0,0,0}}-8 H_{\text{-1,0,0,0}}+16 H_{\text{-1,-1,0,0}}-8 H_{\text{-2,0,0}} + 8 \zeta _3 \left(2 H_{-1}-H_0\right)\notag\\&\hspace{2.5cm}- 4\pi ^2 \left(H_{\text{-1,0}}- 2\,H_{\text{-1,-1}}+H_{-2}\right) -\frac{13 \pi ^4}{45}\,, \label{eq:F4-2L-loop}
\end{align}
where $H$ are the harmonic polylogarithms of argument $x$ \cite{Remiddi:1999ew}.
The expression for the two-loop correction $F_4^{(2)}$ follows from the two-loop negative geometry decomposition \p{eq:Fdecomp2}, 
\begin{align} 
F^{(2)}_4 = - F_4^{(\laddtwopic)} - \frac{1}{2} F_4^{(\laddladdpic)} +\frac{1}{2} F_4^{\left(\looppic\right)} \,.
\end{align}
In the frame $x_0 \to \infty$, the kinematics is that of the four-particle scattering, see \cref{eq:Fx0inf},
\begin{align}
x \to \frac{t}{s} \,,\qquad \lim_{x_0 \to \infty} (x_0^2)^4 F_4^{(0)} = - s t \,,
\end{align}
where $t = x_{13}^2$ and $s=x_{24}^2$\,.


\bibliographystyle{JHEP}
\bibliography{neg_geom.bib}

\providecommand{\href}[2]{#2}\begingroup\raggedright\begin{thebibliography}{10}

\bibitem{Arkani-Hamed:2022rwr}
N.~Arkani-Hamed, L.~J. Dixon, A.~J. McLeod, M.~Spradlin, J.~Trnka and A.~Volovich, \emph{{Solving Scattering in $N$ = 4 Super-Yang-Mills Theory}},  \href{https://arxiv.org/abs/2207.10636}{{\ttfamily 2207.10636}}.

\bibitem{Travaglini:2022uwo}
G.~Travaglini et~al., \emph{{The SAGEX review on scattering amplitudes}}, \href{https://doi.org/10.1088/1751-8121/ac8380}{\emph{J. Phys. A} {\bfseries 55} (2022) 443001} [\href{https://arxiv.org/abs/2203.13011}{{\ttfamily 2203.13011}}].

\bibitem{Hodges:2009hk}
A.~Hodges, \emph{{Eliminating spurious poles from gauge-theoretic amplitudes}}, \href{https://doi.org/10.1007/JHEP05(2013)135}{\emph{JHEP} {\bfseries 05} (2013) 135} [\href{https://arxiv.org/abs/0905.1473}{{\ttfamily 0905.1473}}].

\bibitem{Arkani-Hamed:2013jha}
N.~Arkani-Hamed and J.~Trnka, \emph{{The Amplituhedron}}, \href{https://doi.org/10.1007/JHEP10(2014)030}{\emph{JHEP} {\bfseries 10} (2014) 030} [\href{https://arxiv.org/abs/1312.2007}{{\ttfamily 1312.2007}}].

\bibitem{Britto:2005fq}
R.~Britto, F.~Cachazo, B.~Feng and E.~Witten, \emph{{Direct proof of tree-level recursion relation in Yang-Mills theory}}, \href{https://doi.org/10.1103/PhysRevLett.94.181602}{\emph{Phys. Rev. Lett.} {\bfseries 94} (2005) 181602} [\href{https://arxiv.org/abs/hep-th/0501052}{{\ttfamily hep-th/0501052}}].

\bibitem{Arkani-Hamed:2010wgm}
N.~Arkani-Hamed, J.~L. Bourjaily, F.~Cachazo, A.~Hodges and J.~Trnka, \emph{{A Note on Polytopes for Scattering Amplitudes}}, \href{https://doi.org/10.1007/JHEP04(2012)081}{\emph{JHEP} {\bfseries 04} (2012) 081} [\href{https://arxiv.org/abs/1012.6030}{{\ttfamily 1012.6030}}].

\bibitem{Arkani-Hamed:2014dca}
N.~Arkani-Hamed, A.~Hodges and J.~Trnka, \emph{{Positive Amplitudes In The Amplituhedron}}, \href{https://doi.org/10.1007/JHEP08(2015)030}{\emph{JHEP} {\bfseries 08} (2015) 030} [\href{https://arxiv.org/abs/1412.8478}{{\ttfamily 1412.8478}}].

\bibitem{Herrmann:2020oud}
E.~Herrmann, C.~Langer, J.~Trnka and M.~Zheng, \emph{{Positive Geometries for One-Loop Chiral Octagons}},  \href{https://arxiv.org/abs/2007.12191}{{\ttfamily 2007.12191}}.

\bibitem{Herrmann:2020qlt}
E.~Herrmann, C.~Langer, J.~Trnka and M.~Zheng, \emph{{Positive geometry, local triangulations, and the dual of the Amplituhedron}}, \href{https://doi.org/10.1007/JHEP01(2021)035}{\emph{JHEP} {\bfseries 01} (2021) 035} [\href{https://arxiv.org/abs/2009.05607}{{\ttfamily 2009.05607}}].

\bibitem{Alday:2011ga}
L.~F. Alday, E.~I. Buchbinder and A.~A. Tseytlin, \emph{{Correlation function of null polygonal Wilson loops with local operators}}, \href{https://doi.org/10.1007/JHEP09(2011)034}{\emph{JHEP} {\bfseries 09} (2011) 034} [\href{https://arxiv.org/abs/1107.5702}{{\ttfamily 1107.5702}}].

\bibitem{Engelund:2011fg}
O.~T. Engelund and R.~Roiban, \emph{{On correlation functions of Wilson loops, local and non-local operators}}, \href{https://doi.org/10.1007/JHEP05(2012)158}{\emph{JHEP} {\bfseries 05} (2012) 158} [\href{https://arxiv.org/abs/1110.0758}{{\ttfamily 1110.0758}}].

\bibitem{Engelund:2012re}
O.~T. Engelund and R.~Roiban, \emph{{Correlation functions of local composite operators from generalized unitarity}}, \href{https://doi.org/10.1007/JHEP03(2013)172}{\emph{JHEP} {\bfseries 03} (2013) 172} [\href{https://arxiv.org/abs/1209.0227}{{\ttfamily 1209.0227}}].

\bibitem{Alday:2012hy}
L.~F. Alday, P.~Heslop and J.~Sikorowski, \emph{{Perturbative correlation functions of null Wilson loops and local operators}}, \href{https://doi.org/10.1007/JHEP03(2013)074}{\emph{JHEP} {\bfseries 03} (2013) 074} [\href{https://arxiv.org/abs/1207.4316}{{\ttfamily 1207.4316}}].

\bibitem{Alday:2013ip}
L.~F. Alday, J.~M. Henn and J.~Sikorowski, \emph{{Higher loop mixed correlators in N=4 SYM}}, \href{https://doi.org/10.1007/JHEP03(2013)058}{\emph{JHEP} {\bfseries 03} (2013) 058} [\href{https://arxiv.org/abs/1301.0149}{{\ttfamily 1301.0149}}].

\bibitem{Henn:2019swt}
J.~M. Henn, G.~P. Korchemsky and B.~Mistlberger, \emph{{The full four-loop cusp anomalous dimension in $\mathcal{N}=4$ super Yang-Mills and QCD}}, \href{https://doi.org/10.1007/JHEP04(2020)018}{\emph{JHEP} {\bfseries 04} (2020) 018} [\href{https://arxiv.org/abs/1911.10174}{{\ttfamily 1911.10174}}].

\bibitem{Chicherin:2022bov}
D.~Chicherin and J.~M. Henn, \emph{{Symmetry properties of Wilson loops with a Lagrangian insertion}}, \href{https://doi.org/10.1007/JHEP07(2022)057}{\emph{JHEP} {\bfseries 07} (2022) 057} [\href{https://arxiv.org/abs/2202.05596}{{\ttfamily 2202.05596}}].

\bibitem{Chicherin:2022zxo}
D.~Chicherin and J.~Henn, \emph{{Pentagon Wilson loop with Lagrangian insertion at two loops in $ \mathcal{N} $ = 4 super Yang-Mills theory}}, \href{https://doi.org/10.1007/JHEP07(2022)038}{\emph{JHEP} {\bfseries 07} (2022) 038} [\href{https://arxiv.org/abs/2204.00329}{{\ttfamily 2204.00329}}].

\bibitem{Drummond:2009fd}
J.~M. Drummond, J.~M. Henn and J.~Plefka, \emph{{Yangian symmetry of scattering amplitudes in N=4 super Yang-Mills theory}}, \href{https://doi.org/10.1088/1126-6708/2009/05/046}{\emph{JHEP} {\bfseries 05} (2009) 046} [\href{https://arxiv.org/abs/0902.2987}{{\ttfamily 0902.2987}}].

\bibitem{Dixon:2016apl}
L.~J. Dixon, M.~von Hippel, A.~J. McLeod and J.~Trnka, \emph{{Multi-loop positivity of the planar $ \mathcal{N} $ = 4 SYM six-point amplitude}}, \href{https://doi.org/10.1007/JHEP02(2017)112}{\emph{JHEP} {\bfseries 02} (2017) 112} [\href{https://arxiv.org/abs/1611.08325}{{\ttfamily 1611.08325}}].

\bibitem{Arkani-Hamed:2021iya}
N.~Arkani-Hamed, J.~Henn and J.~Trnka, \emph{{Nonperturbative negative geometries: amplitudes at strong coupling and the amplituhedron}}, \href{https://doi.org/10.1007/JHEP03(2022)108}{\emph{JHEP} {\bfseries 03} (2022) 108} [\href{https://arxiv.org/abs/2112.06956}{{\ttfamily 2112.06956}}].

\bibitem{Drummond:2010cz}
J.~M. Drummond, J.~M. Henn and J.~Trnka, \emph{{New differential equations for on-shell loop integrals}}, \href{https://doi.org/10.1007/JHEP04(2011)083}{\emph{JHEP} {\bfseries 04} (2011) 083} [\href{https://arxiv.org/abs/1010.3679}{{\ttfamily 1010.3679}}].

\bibitem{Brown:2023mqi}
T.~V. Brown, U.~Oktem, S.~Paranjape and J.~Trnka, \emph{{Loops of loops expansion in the amplituhedron}}, \href{https://doi.org/10.1007/JHEP07(2024)025}{\emph{JHEP} {\bfseries 07} (2024) 025} [\href{https://arxiv.org/abs/2312.17736}{{\ttfamily 2312.17736}}].

\bibitem{Korchemskaya:1992je}
I.~A. Korchemskaya and G.~P. Korchemsky, \emph{{On lightlike Wilson loops}}, \href{https://doi.org/10.1016/0370-2693(92)91895-G}{\emph{Phys. Lett. B} {\bfseries 287} (1992) 169}.

\bibitem{Drummond:2007aua}
J.~M. Drummond, G.~P. Korchemsky and E.~Sokatchev, \emph{{Conformal properties of four-gluon planar amplitudes and Wilson loops}}, \href{https://doi.org/10.1016/j.nuclphysb.2007.11.041}{\emph{Nucl. Phys. B} {\bfseries 795} (2008) 385} [\href{https://arxiv.org/abs/0707.0243}{{\ttfamily 0707.0243}}].

\bibitem{Alday:2007hr}
L.~F. Alday and J.~M. Maldacena, \emph{{Gluon scattering amplitudes at strong coupling}}, \href{https://doi.org/10.1088/1126-6708/2007/06/064}{\emph{JHEP} {\bfseries 06} (2007) 064} [\href{https://arxiv.org/abs/0705.0303}{{\ttfamily 0705.0303}}].

\bibitem{Brandhuber:2007yx}
A.~Brandhuber, P.~Heslop and G.~Travaglini, \emph{{MHV amplitudes in N=4 super Yang-Mills and Wilson loops}}, \href{https://doi.org/10.1016/j.nuclphysb.2007.11.002}{\emph{Nucl. Phys. B} {\bfseries 794} (2008) 231} [\href{https://arxiv.org/abs/0707.1153}{{\ttfamily 0707.1153}}].

\bibitem{Eden:2000mv}
B.~Eden, C.~Schubert and E.~Sokatchev, \emph{{Three loop four point correlator in N=4 SYM}}, \href{https://doi.org/10.1016/S0370-2693(00)00515-3}{\emph{Phys. Lett. B} {\bfseries 482} (2000) 309} [\href{https://arxiv.org/abs/hep-th/0003096}{{\ttfamily hep-th/0003096}}].

\bibitem{Arkani-Hamed:2017vfh}
N.~Arkani-Hamed, H.~Thomas and J.~Trnka, \emph{{Unwinding the Amplituhedron in Binary}}, \href{https://doi.org/10.1007/JHEP01(2018)016}{\emph{JHEP} {\bfseries 01} (2018) 016} [\href{https://arxiv.org/abs/1704.05069}{{\ttfamily 1704.05069}}].

\bibitem{DOPT}
L.~Dixon, U.~Oktem, S.~Paranjape and J.~Trnka, \emph{In preparation}, .

\bibitem{THMT}
T.~Brown, J.~M. Henn, E.~Mazzucchelli and J.~Trnka, \emph{In preparation}, .

\bibitem{Arkani-Hamed:2010pyv}
N.~Arkani-Hamed, J.~L. Bourjaily, F.~Cachazo and J.~Trnka, \emph{{Local Integrals for Planar Scattering Amplitudes}}, \href{https://doi.org/10.1007/JHEP06(2012)125}{\emph{JHEP} {\bfseries 06} (2012) 125} [\href{https://arxiv.org/abs/1012.6032}{{\ttfamily 1012.6032}}].

\bibitem{Bourjaily:2013mma}
J.~L. Bourjaily, S.~Caron-Huot and J.~Trnka, \emph{{Dual-Conformal Regularization of Infrared Loop Divergences and the Chiral Box Expansion}}, \href{https://doi.org/10.1007/JHEP01(2015)001}{\emph{JHEP} {\bfseries 01} (2015) 001} [\href{https://arxiv.org/abs/1303.4734}{{\ttfamily 1303.4734}}].

\bibitem{Bourjaily:2015jna}
J.~L. Bourjaily and J.~Trnka, \emph{{Local Integrand Representations of All Two-Loop Amplitudes in Planar SYM}}, \href{https://doi.org/10.1007/JHEP08(2015)119}{\emph{JHEP} {\bfseries 08} (2015) 119} [\href{https://arxiv.org/abs/1505.05886}{{\ttfamily 1505.05886}}].

\bibitem{Bourjaily:2017wjl}
J.~L. Bourjaily, E.~Herrmann and J.~Trnka, \emph{{Prescriptive Unitarity}}, \href{https://doi.org/10.1007/JHEP06(2017)059}{\emph{JHEP} {\bfseries 06} (2017) 059} [\href{https://arxiv.org/abs/1704.05460}{{\ttfamily 1704.05460}}].

\bibitem{Bourjaily:2019iqr}
J.~L. Bourjaily, E.~Herrmann, C.~Langer, A.~J. McLeod and J.~Trnka, \emph{{Prescriptive Unitarity for Non-Planar Six-Particle Amplitudes at Two Loops}}, \href{https://doi.org/10.1007/JHEP12(2019)073}{\emph{JHEP} {\bfseries 12} (2019) 073} [\href{https://arxiv.org/abs/1909.09131}{{\ttfamily 1909.09131}}].

\bibitem{Bourjaily:2020qca}
J.~L. Bourjaily, E.~Herrmann, C.~Langer and J.~Trnka, \emph{{Building bases of loop integrands}}, \href{https://doi.org/10.1007/JHEP11(2020)116}{\emph{JHEP} {\bfseries 11} (2020) 116} [\href{https://arxiv.org/abs/2007.13905}{{\ttfamily 2007.13905}}].

\bibitem{Arkani-Hamed:2010zjl}
N.~Arkani-Hamed, J.~L. Bourjaily, F.~Cachazo, S.~Caron-Huot and J.~Trnka, \emph{{The All-Loop Integrand For Scattering Amplitudes in Planar N=4 SYM}}, \href{https://doi.org/10.1007/JHEP01(2011)041}{\emph{JHEP} {\bfseries 01} (2011) 041} [\href{https://arxiv.org/abs/1008.2958}{{\ttfamily 1008.2958}}].

\bibitem{Badger:2019djh}
S.~Badger, D.~Chicherin, T.~Gehrmann, G.~Heinrich, J.~M. Henn, T.~Peraro et~al., \emph{{Analytic form of the full two-loop five-gluon all-plus helicity amplitude}}, \href{https://doi.org/10.1103/PhysRevLett.123.071601}{\emph{Phys. Rev. Lett.} {\bfseries 123} (2019) 071601} [\href{https://arxiv.org/abs/1905.03733}{{\ttfamily 1905.03733}}].

\bibitem{Henn:2019mvc}
J.~Henn, B.~Power and S.~Zoia, \emph{{Conformal Invariance of the One-Loop All-Plus Helicity Scattering Amplitudes}}, \href{https://doi.org/10.1007/JHEP02(2020)019}{\emph{JHEP} {\bfseries 02} (2020) 019} [\href{https://arxiv.org/abs/1911.12142}{{\ttfamily 1911.12142}}].

\bibitem{Gehrmann:2018yef}
T.~Gehrmann, J.~M. Henn and N.~A. Lo~Presti, \emph{{Pentagon functions for massless planar scattering amplitudes}}, \href{https://doi.org/10.1007/JHEP10(2018)103}{\emph{JHEP} {\bfseries 10} (2018) 103} [\href{https://arxiv.org/abs/1807.09812}{{\ttfamily 1807.09812}}].

\bibitem{Chicherin:2020oor}
D.~Chicherin and V.~Sotnikov, \emph{{Pentagon Functions for Scattering of Five Massless Particles}}, \href{https://doi.org/10.1007/JHEP12(2020)167}{\emph{JHEP} {\bfseries 20} (2020) 167} [\href{https://arxiv.org/abs/2009.07803}{{\ttfamily 2009.07803}}].

\bibitem{Gehrmann:2015bfy}
T.~Gehrmann, J.~M. Henn and N.~A. Lo~Presti, \emph{{Analytic form of the two-loop planar five-gluon all-plus-helicity amplitude in QCD}}, \href{https://doi.org/10.1103/PhysRevLett.116.062001}{\emph{Phys. Rev. Lett.} {\bfseries 116} (2016) 062001} [\href{https://arxiv.org/abs/1511.05409}{{\ttfamily 1511.05409}}].

\bibitem{Henn:2013pwa}
J.~M. Henn, \emph{{Multiloop integrals in dimensional regularization made simple}}, \href{https://doi.org/10.1103/PhysRevLett.110.251601}{\emph{Phys. Rev. Lett.} {\bfseries 110} (2013) 251601} [\href{https://arxiv.org/abs/1304.1806}{{\ttfamily 1304.1806}}].

\bibitem{Henn:2014qga}
J.~M. Henn, \emph{{Lectures on differential equations for Feynman integrals}}, \href{https://doi.org/10.1088/1751-8113/48/15/153001}{\emph{J. Phys. A} {\bfseries 48} (2015) 153001} [\href{https://arxiv.org/abs/1412.2296}{{\ttfamily 1412.2296}}].

\bibitem{Tkachov:1981wb}
F.~V. Tkachov, \emph{{A theorem on analytical calculability of 4-loop renormalization group functions}}, \href{https://doi.org/10.1016/0370-2693(81)90288-4}{\emph{Phys. Lett. B} {\bfseries 100} (1981) 65}.

\bibitem{Chetyrkin:1981qh}
K.~G. Chetyrkin and F.~V. Tkachov, \emph{{Integration by parts: The algorithm to calculate $\beta$-functions in 4 loops}}, \href{https://doi.org/10.1016/0550-3213(81)90199-1}{\emph{Nucl. Phys. B} {\bfseries 192} (1981) 159}.

\bibitem{Lee:2012cn}
R.~N. Lee, \emph{{Presenting LiteRed: a tool for the Loop InTEgrals REDuction}},  \href{https://arxiv.org/abs/1212.2685}{{\ttfamily 1212.2685}}.

\bibitem{Lee:2013mka}
R.~N. Lee, \emph{{LiteRed 1.4: a powerful tool for reduction of multiloop integrals}}, \href{https://doi.org/10.1088/1742-6596/523/1/012059}{\emph{J. Phys. Conf. Ser.} {\bfseries 523} (2014) 012059} [\href{https://arxiv.org/abs/1310.1145}{{\ttfamily 1310.1145}}].

\bibitem{Smirnov:2019qkx}
A.~V. Smirnov and F.~S. Chuharev, \emph{{FIRE6: Feynman Integral REduction with Modular Arithmetic}}, \href{https://doi.org/10.1016/j.cpc.2019.106877}{\emph{Comput. Phys. Commun.} {\bfseries 247} (2020) 106877} [\href{https://arxiv.org/abs/1901.07808}{{\ttfamily 1901.07808}}].

\bibitem{Henn:2014lfa}
J.~M. Henn, K.~Melnikov and V.~A. Smirnov, \emph{{Two-loop planar master integrals for the production of off-shell vector bosons in hadron collisions}}, \href{https://doi.org/10.1007/JHEP05(2014)090}{\emph{JHEP} {\bfseries 05} (2014) 090} [\href{https://arxiv.org/abs/1402.7078}{{\ttfamily 1402.7078}}].

\bibitem{Henn:2020lye}
J.~Henn, B.~Mistlberger, V.~A. Smirnov and P.~Wasser, \emph{{Constructing d-log integrands and computing master integrals for three-loop four-particle scattering}}, \href{https://doi.org/10.1007/JHEP04(2020)167}{\emph{JHEP} {\bfseries 04} (2020) 167} [\href{https://arxiv.org/abs/2002.09492}{{\ttfamily 2002.09492}}].

\bibitem{Chen:1977oja}
K.-T. Chen, \emph{{Iterated path integrals}}, \href{https://doi.org/10.1090/S0002-9904-1977-14320-6}{\emph{Bull. Am. Math. Soc.} {\bfseries 83} (1977) 831}.

\bibitem{Liu:2022chg}
X.~Liu and Y.-Q. Ma, \emph{{AMFlow: A Mathematica package for Feynman integrals computation via auxiliary mass flow}}, \href{https://doi.org/10.1016/j.cpc.2022.108565}{\emph{Comput. Phys. Commun.} {\bfseries 283} (2023) 108565} [\href{https://arxiv.org/abs/2201.11669}{{\ttfamily 2201.11669}}].

\bibitem{Goncharov:2010jf}
A.~B. Goncharov, M.~Spradlin, C.~Vergu and A.~Volovich, \emph{{Classical Polylogarithms for Amplitudes and Wilson Loops}}, \href{https://doi.org/10.1103/PhysRevLett.105.151605}{\emph{Phys. Rev. Lett.} {\bfseries 105} (2010) 151605} [\href{https://arxiv.org/abs/1006.5703}{{\ttfamily 1006.5703}}].

\bibitem{Chicherin:2017dob}
D.~Chicherin, J.~Henn and V.~Mitev, \emph{{Bootstrapping pentagon functions}}, \href{https://doi.org/10.1007/JHEP05(2018)164}{\emph{JHEP} {\bfseries 05} (2018) 164} [\href{https://arxiv.org/abs/1712.09610}{{\ttfamily 1712.09610}}].

\bibitem{Fevola:2023kaw}
C.~Fevola, S.~Mizera and S.~Telen, \emph{{Landau Singularities Revisited: Computational Algebraic Geometry for Feynman Integrals}}, \href{https://doi.org/10.1103/PhysRevLett.132.101601}{\emph{Phys. Rev. Lett.} {\bfseries 132} (2024) 101601} [\href{https://arxiv.org/abs/2311.14669}{{\ttfamily 2311.14669}}].

\bibitem{Fevola:2023fzn}
C.~Fevola, S.~Mizera and S.~Telen, \emph{{Principal Landau determinants}}, \href{https://doi.org/10.1016/j.cpc.2024.109278}{\emph{Comput. Phys. Commun.} {\bfseries 303} (2024) 109278} [\href{https://arxiv.org/abs/2311.16219}{{\ttfamily 2311.16219}}].

\bibitem{Jiang:2024eaj}
X.~Jiang, J.~Liu, X.~Xu and L.~L. Yang, \emph{{Symbol letters of Feynman integrals from Gram determinants}},  \href{https://arxiv.org/abs/2401.07632}{{\ttfamily 2401.07632}}.

\bibitem{Peraro:2019svx}
T.~Peraro, \emph{{$\text{FiniteFlow}$: multivariate functional reconstruction using finite fields and dataflow graphs}}, \href{https://doi.org/10.1007/JHEP07(2019)031}{\emph{JHEP} {\bfseries 07} (2019) 031} [\href{https://arxiv.org/abs/1905.08019}{{\ttfamily 1905.08019}}].

\bibitem{Hidding:2020ytt}
M.~Hidding, \emph{{DiffExp, a Mathematica package for computing Feynman integrals in terms of one-dimensional series expansions}}, \href{https://doi.org/10.1016/j.cpc.2021.108125}{\emph{Comput. Phys. Commun.} {\bfseries 269} (2021) 108125} [\href{https://arxiv.org/abs/2006.05510}{{\ttfamily 2006.05510}}].

\bibitem{CHMTYZ}
D.~Chicherin, J.~M. Henn, E.~Mazzucchelli, J.~Trnka, Q.~Yang and S.-Q. Zhang, \emph{In preparation}, .

\bibitem{Henn:2024qwe}
J.~Henn and P.~Raman, \emph{{Positivity properties of scattering amplitudes}},  \href{https://arxiv.org/abs/2407.05755}{{\ttfamily 2407.05755}}.

\bibitem{Henn:2024ngj}
J.~M. Henn, A.~Matija\v{s}i\'c, J.~Miczajka, T.~Peraro, Y.~Xu and Y.~Zhang, \emph{{A computation of two-loop six-point Feynman integrals in dimensional regularization}}, \href{https://doi.org/10.1007/JHEP08(2024)027}{\emph{JHEP} {\bfseries 08} (2024) 027} [\href{https://arxiv.org/abs/2403.19742}{{\ttfamily 2403.19742}}].

\bibitem{Abreu:2024flk}
S.~Abreu, D.~Chicherin, V.~Sotnikov and S.~Zoia, \emph{{Two-Loop Five-Point Two-Mass Planar Integrals and Double Lagrangian Insertions in a Wilson Loop}},  \href{https://arxiv.org/abs/2408.05201}{{\ttfamily 2408.05201}}.

\bibitem{Henn:2023pkc}
J.~M. Henn, M.~Lagares and S.-Q. Zhang, \emph{{Integrated negative geometries in ABJM}}, \href{https://doi.org/10.1007/JHEP05(2023)112}{\emph{JHEP} {\bfseries 05} (2023) 112} [\href{https://arxiv.org/abs/2303.02996}{{\ttfamily 2303.02996}}].

\bibitem{He:2023exb}
S.~He, C.-K. Kuo, Z.~Li and Y.-Q. Zhang, \emph{{Emergent unitarity, all-loop cuts and integrations from the ABJM amplituhedron}}, \href{https://doi.org/10.1007/JHEP07(2023)212}{\emph{JHEP} {\bfseries 07} (2023) 212} [\href{https://arxiv.org/abs/2303.03035}{{\ttfamily 2303.03035}}].

\bibitem{Lagares:2024epo}
M.~Lagares and S.-Q. Zhang, \emph{{Higher-loop integrated negative geometries in ABJM}}, \href{https://doi.org/10.1007/JHEP05(2024)142}{\emph{JHEP} {\bfseries 05} (2024) 142} [\href{https://arxiv.org/abs/2402.17432}{{\ttfamily 2402.17432}}].

\bibitem{Li:2024lbw}
Z.~Li, \emph{{Integrating the full four-loop negative geometries and all-loop ladder-type negative geometries in ABJM theory}},  \href{https://arxiv.org/abs/2402.17023}{{\ttfamily 2402.17023}}.

\bibitem{Abreu:2018aqd}
S.~Abreu, L.~J. Dixon, E.~Herrmann, B.~Page and M.~Zeng, \emph{{The two-loop five-point amplitude in $\mathcal{N} =4$ super-Yang-Mills theory}}, \href{https://doi.org/10.1103/PhysRevLett.122.121603}{\emph{Phys. Rev. Lett.} {\bfseries 122} (2019) 121603} [\href{https://arxiv.org/abs/1812.08941}{{\ttfamily 1812.08941}}].

\bibitem{Chicherin:2018yne}
D.~Chicherin, T.~Gehrmann, J.~M. Henn, P.~Wasser, Y.~Zhang and S.~Zoia, \emph{{Analytic result for a two-loop five-particle amplitude}}, \href{https://doi.org/10.1103/PhysRevLett.122.121602}{\emph{Phys. Rev. Lett.} {\bfseries 122} (2019) 121602} [\href{https://arxiv.org/abs/1812.11057}{{\ttfamily 1812.11057}}].

\bibitem{Chicherin:2019xeg}
D.~Chicherin, T.~Gehrmann, J.~M. Henn, P.~Wasser, Y.~Zhang and S.~Zoia, \emph{{The two-loop five-particle amplitude in $ \mathcal{N} $ = 8 supergravity}}, \href{https://doi.org/10.1007/JHEP03(2019)115}{\emph{JHEP} {\bfseries 03} (2019) 115} [\href{https://arxiv.org/abs/1901.05932}{{\ttfamily 1901.05932}}].

\bibitem{Abreu:2019rpt}
S.~Abreu, L.~J. Dixon, E.~Herrmann, B.~Page and M.~Zeng, \emph{{The two-loop five-point amplitude in $ \mathcal{N} $ = 8 supergravity}}, \href{https://doi.org/10.1007/JHEP03(2019)123}{\emph{JHEP} {\bfseries 03} (2019) 123} [\href{https://arxiv.org/abs/1901.08563}{{\ttfamily 1901.08563}}].

\bibitem{Caron-Huot:2020vlo}
S.~Caron-Huot, D.~Chicherin, J.~Henn, Y.~Zhang and S.~Zoia, \emph{{Multi-Regge Limit of the Two-Loop Five-Point Amplitudes in $\mathcal{N} = 4$ Super Yang-Mills and $\mathcal{N} = 8$ Supergravity}}, \href{https://doi.org/10.1007/JHEP10(2020)188}{\emph{JHEP} {\bfseries 10} (2020) 188} [\href{https://arxiv.org/abs/2003.03120}{{\ttfamily 2003.03120}}].

\bibitem{wasow1965asymptotic}
W.~Wasow, \emph{Asymptotic expansions for ordinary differential equations}, Pure and Applied Mathematics, Vol. XIV. Interscience Publishers John Wiley \& Sons, Inc., New York-London-Sydney, 1965.

\bibitem{Remiddi:1999ew}
E.~Remiddi and J.~A.~M. Vermaseren, \emph{{Harmonic polylogarithms}}, \href{https://doi.org/10.1142/S0217751X00000367}{\emph{Int. J. Mod. Phys. A} {\bfseries 15} (2000) 725} [\href{https://arxiv.org/abs/hep-ph/9905237}{{\ttfamily hep-ph/9905237}}].

\end{thebibliography}\endgroup
\end{document}